\documentclass[aps,pra,reprint,showkeys,amssymb,floatfix,longbibliography,superscriptaddress]{revtex4-2}
\usepackage{graphicx}
\usepackage{amsmath,amsfonts,amssymb}
\usepackage[colorlinks=true,urlcolor=blue,linkcolor=blue,citecolor=blue,breaklinks=true,pdfauthor={Hannes Lagemann}]{hyperref}
\usepackage{braket}
\usepackage[svgnames]{xcolor}
\usepackage{tikz}
\usepackage{color}
\usepackage{enumerate}
\usepackage[caption=false]{subfig}
\usepackage{multirow}
\usepackage{qcircuit}
\usepackage[normalem]{ulem}
\usepackage{placeins}
\usepackage{url}
\usepackage{dsfont}
\usepackage{braket}
\usepackage[multidot]{grffile}
\usepackage{float}
\usepackage{lipsum}
\usepackage{afterpage}
\usepackage{epstopdf}
\usepackage{silence}
\usepackage{longtable}
\usetikzlibrary{decorations.markings}
\usetikzlibrary{shapes.geometric}
\pgfdeclarelayer{edgelayer}
\pgfdeclarelayer{nodelayer}
\pgfsetlayers{edgelayer,nodelayer,main}

\tikzstyle{none}=[inner sep=0pt]

\tikzstyle{rn}=[circle,fill=Red,draw=Black,line width=0.8 pt]
\tikzstyle{gn}=[circle,fill=Lime,draw=Black,line width=0.8 pt]
\tikzstyle{yn}=[circle,fill=Yellow,draw=Black,line width=0.8 pt]

\tikzstyle{simple}=[-,draw=Black,line width=2.000]
\tikzstyle{arrow}=[-,draw=Black,postaction={decorate},decoration={markings,mark=at position .5 with {\arrow{>}}},line width=2.000]
\tikzstyle{tick}=[-,draw=Black,postaction={decorate},decoration={markings,mark=at position .5 with {\draw (0,-0.1) -- (0,0.1);}},line width=2.000]

\tikzstyle{newstyle}=[
  -,draw=Green,line width=2.000]

\newcommand{\appref}[1]{Appendix~\ref{#1}}

\newcommand{\secref}[1]{Sec.~\ref{#1}}
\newcommand{\secaref}[2]{Secs.~\ref{#1} and \ref{#2}}
\newcommand{\ssecref}[1]{Section~\ref{#1}}
\newcommand{\figref}[1]{Fig.~\ref{#1}}
\newcommand{\figsref}[1]{Figs.~\ref{#1}}
\newcommand{\tabref}[1]{Table~\ref{#1}}
\newcommand{\tabsref}[2]{Tables~\ref{#1} and \ref{#2}}
\newcommand{\equref}[1]{Eq.~\eqref{#1}}
\newcommand{\equaref}[2]{Eqs.~\eqref{#1} and \eqref{#2}}
\newcommand{\equsref}[2]{Eqs.~\eqref{#1}--\eqref{#2}}

\newcommand{\REF}{Ref.~}
\newcommand{\REFS}{Refs.~}

\newcommand{\ie}{i.e.~}
\newcommand{\eg}{e.g.~}
\newcommand{\listcite}{\cite{McKay16,Yan18,Rol19,Blais2020circuit,Ganzhorn20,Gu21}}
\newcommand{\listciteone}{\cite{McKay16,Roth19,Ganzhorn20,Gu21}}
\newcommand{\listcitetwo}{\cite{Roth19,Roth20,Ganzhorn20,McKay16,Ganzhorn20,Gu21,Baker22}}
\newcommand{\listadia}{\cite{McKay16,Yan18,Roth19,Gu21,Baker22}}
\newcommand{\archone}{\cite{McKay16,Roth19,Ganzhorn20,Bengtsson2020,Gu21}}
\newcommand{\archtwo}{\cite{Rol19,Blais2020circuit,Lacroix2020,Krinner2020}}
\newcommand{\width}{0.875}

\newcommand{\change}[1]{\textcolor{black}{#1}}

\begin{document}
\title{Numerical analysis of effective models for flux-tunable transmon systems}

\author{H. Lagemann}
\affiliation{Institute for Advanced Simulation,
  J\"ulich Supercomputing Centre,\\
  Forschungszentrum J\"ulich, D-52425 J\"ulich, Germany}
\affiliation{RWTH Aachen University, D-52056 Aachen, Germany}
\author{D. Willsch}
\affiliation{Institute for Advanced Simulation,
  J\"ulich Supercomputing Centre,\\
  Forschungszentrum J\"ulich, D-52425 J\"ulich, Germany}
\author{M. Willsch}
\affiliation{Institute for Advanced Simulation,
  J\"ulich Supercomputing Centre,\\
  Forschungszentrum J\"ulich, D-52425 J\"ulich, Germany}
\author{F. Jin}
\affiliation{Institute for Advanced Simulation,
  J\"ulich Supercomputing Centre,\\
  Forschungszentrum J\"ulich, D-52425 J\"ulich, Germany}
\author{H. De Raedt}
\affiliation{Institute for Advanced Simulation,
  J\"ulich Supercomputing Centre,\\
  Forschungszentrum J\"ulich, D-52425 J\"ulich, Germany}
\affiliation{Zernike Institute for Advanced Materials,\\
University of Groningen, Nijenborgh 4, NL-9747 AG Groningen, The Netherlands}
\author{K. Michielsen}
\affiliation{Institute for Advanced Simulation,
  J\"ulich Supercomputing Centre,\\
  Forschungszentrum J\"ulich, D-52425 J\"ulich, Germany}
\affiliation{RWTH Aachen University, D-52056 Aachen, Germany}

\date{\today}
\begin{abstract}
Simulations and analytical calculations that aim to describe flux-tunable transmons are usually based on effective models of the corresponding lumped-element model. However, when a control pulse is applied, in most cases it is not known how much the predictions made with the effective models deviate from the predictions made with the original lumped-element model. In this work we compare the numerical solutions of the time-dependent Schrödinger equation for both the effective and the lumped-element models, for microwave and unimodal control pulses (external fluxes). These control pulses are used to model single-qubit (X) and two-qubit gate (Iswap and Cz) transitions. First, we derive a non-adiabatic effective Hamiltonian for a single flux-tunable transmon and compare the pulse response of this model to the one of the corresponding circuit Hamiltonian. Here we find that both models predict similar outcomes for similar control pulses. Then, we study how different approximations affect single-qubit (X) and two-qubit gate (Iswap and Cz) transitions in two different two-qubit systems. For this purpose we consider three different systems in total: a single flux-tunable transmon and two two-qubit systems. In summary, we find that a series of commonly applied approximations (individually and/or in combination) can change the response of a system substantially, when a control pulse is applied.
\end{abstract}
\maketitle

\section{Introduction}
The successful construction of a fully functioning universal quantum computer comes with the promise of allowing us to solve certain computational problems faster (potentially exponentially faster) than with a classical computer. However, the construction of a universal quantum computer comes with its own challenges, i.e.~the task to understand the dynamic behaviour of quantum systems.

Many experimental prototypes, which aim to realise a universal quantum computer, are based on superconducting circuits. Theoretical descriptions of these systems often use a so-called circuit Hamiltonian model. Here we make a lumped-element approximation \cite[Section 1.4]{Balanis12} to derive a Hamiltonian, see for example Ref.~\cite{DV97}, which approximately describes the behaviour of a particular superconducting circuit.

Unfortunately, it is usually the case that the circuit Hamiltonian model is still too complicated to be treated analytically. Therefore, in most cases additional simplifications are made so that an approximant of the circuit Hamiltonian can be derived. These approximants usually do not come with an estimation of the corresponding approximation error.

In this work we numerically study several instances of such approximants, i.e. effective Hamiltonians, by comparing them to their circuit Hamiltonian counterparts. To this end, we solve the time-dependent Schrödinger equation (TDSE) for both models. This allows us to compare the corresponding solutions and to filter out differences. Furthermore, we also compare the spectra of selected models, see \appref{sec: Effective Hamiltonians appendix}.

Since the number of different superconducting circuits is vast, we will focus on three different circuit Hamiltonians and their corresponding effective Hamiltonians. \change{Two of these Hamiltonians are designed to model existing experimental systems; see \REFS\cite{Ganzhorn20,Lacroix2020}.} Note that for a particular circuit Hamiltonian there might exist a vast amount of different effective models.

\renewcommand{\width}{1.0}
\begin{figure}[!tbp]
  \begin{minipage}{0.45\textwidth}
    \centering
    \includegraphics[width=\width\textwidth]{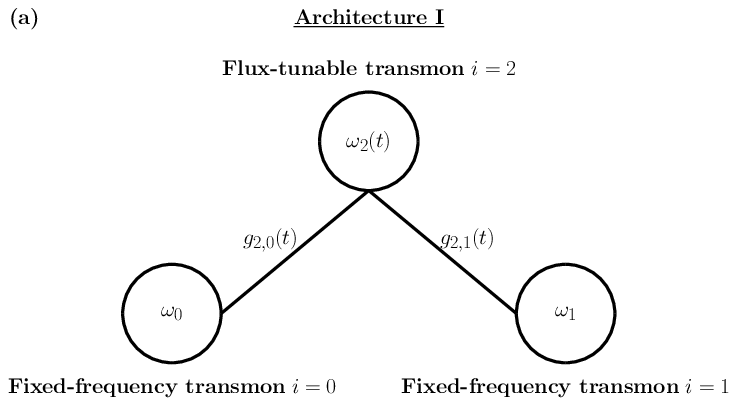} %
  \end{minipage}

  \vspace{1 cm}

  \begin{minipage}{0.45\textwidth}
    \centering
    \includegraphics[width=\width\textwidth]{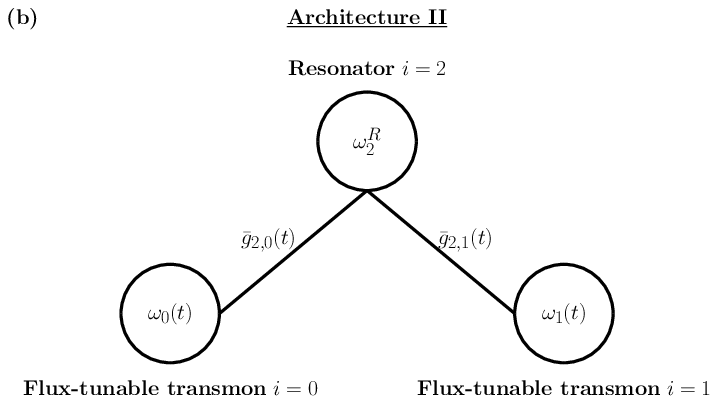} %
  \end{minipage}
  \caption{Sketches of the circuit architectures I(a) and II(b). Both types of architectures use flux-tunable transmon qubits to activate two-qubit gate transitions, see \REF\cite{DiCarlo09} for (a) and \REF\cite{McKay16} for (b). We use the circuit and effective Hamiltonians given by \equaref{eq:architecture I}{eq:architecture I effective} and the device parameters listed in \tabsref{tab:device_parameter_flux_tunable_coupler_chip}{tab:device_parameter_flux_tunable_coupler_chip_effective} to perform simulations of Iswap and Cz two-qubit gate transitions for architecture I. Similarly, we use the circuit and effective Hamiltonians given by \equaref{eq:architecture II}{eq:architecture II effective} and the device parameters listed in \tabsref{tab:device_parameter_resonator_coupler_chip}{tab:device_parameter_resonator_coupler_chip_effective} to perform simulations of Iswap and Cz two-qubit gate transitions for architecture II.\label{fig:arch_sketch}}
\end{figure}

\change{Considering only systems based on transmon qubits}, one might divide the different circuit architectures into two categories: architectures which only use fixed-frequency transmon qubits\change{, an architecture which is primarily studied by IBM,} and those using flux-tunable transmon qubits to implement their two-qubit gates. In this work, we focus on circuits which use flux-tunable transmons to implement two-qubit gates. Additionally, we restrict our analysis to systems which only contain one or two qubits, as this suffices to show where the models deviate from each other.

We look at three different systems. The first system is a single flux-tunable transmon. The second system\change{, architecture I,} consists of two fixed-frequency transmons, coupled to a flux-tunable transmon. \change{The flux-tunable transmon works as a coupler only, see \figref{fig:arch_sketch}(a).} The third system\change{, architecture II,} is made up of two flux-tunable transmons, coupled to a transmission line resonator. \change{Here the resonator functions only as a coupler element, see \figref{fig:arch_sketch}(b).}

This work is structured as follows. In \secref{sec:Circuit Hamiltonians} we introduce the different circuit Hamiltonian models for the three systems we simulate in this work. Next, in \secref{sec:FromCircuitToEffective Hamiltonians} we derive a non-adiabatic effective Hamiltonian for a single flux-tunable transmon. Then, in \secref{sec:Effective Hamiltonians}, we define two effective two-qubit Hamiltonians, one for architecture I and one for architecture II. The flux-tunable transmons in both these systems are modelled with the adiabatic approximation. In \secref{sec:control pulse} we define a simple control pulse (for the external flux) which allows us to model microwave and unimodal pulses. This pulse (the external flux) enables us to activate resonant and non-adiabatic transitions between the states of the systems we consider. \ssecref{sec:Results} contains the main results of this work. First, in \secref{sec: Revised derivation of the effective Hamiltonian for flux-tunable transmons}, we study a single flux-tunable transmon. Here we model resonant transitions activated by microwave pulses and non-adiabatic transitions activated by unimodal pulses with the circuit Hamiltonian model and the non-adiabatic effective model we derive in \secref{sec:FromCircuitToEffective Hamiltonians}. Here we compare how well both models agree with one another. Next, in \secref{sec:Single-qubit operations effective}, we identify several transitions (interactions) which are suppressed in the effective model for architecture I, see \figref{fig:arch_sketch}(a), by the adiabatic approximation we use to model the flux-tunable transmons in the effective two-qubit model. Finally, in \secref{sec: two-qubit operations effective}, we study two-qubit gate transitions, \ie transitions which can be used to implement two-qubit gates \change{with the corresponding architectures I and II}. For architecture I we use a microwave pulse to activate resonant two-qubit Iswap and Cz interactions. Similarly, for architecture II we use a unimodal pulse to activate non-adiabatic transitions which allow us to model Iswap and Cz gates. Here we focus on the often made approximation, see \REFS\listciteone, that the effective interaction strength is of static nature. Additionally, we test whether or not non-adiabatic effects play a role. A summary and conclusions drawn from our analysis are presented in \secref{sec:SummaryAndConclusions}.

\change{To assist the reader in navigating through the material, we list the main findings:}
\begin{enumerate}
  \item \change{We provide a derivation of a non-adiabatic effective Hamiltonian for flux-tunable transmons, see \secref{sec:FromCircuitToEffective Hamiltonians}. Furthermore, we compare the dynamics of the adiabatic and the non-adiabatic effective model with the ones of the associated circuit model by solving the TDSE for the different model Hamiltonians. This is done for a single flux-tunable transmon, see \secref{sec: Revised derivation of the effective Hamiltonian for flux-tunable transmons} and two two-qubit systems, see \figref{fig:arch_sketch}(a-b) and \secref{sec: two-qubit operations effective}.}

  \item \change{We identify transitions which are suppressed in the adiabatic effective two-qubit model for the system illustrated in \figref{fig:arch_sketch}(a), see \secref{sec:Single-qubit operations effective}.}

  \item \change{We show that neglecting nominal small time-dependent oscillations of the interaction strength in an effective model for the two-qubit system shown in \figref{fig:arch_sketch}(a), can lead to substantial shifts in the duration of a control pulse, see \secref{sec:two-qubitArchitectureI}. Conversely, we also show that neglecting a nominal much larger time-dependent square pulse like reduction of the interaction strength in an effective model for the two-qubit system shown in \figref{fig:arch_sketch}(b), can lead to modest shifts in the duration of a control pulse, see \secref{sec:two-qubitArchitectureII}.}
\end{enumerate}



\section{Model}\label{sec: Model}
In this section, we introduce the circuit Hamiltonian models, see \secref{sec:Circuit Hamiltonians}, we derive effective Hamiltonians for a single fixed-frequency and a single flux-tunable transmon, see \secref{sec:FromCircuitToEffective Hamiltonians}, and we discuss effective Hamiltonian models which are commonly used to model two-qubit systems, see \secref{sec:Effective Hamiltonians}. Furthermore, in \secref{sec:control pulse} we define a control pulse which can be used to implement single- and two-qubit gate transitions with an external flux. \change{Note that throughout this work we use $\hbar=1$.}

\subsection{Circuit Hamiltonians}\label{sec:Circuit Hamiltonians}
The systems we model in this work consist of two different types of transmons as well as transmission line resonators. The couplings between the different subsystems are modelled as dipole-dipole interactions.

The first type of transmon is the so-called fixed-frequency transmon (see \REF\cite{Koch}), described by the circuit Hamiltonian
\begin{equation}\label{eq:fixed-frequency transmon}
 \hat{H}_{\text{Fix}} = E_{C} \hat{n}^2 - E_{J} \cos(\hat{\varphi}),
\end{equation}
where $E_{C}$ denotes the capacitive energy and $E_{J}$ is the Josephson energy. The Hamiltonian is defined in terms of the charge $\hat{n}$ and the flux $\hat{\varphi}$ operators. Note that the Hamiltonian in \equref{eq:fixed-frequency transmon} is often expressed with a factor $4 E_{C}$ instead of $E_{C}$. In this work we
adopt the convention used in \REF\cite{Wi17} and not the one of \REF\cite{Koch}.

The second type of transmon is the flux-tunable transmon. This type of transmon is the main object of our investigation. If the capacitances of both Josephson junctions are equal, we can define the corresponding circuit Hamiltonian which fulfils the irrotational constraint (cf.~\cite{You,Riwar2022CircuitQuantizationTimeDependentMagneticField}) as
\begin{equation}\label{eq:flux-tunable transmon}
 \hat{H}_{\text{Tun}} = E_{C} \hat{n}^2 - E_{J,1} \cos(\hat{\varphi}+\frac{\varphi(t)}{2}) - E_{J,2} \cos(\hat{\varphi}-\frac{\varphi(t)}{2}).
\end{equation}
This system is characterised by two Josephson energies $E_{J,1}$ and $E_{J,2}$ and another time-dependent variable $\varphi(t)$, which represents an external flux. This external flux is dimensionless
\begin{equation}\label{eq:external_flux}
  \varphi(t)=\Phi(t)/\phi_{0},
\end{equation}
where $\Phi(t)$ has the dimension of flux and $\phi_{0}$ is the flux quantum. Furthermore, since the Hamiltonian is $2 \pi$ periodic, $\varphi(t)$ is usually given in units of $2 \pi$. We adopt this convention too.

The circuit Hamiltonians in \equaref{eq:fixed-frequency transmon}{eq:flux-tunable transmon} are usually only referred to as transmons if $E_{J}/E_{C} \gg 1$ and $(E_{J,1}+E_{J,2})/E_{C} \gg 1$. Therefore, in this work, we assume that this is true.

Individual transmons can be coupled directly, or indirectly, or both. In this paper, we only consider indirect couplings. This means interactions between individual transmons are conveyed by an additional circuit element, often called a coupler. This coupler can be a transmon itself or a transmission line resonator.

Transmission line resonators are described by the Hamiltonian
\begin{equation}\label{eq: transmission line resonators}
  \hat{H}_{\text{Res}}=\omega^{R} \hat{a}^{\dagger} \hat{a},
\end{equation}
where $\omega^{R}$ is the resonator frequency. The operators $\hat{a}^{\dagger}$ and $\hat{a}$ are the bosonic number operators.

We describe the dipole-dipole coupling between two arbitrary transmons $i$ and $j$ by means of the interaction operator
\begin{equation}\label{eq:int_I}
\hat{V}_{i,j} = G_{i,j} \hat{n}_{i} \hat{n}_{j},
\end{equation}
where $G_{i,j}$ is the interaction strength. Similarly, we model the coupling between an arbitrary resonator $j$ and an arbitrary transmon $i$ with the operator
\begin{equation}\label{eq:int_II}
\hat{W}_{j,i} = G_{j,i} (\hat{a}^{\dagger}+\hat{a})_{j} \hat{n}_{i}.
\end{equation}

We can use the different subsystems and the corresponding interaction terms to construct different circuit architectures. In this work, we consider two different architectures, which use flux-tunable transmons to implement the Iswap and Cz two-qubit gates. Architecture I, which is discussed in \REFS\archone, is described by the circuit Hamiltonian
\begin{equation}\label{eq:architecture I}
  \hat{H}_{I}=\hat{H}_{\text{Fix},0}+\hat{H}_{\text{Fix},1}+\hat{H}_{\text{Tun},2}+\hat{V}_{2,1}+\hat{V}_{2,0},
\end{equation}
and architecture II, which is discussed in \REFS\archtwo, is described by
\begin{equation}\label{eq:architecture II}
  \hat{H}_{II}=\hat{H}_{\text{Tun},0}+\hat{H}_{\text{Tun},1}+H_{\text{Res},2}+\hat{W}_{2,1}+\hat{W}_{2,0}.
\end{equation}
In the first case, we use a flux-tunable transmon to indirectly couple two fixed-frequency transmons, see \figref{fig:arch_sketch}(a). In the second case, we use a resonator as a coupler between two flux-tunable transmons, see \figref{fig:arch_sketch}(b). The device parameters that we use in our simulations to obtain the results in \secref{sec:Results}, are listed in \tabref{tab:device_parameter_flux_tunable_coupler_chip} for architecture I and \tabref{tab:device_parameter_resonator_coupler_chip} for architecture II, respectively.

\begin{table}[!tbp]
\caption{\label{tab:device_parameter_flux_tunable_coupler_chip} Device parameters for a tunable coupler architecture\change{, \ie architecture I.} \change{Note that throughout this work we use $\hbar=1$}. The parameter $\omega=E_{1}-E_{0}$ denotes the qubit frequency and $\alpha=(E_{2}-E_{1})-(E_{1}-E_{0})$ is the so-called qubit anharmonicity. All (angular) frequencies are in GHz except the flux offset parameter $\varphi_{0}=\varphi(0)$ which is given in units of the flux quantum $\phi_{0}$, see \equaref{eq:external_flux}{eq:pulse}. These parameters are motivated by experiments performed by the authors of \REF\cite{Ganzhorn20}.}
\begin{ruledtabular}
\begin{tabular}{ccccccccc}
$i$ &$\omega/2 \pi$&$\alpha/2\pi$ &$E_{C}$&$E_{J,1}$&$E_{J,2}$ & $\varphi_{0}/2\pi$ & $G_{2,i}/2\pi$\\
\hline
0 & 5.100 & -0.310 & 6.777 & 84.482  & n/a & n/a & 0.085\\
1 & 6.200 & -0.285 & 6.453 & 127.992  & n/a & n/a & 0.085\\
2 & 8.100 & -0.235 & 5.529 & 112.450  & 134.999 & 0.15 & n/a\\
\end{tabular}
\end{ruledtabular}
\caption{\label{tab:device_parameter_resonator_coupler_chip} Device parameters for architecture II in the same units as the parameters in \tabref{tab:device_parameter_flux_tunable_coupler_chip}. These device parameters are motivated by experiments performed by the authors of \REF\cite{Lacroix2020}.}
\begin{ruledtabular}
\begin{tabular}{ccccccccc}
$i$ & $\omega^{R}/2 \pi$ &$\omega/2 \pi$&$\alpha/2\pi$ &$E_{C}$&$E_{J,1}$&$E_{J,2}$ & $\varphi_{0}/2\pi$ & $G_{2,i}/2\pi$\\
\hline
0 & n/a & 4.200& -0.320 & 6.712 & 19.728  & 59.184 & 0 & 0.300\\
1 & n/a & 5.200& -0.295 & 6.512 & 30.265  & 60.529 & 0 & 0.300\\
2 & 45.000 & n/a & n/a & n/a & n/a & n/a & n/a & n/a\\
\end{tabular}
\end{ruledtabular}
\end{table}

\subsection{From circuit to effective Hamiltonians}\label{sec:FromCircuitToEffective Hamiltonians}
\change{In this section, we provide the derivation of a non-adiabatic effective Hamiltonian for flux-tunable transmons. A more detailed discussion, written for readers who are unfamiliar with transmon qubits, is given in \appref{app:DiscussionOfCosineExpansionArgument}.}

In case of the fixed-frequency transmon, we use the harmonic basis states
\begin{equation}
  \mathcal{B}=\{\ket{m}\}_{m\in \mathbb{N}},
\end{equation}
to model the dynamics of the system with an effective Hamiltonian. First, we expand the cosine in \equref{eq:fixed-frequency transmon} to the quartic order. Then, we decompose the term
\begin{equation}
  \frac{E_{J}}{4!}\hat{\varphi}^{4}=\frac{E_{C}}{48}\left(\hat{D}+\hat{V}\right),
\end{equation}
into a part $\hat{D}$ which is diagonal in the basis $\mathcal{B}$ and one $\hat{V}$ which is \change{off-diagonal} in $\mathcal{B}$. We use the diagonal part to define the Hamiltonian
\begin{equation}\label{eq:fixed-frequency transmon eff def}
  \hat{H}_{\text{fix}}=\omega \hat{b}^{\dagger} \hat{b} -\frac{E_{C}}{48} \hat{D},
\end{equation}
where $\omega=\sqrt{2 E_{C} E_{J}}$. Here $\hat{b}^{\dagger}$ and $\hat{b}$ are the bosonic number operators which can be defined in terms of their action on the basis states $\ket{m} \in \mathcal{B}$. The Hamiltonian can be expressed as
\begin{equation}\label{eq:fixed-frequency transmon eff}
  \hat{H}_{\text{fix}}=\omega^{\prime} \hat{b}^{\dagger} \hat{b} + \frac{\alpha}{2} \hat{b}^{\dagger} \hat{b} \left(\hat{b}^{\dagger} \hat{b} -\hat{I}\right),
\end{equation}
where $\omega^{\prime}=\sqrt{2 E_{C} E_{J}} + \alpha$ denotes the transmon qubit frequency and $\alpha=-E_{C}/4$ is referred to as the transmon's anharmonicity. The spectrum of the Hamiltonian in \equref{eq:fixed-frequency transmon eff} is in agreement, up to a constant factor, with the results in \REF\cite[Appendix C]{Koch}. The corresponding results are obtained by means of time-independent perturbation theory. Note that the derivation of \equref{eq:fixed-frequency transmon eff} provided in this section is similar but not equivalent to the one presented in \REF\cite[Section B 4.1.3]{DiVincenzo13}.

For the flux-tunable transmon, one can make use of the fact that the Hamiltonian given by \equref{eq:flux-tunable transmon} can be expressed as
\begin{equation}\label{eq:flux-tunable transmon recast}
 \hat{H}_{\text{Tun}} = E_{C} \hat{n}^2 -  E_{J,\text{eff}}(t) \cos(\hat{\varphi}-\varphi_{\text{eff}}(t)),
\end{equation}
with the effective Josephson energy
\begin{equation}\label{eq: Josephson energy flux}
  E_{J,\text{eff}}(t)=E_{\Sigma} \sqrt{\cos\left(\frac{\varphi(t)}{2}\right)^{2}+d^{2} \sin\left(\frac{\varphi(t)}{2}\right)^{2}},
\end{equation}
and the effective external flux
\begin{equation}
  \varphi_{\text{eff}}(t)=\arctan\left(d \tan\left(\frac{\varphi(t)}{2}\right)\right).
\end{equation}
Here, we introduced the new parameters $E_{\Sigma}=(E_{J,1}+E_{J,2})$ and $d=(E_{J,2}-E_{J,1})/(E_{J,2}+E_{J,1})$. The latter one is usually referred to as the asymmetry factor, see \REF\cite{Koch}.

We want to repeat the quartic-order cosine expansion argumentation that we provided for the fixed-frequency transmon. However, since there is a time dependence in the cosine function in \equref{eq:flux-tunable transmon recast}, we need to use the time-dependent harmonic basis states
\begin{equation}
  \mathcal{B}(t)=\{\ket{m(t)}\}_{m\in \mathbb{N}},
\end{equation}
to model the dynamics of the system. The TDSE for the state vector
\begin{equation}\label{eq:td_basis}
  \ket{\Psi^{*}(t)}=\hat{\mathcal{W}}(t)\ket{\Psi(t)},
\end{equation}
where $\hat{\mathcal{W}}(t)$ denotes the unitary transformation which maps the basis states $\mathcal{B}(0)$ to the basis states $\mathcal{B}(t)$, only stays form invariant, if we use the transformed Hamiltonian
\begin{equation}\label{eq:HamTrafo_main}
  \hat{H}_{\text{tun}}^{*}(t)=\hat{\mathcal{W}}(t)\hat{H}(t)\hat{\mathcal{W}}^{\dagger}(t) - i\hat{\mathcal{W}}(t)\partial_{t}\hat{\mathcal{W}}^{\dagger}(t).
\end{equation}
Here $\hat{H}(t)$ denotes the fourth-order Hamiltonian which is diagonal in the basis $\mathcal{B}(t)$, \ie we expand the cosine in \equref{eq:flux-tunable transmon recast} to the quartic order and only keep the contributions which are diagonal in the basis $\mathcal{B}(t)$, as for the fixed-frequency transmons in the basis $\mathcal{B}$. Therefore, we can determine the first term in \equref{eq:HamTrafo_main} to be
\begin{equation}\label{eq:diagonal_main}
\hat{\mathcal{W}}(t)\hat{H}(t)\hat{\mathcal{W}}^{\dagger}(t)=\omega^{\prime}(t) \hat{b}^{\dagger} \hat{b} + \frac{\alpha}{2} \hat{b}^{\dagger} \hat{b} \left(\hat{b}^{\dagger} \hat{b}-\hat{I}\right),
\end{equation}
where $\omega^{\prime}(t)=\omega(t)+\alpha$ and
\begin{equation}\label{eq:tunable frequency}
  \omega(t)=\sqrt{2 E_{C} E_{\Sigma}} \sqrt[4]{\cos\left(\frac{\varphi(t)}{2}\right)^{2} + d^{2} \sin\left(\frac{\varphi(t)}{2}\right)^{2}},
\end{equation}
denotes the tunable frequency.

We can make use of the fact that the harmonic basis states $\mathcal{B}(t)$ can be expressed analytically in the $\varphi$-space, this enables us to determine the second term in \equref{eq:HamTrafo_main}. The result reads
\begin{equation}\label{eq:basis_trafo_main}
      \begin{split}
     \change{- i\hat{\mathcal{W}}(t)\partial_{t}\hat{\mathcal{W}}^{\dagger}(t)}=&-i \sqrt{\frac{\xi(t)}{2}} \dot{\varphi_{\text{eff}}}(t)  \left(\hat{b}^{\dagger}-\hat{b}\right) \\
     &+\frac{i}{4}\frac{\dot{\xi}(t)}{\xi(t)}\left(\hat{b}^{\dagger}\hat{b}^{\dagger}-\hat{b}\hat{b}\right),
     \end{split}
\end{equation}
where $\xi(t)=\sqrt{E_{J,\text{eff}}(t)/(2 E_{C})}$ and we assume that $\xi(t)\neq0$ for all times $t$. Additionally, we find
\begin{equation}
  \dot{\varphi_{\text{eff}}}(t)=\dot{\varphi}(t)\frac{d}{2 \left(\cos\left(\frac{\varphi(t)}{2}\right)^{2}+d^{2} \sin\left(\frac{\varphi(t)}{2}\right)^{2}\right)}
\end{equation}
and
\begin{equation}
  \frac{\dot{\xi}(t)}{\xi(t)}=\dot{\varphi}(t)\frac{(d^{2}-1) \sin(\varphi(t))}{8 \left(\cos\left(\frac{\varphi(t)}{2}\right)^{2}+d^{2} \sin\left(\frac{\varphi(t)}{2}\right)^{2}\right)},
\end{equation}
so that the first (second) drive term in \equref{eq:basis_trafo_main} disappears if $d=0$ ($d=1$). Consequently, we see that both drive terms in \equref{eq:basis_trafo_main} are not necessarily periodic in $\varphi(t)$, see the factor $\dot{\varphi}(t)$.

So far we did not discuss whether or not it is justified to drop the higher-order terms in the cosine expansion. We investigate this question in \secref{sec: Revised derivation of the effective Hamiltonian for flux-tunable transmons}, \ie we compare the results for the effective Hamiltonian model with the ones of the circuit Hamiltonian model by solving the TDSE for both Hamiltonians numerically.

\subsection{Effective Hamiltonians}\label{sec:Effective Hamiltonians}
The circuit Hamiltonian in \equref{eq:fixed-frequency transmon} for a fixed-frequency transmon was analytically discussed by the authors of \REF\cite{Koch}. This work motivated several studies, see for example \REFS\listcite, where fixed-frequency and/or flux-tunable transmons are modelled as anharmonic oscillators with fixed or tunable frequencies.

In practice, only a few basis states are used to model the dynamics of a transmon. Furthermore, presumably for simplicity one often uses a simpler choice for the parametrisation of the model. For the fixed-frequency transmon the corresponding effective Hamiltonian can be expressed as
\begin{equation}\label{eq:fixed-frequency transmon effective}
  \hat{H}_{\text{fix}}=\sum_{m=0,1,2,3} \left(m \omega + \frac{\alpha}{2} m(m-1)\right) \ket{m}\!\bra{m},
\end{equation}
where the qubit frequency $\omega=(E^{(1)}-E^{(0)})$ and anharmonicity $\alpha=(E^{(2)}-E^{(0)}) - 2 \omega$ might be directly fitted to the first and second energy gaps. Obviously, this approach is preferable when detailed knowledge of the capacitive and Josephson energies is not available.

Similarly, in practice flux-tunable transmons are often modelled with the effective Hamiltonian
\begin{equation}\label{eq:flux-tunable transmon effective}
  \hat{H}_{\text{tun}}(t)=\sum_{m=0,1,2,3} \left(m \omega(t) + \frac{\alpha}{2} m(m-1)\right) \ket{m}\!\bra{m},
\end{equation}
where $\omega(t)$ is given by \equref{eq:tunable frequency}. In this model, the parameters $\omega(0)$, $\alpha$ and $d$ are used to characterise the flux-tunable transmon qubit. We emphasise that using the tunable frequency given by \equref{eq:tunable frequency} to approximate the spectrum of the circuit Hamiltonian in \equref{eq:flux-tunable transmon} does not always lead to accurate results. We explore this issue in \appref{sec: Effective Hamiltonians appendix}.

\change{The Hamiltonian in \equref{eq:flux-tunable transmon effective} is often stated with reference to \REF\cite{Koch} but there is no mentioning of non-adiabatic effects, see for example \REFS\listadia. Furthermore, note that in \equref{eq:flux-tunable transmon effective} the time dependence of the basis states is not made explicit, this seems to be common practice when working with this model.} We simply state the Hamiltonian in \equref{eq:flux-tunable transmon effective} and do not advocate its use. In fact, we are interested in the question to what extent this effective Hamiltonian deviates from its circuit Hamiltonian counterpart in \equref{eq:flux-tunable transmon} and the effective Hamiltonian given by \equref{eq:HamTrafo_main}, see \secref{sec: Revised derivation of the effective Hamiltonian for flux-tunable transmons}.

The Hamiltonian in \equref{eq:flux-tunable transmon effective} is so simple that we can determine the formal solution of the TDSE for all pulses $\varphi(t)$. If we initialise the system in some arbitrary state
\begin{equation}
  \ket{\Psi^{\text{tun}}(t_{0})}=\sum_{m =0, 1, 2, 3} c_{m}(t_{0}) \ket{m},
\end{equation}
we obtain
\begin{equation}\label{eq:statevectorAC}
  \ket{\Psi^{\text{tun}}(t)}=\sum_{m =0, 1, 2, 3} e^{ -i \int_{t_{0}}^{t} E^{(m)}(t^{\prime}) dt^{\prime}}c_{m}(t_{0}) \ket{m},
\end{equation}
as the formal solution of the TDSE. As one can see, the state population cannot change, no matter how we modulate the external flux $\varphi(t)$.

In \secref{sec:FromCircuitToEffective Hamiltonians}, we derive the model of a time-dependent anharmonic oscillator, see \equsref{eq:HamTrafo_main}{eq:basis_trafo_main}. Here we find that the non-adiabatic drive term in \equref{eq:basis_trafo_main} is proportional to the derivative $\dot{\varphi}(t)$ of the external flux. \change{Consequently, the Hamiltonian in \equref{eq:flux-tunable transmon effective} can generate the correct dynamics if the external flux is varied sufficiently slowly such that $\dot{\varphi}(t) \rightarrow 0$ and the system is described in a time-dependent basis, see \equaref{eq:td_basis}{eq:HamTrafo_main}.} Note that this result is in agreement with the adiabatic theorem, see \REFS\cite{Weinberg2015,Amin}.

The model Hamiltonian for the transmission line resonator given by \equref{eq: transmission line resonators} is already diagonal in the harmonic basis. Therefore, no further approximations are necessary. However, if we intend to derive effective Hamiltonians for the circuit Hamiltonians in \equaref{eq:architecture I}{eq:architecture II}, we also have to consider the interaction operators. This means we have to replace the charge operator $\hat{n}$ by an effective operator $\hat{n}_{\text{eff}}$. In this work we use the operator
\begin{equation}
  \hat{n}_{\text{eff}} = \sqrt[4]{\frac{E_{J}}{8 E_{C}}} \sum_{m=0,1,2,3}\sqrt{m+1} \left(\ket{m}\!\bra{m+1} + \ket{m+1}\!\bra{m}\right),
\end{equation}
which was also discussed in \REF\cite{Koch}. If we couple flux-tunable transmons, we perform the substitution $E_{J} \rightarrow E_{J\text{eff}}(t)$. The effective interaction strength for a coupling between a fixed-frequency transmon $i$ and a flux-tunable transmon $j$ is given by
\begin{equation}\label{eq:eff_int_I}
  g_{j,i}(t)=G_{j,i} \sqrt[4]{\frac{E_{J_{j}\text{eff}}(t)}{8 E_{C_{j}}}}\sqrt[4]{\frac{E_{J_{i}}}{8 E_{C_{i}}}},
\end{equation}
where $G_{j,i}$ is the original coupling strength, see \equaref{eq:int_I}{eq:int_II}. Similarly, the effective interaction strength, between a resonator $j$ and a flux-tunable transmon $i$, reads
\begin{equation}\label{eq:eff_int_II}
  \bar{g}_{j,i}(t)=G_{j,i} \sqrt[4]{\frac{E_{J_{i}\text{eff}}(t)}{8 E_{C_{i}}}}.
\end{equation}
We find that the effective interaction strength is now time dependent. Note that this time dependence is frequently neglected, see \REFS\listcitetwo.

It is often the case that the complete effective Hamiltonian is expressed solely in terms of bosonic number operators. In this representation the effective model Hamiltonian for architecture I reads
\begin{equation}\label{eq:architecture I effective}
\begin{split}
  \hat{H}_{I}^{\text{eff}}&= \omega_{0} \hat{b}_{0}^{\dagger} \hat{b}_{0} + \frac{\alpha_{0}}{2} \hat{b}_{0}^{\dagger}\hat{b}_{0}(\hat{b}_{0}^{\dagger}\hat{b}_{0}-\hat{I})\\
                   &+ \omega_{1} \hat{b}_{1}^{\dagger} \hat{b}_{1} + \frac{\alpha_{1}}{2} \hat{b}_{1}^{\dagger}\hat{b}_{1}(\hat{b}_{1}^{\dagger}\hat{b}_{1}-\hat{I})\\
                   &+ \omega_{2}(t)\hat{b}_{2}^{\dagger} \hat{b}_{2} + \frac{\alpha_{2}}{2} \hat{b}_{2}^{\dagger}\hat{b}_{2}(\hat{b}_{2}^{\dagger}\hat{b}_{2}-\hat{I})\\
                   &+ g_{2,1}(t) (\hat{b}_{2}^{\dagger}+\hat{b}_{2})(\hat{b}_{1}^{\dagger}+\hat{b}_{1}) + g_{2,0}(t) (\hat{b}_{2}^{\dagger}+\hat{b}_{2}) (\hat{b}_{0}^{\dagger}+\hat{b}_{0}).
\end{split}
\end{equation}
Similarly, the effective model Hamiltonian for architecture II can be expressed as
\begin{equation}\label{eq:architecture II effective}
\begin{split}
  \hat{H}_{II}^{\text{eff}}&= \omega_{0}(t) \hat{b}_{0}^{\dagger}\hat{b}_{0} + \frac{\alpha_{0}}{2} \hat{b}_{0}^{\dagger}\hat{b}_{0}(\hat{b}_{0}^{\dagger}\hat{b}_{0}-\hat{I})\\
                   &+ \omega_{1}(t) \hat{b}_{1}^{\dagger}\hat{b}_{1} + \frac{\alpha_{1}}{2} \hat{b}_{1}^{\dagger}\hat{b}_{1}(\hat{b}_{1}^{\dagger}\hat{b}_{1}-\hat{I})\\
                   &+ \omega_{2}^{R} \hat{a}_{2}^{\dagger}\hat{a}_{2}\\
                    &+ \bar{g}_{2,1}(t) (\hat{a}_{2}^{\dagger}+\hat{a}_{2}) (\hat{b}_{1}^{\dagger}+\hat{b}_{1}) + \bar{g}_{2,0}(t) (\hat{a}_{2}^{\dagger}+\hat{a}_{2}) (\hat{b}_{0}^{\dagger}+\hat{b}_{0}).
\end{split}
\end{equation}
The device parameters that we use in our simulations to obtain the results in \secref{sec:Results}, are listed in \tabref{tab:device_parameter_flux_tunable_coupler_chip_effective} for architecture I and \tabref{tab:device_parameter_resonator_coupler_chip_effective} for architecture II, respectively. \change{Note that the Hamiltonians in \equaref{eq:architecture I effective}{eq:architecture II effective} both lack the drive term given by \equref{eq:basis_trafo_main}. Consequently, here we model the flux-tunable transmons adiabatically. In \secaref{sec:Single-qubit operations effective}{sec: two-qubit operations effective} we simulate both Hamiltonians with and without the drive term and compare the results.}

\begin{table}[!tbp]
\caption{\label{tab:device_parameter_flux_tunable_coupler_chip_effective} Parameters for an effective Hamiltonian model of architecture I, see \tabref{tab:device_parameter_flux_tunable_coupler_chip} and \equref{eq:architecture I} for details and units.}
\begin{ruledtabular}
\begin{tabular}{ccccccccc}
$i$ &$\omega/2 \pi$&$\alpha/2\pi$  & $\varphi_{0}/2\pi$ & $g_{2,i}(\varphi_{0})/2\pi$\\
\hline
  0 & 5.100 & -0.310 &  n/a & 0.146\\
  1 & 6.200 & -0.285 &  n/a & 0.164\\
  2 & 8.100 & -0.235 &  0.15 & n/a\\
\end{tabular}
\end{ruledtabular}
\caption{\label{tab:device_parameter_resonator_coupler_chip_effective} Parameters for an effective Hamiltonian model of architecture II, see \tabref{tab:device_parameter_resonator_coupler_chip} and \equref{eq:architecture II} for details and units.}
\begin{ruledtabular}
\begin{tabular}{ccccccccc}
$i$ & $\omega^{R}/2 \pi$ &$\omega/2 \pi$&$\alpha/2\pi$ & $\varphi_{0}/2\pi$ & $ g_{2,i}(\varphi_{0})/2\pi$\\
\hline
  0 & n/a & 4.200& -0.320 & 0 &0.307\\
  1 & n/a & 5.200& -0.295 & 0 &0.344\\
  2 & 45.000 & n/a & n/a & n/a & n/a\\
\end{tabular}
\end{ruledtabular}
\end{table}

\subsection{Control Pulse}\label{sec:control pulse}
All simulations in this work are performed with a control pulse (external flux) of the form
\begin{equation}\label{eq:pulse}
\varphi(t)=\varphi_{0}+\delta e(t) \cos(\omega^{D}t),
\end{equation}
where the real valued parameters $\varphi_{0}$, $\delta$ and $\omega^{D}$ denote the flux offset, the pulse amplitude and the drive frequency, respectively. The envelope function $e(t)$ is taken to be of the form
\begin{equation}
e(t) = \begin{cases}
\sin(\lambda t) &\text{if $0 \leq t < T_{\text{r/f}}$}\\
1     &\text{if $T_{\mathrm{r/f}} \leq t \leq \Delta T $}\\
\sin(\frac{\pi}{2}+\lambda (t-\Delta T)) &\text{if $ \Delta T < t \leq T_{\mathrm{d}} $.}
\end{cases}
\end{equation}
Here $T_{\text{r/f}}$ denotes the rise and fall time, $T_{\mathrm{d}}$ is the control pulse duration and $\Delta T=(T_{\mathrm{d}}-T_{\text{r/f}})$. The parameter $\lambda=\pi/(2 T_{\text{r/f}})$ is determined by the rise and fall time. This generic flux pulse allows us to control various transitions between states of the systems.
\renewcommand{\width}{1.0}
\begin{figure}[!tbp]
  \begin{minipage}{0.45\textwidth}
    \centering
    \includegraphics[width=\width\textwidth]{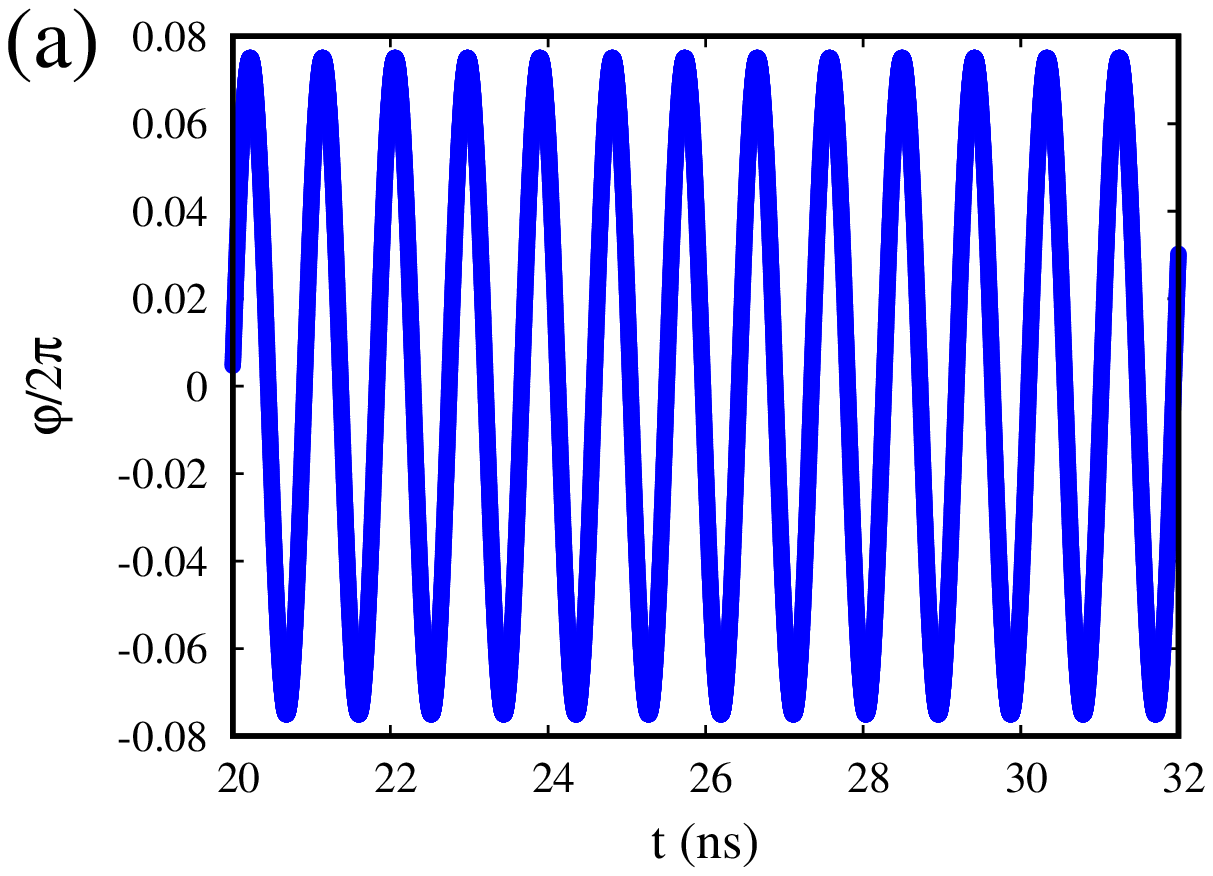}
  \end{minipage}
  \begin{minipage}{0.45\textwidth}
    \centering
    \includegraphics[width=\width\textwidth]{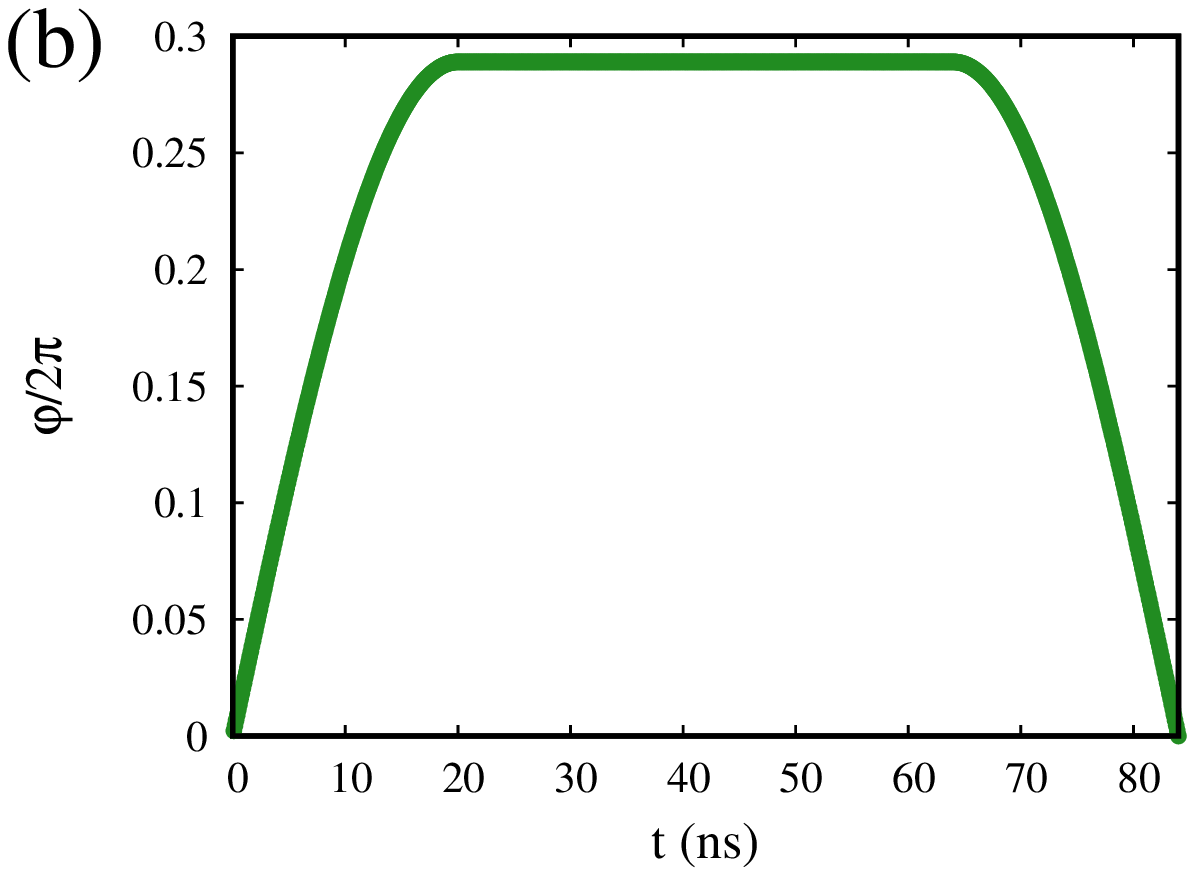}
  \end{minipage}
  \caption{(Color online) External flux $\varphi/2\pi$ as a function of time for two different flux control pulses. Figure~\ref{fig:pulse_time_evo}(a): microwave pulse using \equref{eq:pulse}, amplitude $\delta/2\pi=0.075$, drive frequency $\omega^{D}/2\pi=1.089$ GHz, a rise and fall time of $T_{\text{r/f}}=13$ ns and pulse duration $T_{\mathrm{d}}=205.4$ ns. Figure~\ref{fig:pulse_time_evo}(b): unimodal pulse using \equref{eq:pulse}, amplitude $\delta/2\pi=0.297$, drive frequency $\omega^{D}/2\pi=0$ ns, a rise and fall time of $T_{\text{r/f}}=20$ ns and pulse duration $T_{\mathrm{d}}=84$ ns.}
  \label{fig:pulse_time_evo}
\end{figure}

Figures~\ref{fig:pulse_time_evo}(a-b) show the external flux $\varphi/2\pi$ as functions of time $t$ for the two different types of flux control pulses we use in this work. Figure~\ref{fig:pulse_time_evo}(a) shows a microwave pulse. Here we use \equref{eq:pulse}, the amplitude $\delta/2\pi=0.075$, the drive frequency $\omega^{D}/2\pi=1.089$ GHz, a rise and fall time $T_{\text{r/f}}=13$ ns and the pulse duration $T_{\mathrm{d}}=205.4$ ns. This type of control pulse is used for architecture I. Figure~\ref{fig:pulse_time_evo}(b) shows a unimodal pulse. Here we use \equref{eq:pulse}, the amplitude $\delta/2\pi=0.297$, the drive frequency $\omega^{D}/2\pi=0$ GHz, a rise and fall time $T_{\text{r/f}}=20$ ns and the pulse duration $T_{\mathrm{d}}=84$ ns. This type of control pulse is used to implement non-adiabatic gates, see \REF\cite{DiCarlo09}, with architecture II.


\section{Results}\label{sec:Results}

In this section we present our findings. First, in \secref{sec: Revised derivation of the effective Hamiltonian for flux-tunable transmons}, we consider a single flux-tunable transmon. Here we focus on the transition dynamics and compare the effective Hamiltonians in \equaref{eq:HamTrafo_main}{eq:flux-tunable transmon effective} with the circuit Hamiltonian given by \equref{eq:flux-tunable transmon}. Next, in \secref{sec:Single-qubit operations effective}, we identify transitions (interactions) which seem to be suppressed in the effective model of architecture I given by \equref{eq:architecture I effective}. Finally, in \secref{sec: two-qubit operations effective}, we study how different approximations affect the unsuppressed transitions which are often used to implement two-qubit gates with architectures I and II.

A detailed discussion of the simulation results for the circuit Hamiltonian (where we do not make approximations to solve the TDSE) is provided in \appref{sec:CircuitHamiltonianSimulations}. A summary of the simulation results for the circuit Hamiltonian can be found in \tabref{tab:summary_circuit_hamiltonian_results}. Here we use the device parameters listed in \change{ \tabref{tab:device_parameter_flux_tunable_coupler_chip} (\tabref{tab:device_parameter_resonator_coupler_chip}) to obtain the results for architecture I (architecture II).} In the following sections we compare these results with the ones we obtain by simulating the effective models. A summary of the results for the effective models can be found in \tabref{tab:summary_effective_hamiltonian_results}. \change{\appref{sec:Methodology} introduces the simulation algorithm we use to obtain the results in this section. Note that throughout this work we use $\hbar=1$.}

\subsection{Simulations of a single flux-tunable transmon}\label{sec: Revised derivation of the effective Hamiltonian for flux-tunable transmons}
In this section, we compare the pulse response of the circuit Hamiltonian given by \equref{eq:flux-tunable transmon} with the one of the effective Hamiltonians in \equaref{eq:HamTrafo_main}{eq:flux-tunable transmon effective}. Note that we do not need to simulate the effective Hamiltonian given by \equref{eq:flux-tunable transmon effective}. The formal solution of its TDSE is given by \equref{eq:statevectorAC} in \secref{sec:Effective Hamiltonians}.

For the simulations in this section we use the device parameters listed in \tabref{tab:device_parameter_flux_tunable_coupler_chip}, row $i=2$ and the pulse $\varphi(t)$ in \equref{eq:pulse}. We consider two cases. First, we consider resonant transitions driven by a microwave pulse, see \figref{fig:pulse_time_evo}(a), whose drive frequency $\omega^{D}$ coincides with the energy gap $E^{(1)}-E^{(0)}$ of the flux-tunable transmon system. The results are presented in \figsref{fig:ResMapComp}(a-b) and \figsref{fig:TimeEvolComp}(a-b). Second, we consider non-adiabatic transitions driven by a unimodal pulse, see \figref{fig:pulse_time_evo}(b), with the drive frequency  $\omega^{D}=0$. The corresponding results are displayed in \figsref{fig:leakage}(a-h).

\renewcommand{\width}{1.0}
\begin{figure}[!tbp]
    \centering
    \begin{minipage}{0.45\textwidth}
        \centering
        \includegraphics[width=\width\textwidth]{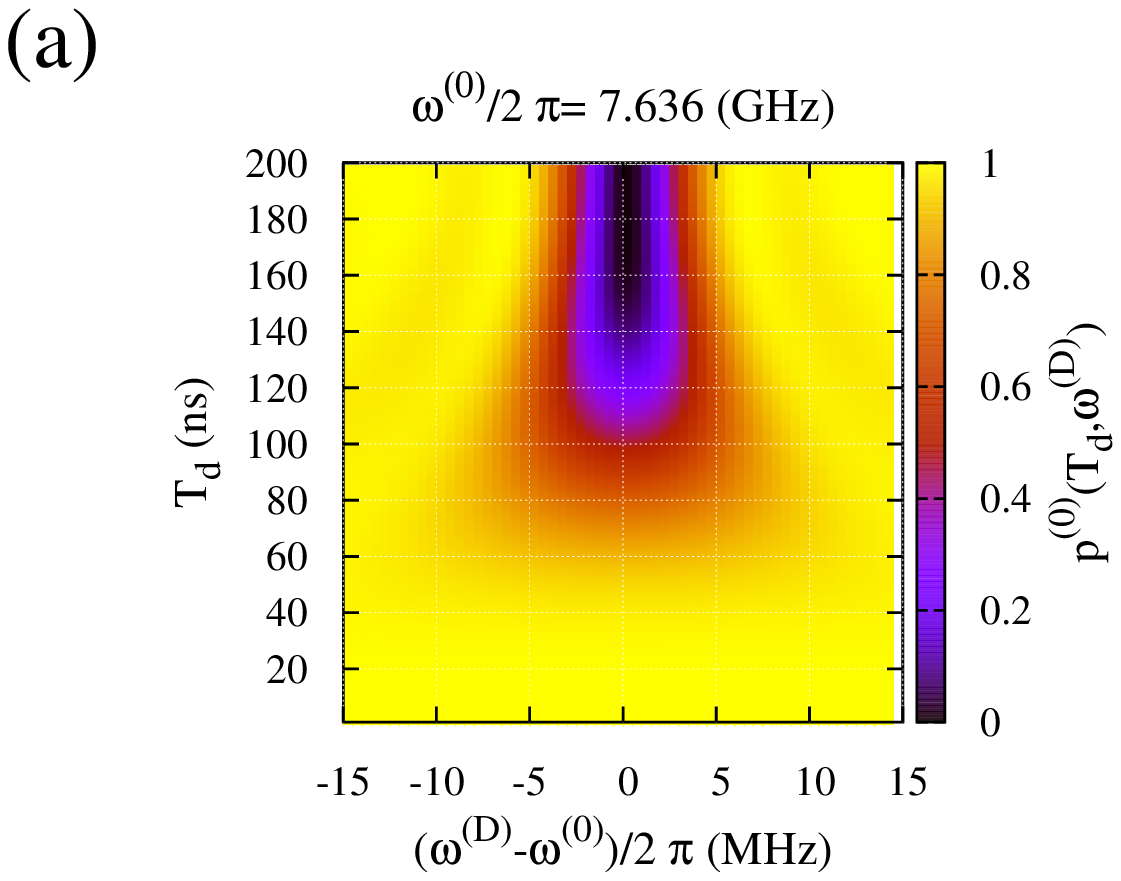} 
    \end{minipage}
    \begin{minipage}{0.45\textwidth}
        \centering
        \includegraphics[width=\width\textwidth]{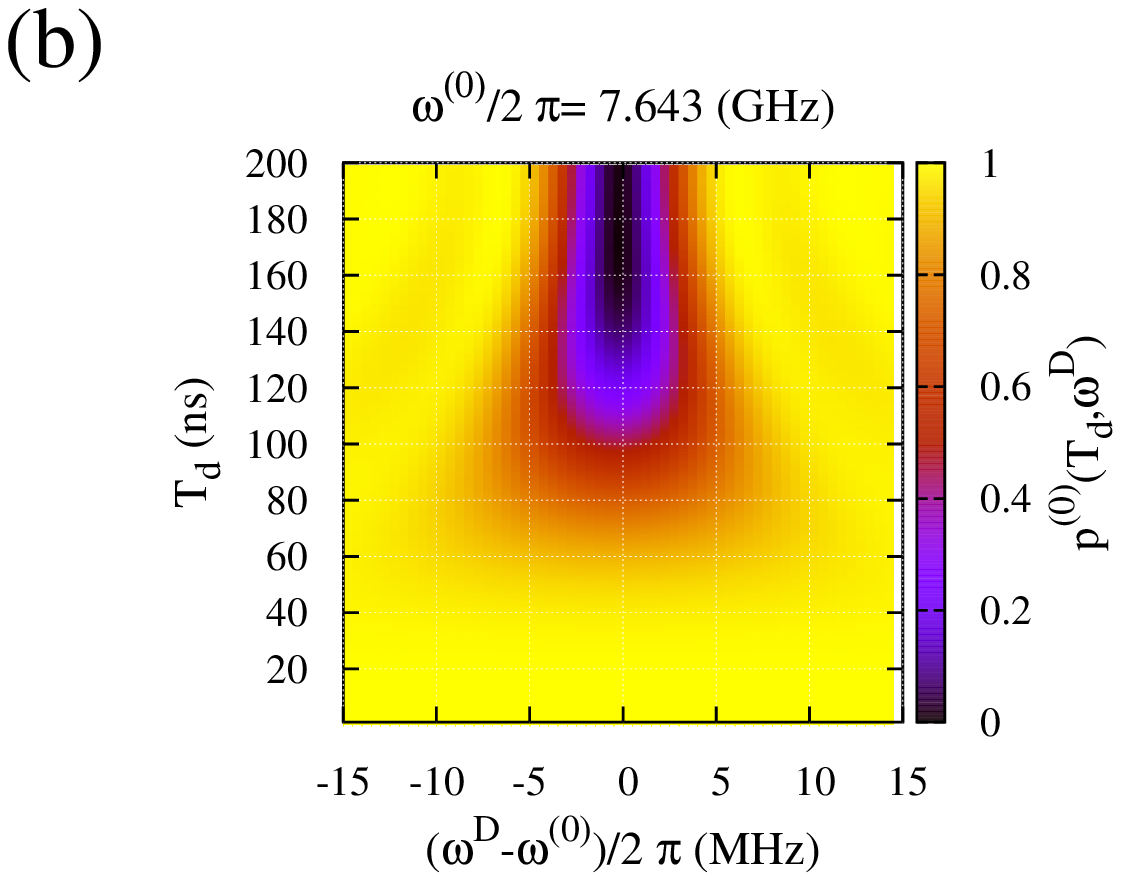} 
    \end{minipage}
    \caption{(Color online) Ground-state probabilities $p^{(0)}$ as functions of the pulse duration $T_{\mathrm{d}}$ and the drive frequency $\omega^{D}$. We use the device parameters for a single flux-tunable transmon listed in \tabref{tab:device_parameter_flux_tunable_coupler_chip}, row $i=2$ and the pulse given by \equref{eq:pulse} with $T_{\mathrm{r/f}}=T_{\mathrm{d}}/2$ and the pulse amplitude $\delta/2\pi=0.001$, see \figref{fig:pulse_time_evo}(a), to obtain the results. The results in (a) are obtained by solving the TDSE for the circuit Hamiltonian in \equref{eq:flux-tunable transmon}. Similarly, the results in (b) are obtained by solving the TDSE for the non-adiabatic effective Hamiltonian in \equref{eq:HamTrafo_main}. At time $t=0$ the systems are initialised in the corresponding eigenstates $p^{(0)}(0)=1$. Here we model Rabi transitions between the ground state and the first excited state. Note that (a) and (b) are centred around the frequency $\omega^{(0)}$ which corresponds to energy difference $E^{(1)}-E^{(0)}$ in the corresponding model, \ie the circuit or the effective model. We see that apart from the shift in the transition frequency both models show a similar qualitative and quantitative behaviour. However, the effective model given by the Hamiltonian in \equref{eq:flux-tunable transmon effective} does not allow us to model these transitions.}
    \label{fig:ResMapComp}
\end{figure}
\renewcommand{\width}{1.0}
\begin{figure}[!tbp]
    \centering
    \centering
    \begin{minipage}{0.45\textwidth}
        \centering
        \includegraphics[width=\width\textwidth]{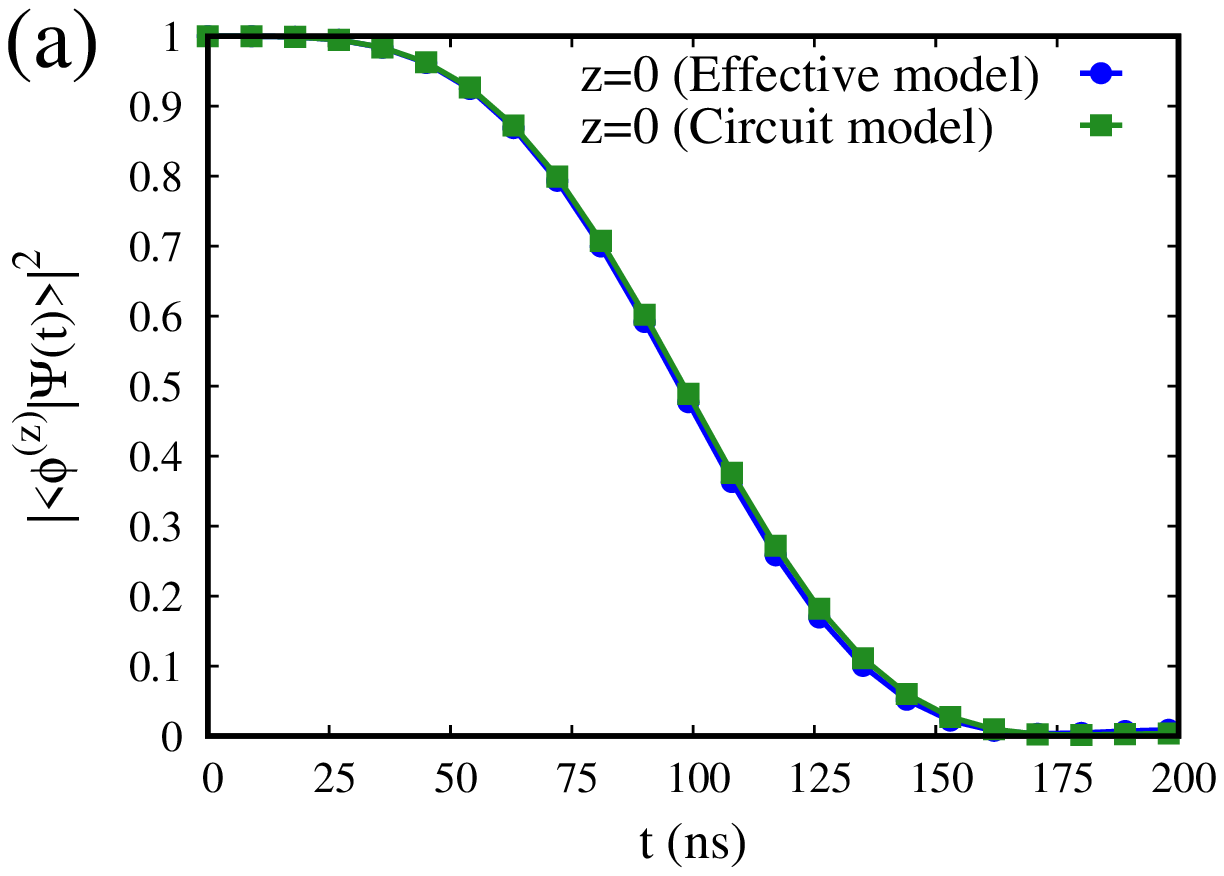} 
    \end{minipage}
    \begin{minipage}{0.45\textwidth}
        \centering
        \includegraphics[width=\width\textwidth]{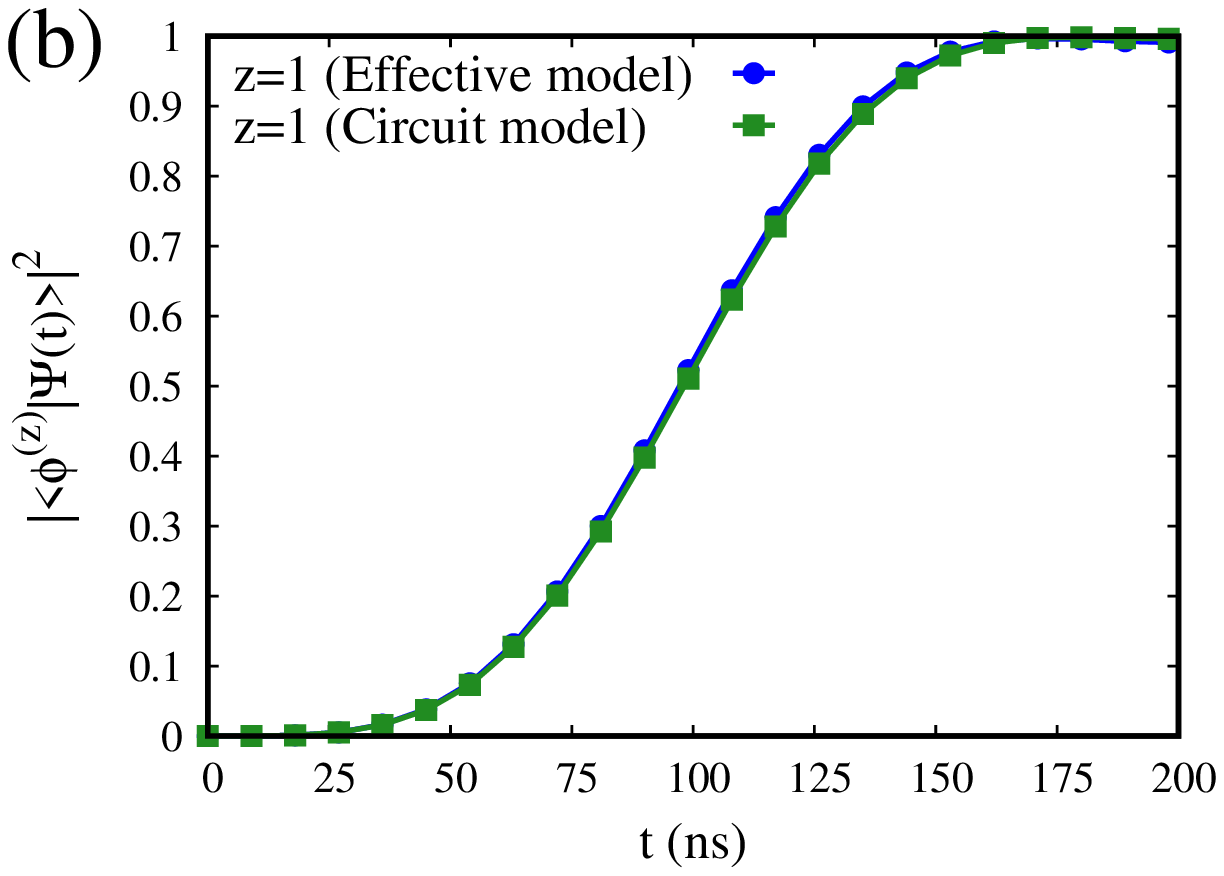} 
    \end{minipage}
    \caption{\change{(Color online) Probabilities $p^{(0)}(t)$ (a) and $p^{(1)}(t)$ (b) as functions of time $t$ obtained with the effective model (blue circles) and the circuit model (green squares). We use the pulse given by \equref{eq:pulse} with the pulse amplitude $\delta/2\pi=0.001$, the pulse duration $T_{\mathrm{d}}=200$ ns and the rise and fall time $T_{\mathrm{r/f}}=100$ ns and the device parameters for a single flux-tunable transmon listed in \tabref{tab:device_parameter_flux_tunable_coupler_chip}, row $i=2$, to obtain the results. We use the drive frequency $\omega^{D}=7.636$ GHz to obtain the results with the the circuit Hamiltonian \equref{eq:flux-tunable transmon}. Similarly, we use the drive frequency $\omega^{D}=7.643$ GHz to obtain the results with the effective Hamiltonian \equref{eq:HamTrafo_main}. The systems are initialised in the ground state $p^{(0)}=1$ at time $t=0$. Note that in (a-b) we use the frequencies which cut through the centres of the chevron patterns in \figsref{fig:ResMapComp}(a-b). As one can see, (a) and (b) show qualitatively and quantitatively similar behaviour with respect to the time evolution.}}
    \label{fig:TimeEvolComp}
\end{figure}

Figures \ref{fig:ResMapComp}(a-b) show the ground-state probabilities $p^{(0)}$ as functions of the pulse duration $T_{\mathrm{d}}$ and the drive frequency $\omega^{D}$. We use the pulse amplitude $\delta/2 \pi=0.001$ and the rise and fall time $T_{\mathrm{r/f}}=T_{\mathrm{d}}/2$ to obtain the results. For (a) we solve the TDSE for the circuit Hamiltonian given by \equref{eq:flux-tunable transmon} and centre the results around the transition frequency $\omega^{(0)}=7.636$ GHz. Similarly, for (b) we solve the TDSE for the effective Hamiltonian in \equref{eq:HamTrafo_main} and centre the results around the transition frequency $\omega^{(0)}=7.643$ GHz. The $7$ MHz difference in terms of the transition frequency stems from the fact that the fourth-order expansion does not lead to the exact same spectrum.

We also simulated the effective model given by \equref{eq:HamTrafo_main} with higher-order terms (data not shown), see \secref{sec:FromCircuitToEffective Hamiltonians}. Here we find that the chevron pattern in \figref{fig:ResMapComp}(b) stays the same but the transition frequency changes due to the higher-order terms. If we add enough terms to the cosine expansion, the results converge. Note that we simulated the model up to the 60th order.

Furthermore, in \appref{sec: Effective Hamiltonians appendix} we numerically investigate how well the spectrum of the circuit Hamiltonian can be approximated by the tunable frequency given by \equref{eq:tunable frequency}. We find that the deviations increase with the flux $\varphi/2\pi \rightarrow 0.5$. For the fourth-order expansion and the operating point $\varphi_{0}/2\pi=0.15$, deviations of the order of $10$ MHz are characteristic.

Clearly, the results in Figures \ref{fig:ResMapComp}(a-b) show a similar qualitative and quantitative behaviour. \change{Furthermore, \figsref{fig:TimeEvolComp}(a-b) show the time evolution of the probabilities $p^{(0)}(t)$(a) and $p^{(1)}(t)$(b) obtained with the effective and the circuit model. Here we use the frequencies which cut through the centres of the chevron patterns in \figsref{fig:ResMapComp}(a-b) and add the data for the first excited state $p^{(1)}(t)$, see \figsref{fig:TimeEvolComp}(b). One can observe that the time evolutions of the probabilities are qualitatively and quantitatively very similar.}

The time evolution of the effective Hamiltonian given by \equref{eq:flux-tunable transmon effective} for this scenario is trivial, \ie the system simply remains in its initial state.

\renewcommand{\width}{1.0}
\begin{figure}[!tbp]
    \centering
    \begin{minipage}{0.45\textwidth}
        \centering
        \includegraphics[width=\width\textwidth]{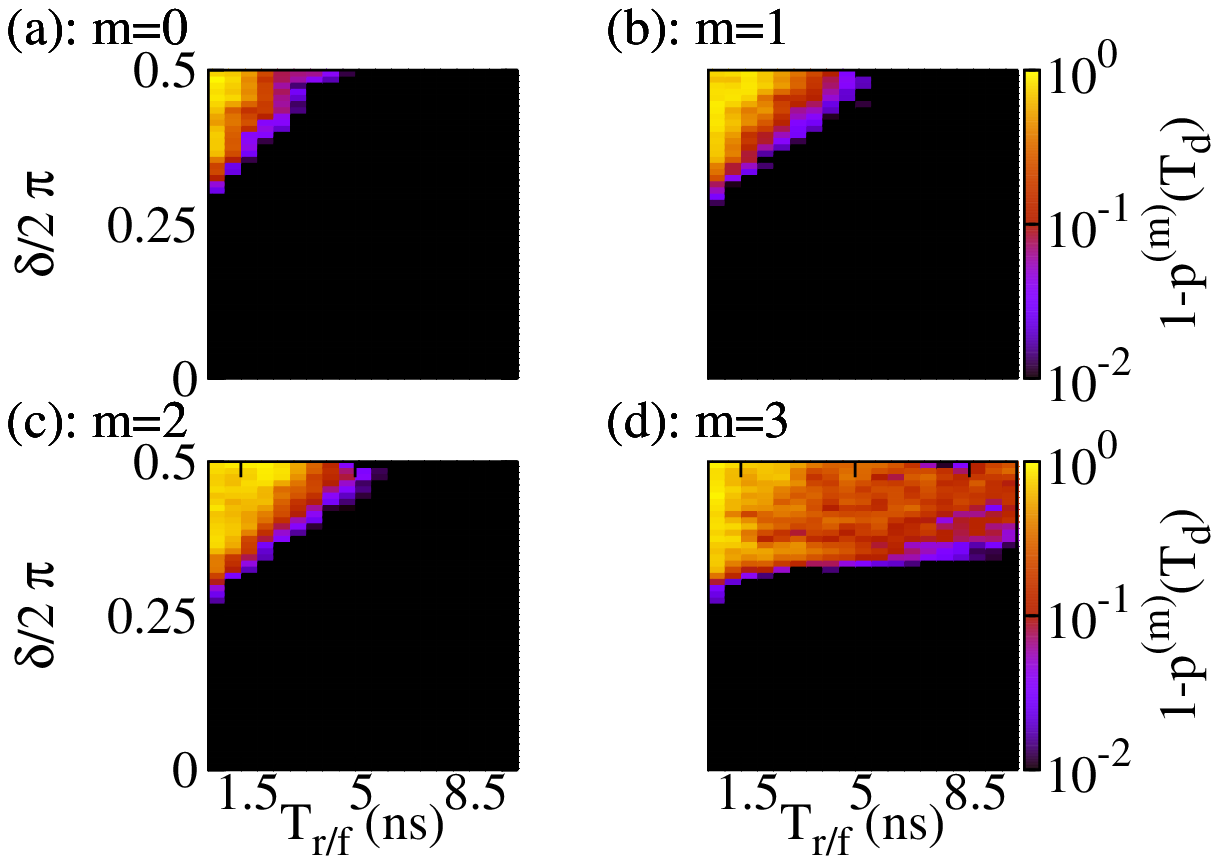} 
    \end{minipage}
    \begin{minipage}{0.45\textwidth}
        \centering
        \includegraphics[width=\width\textwidth]{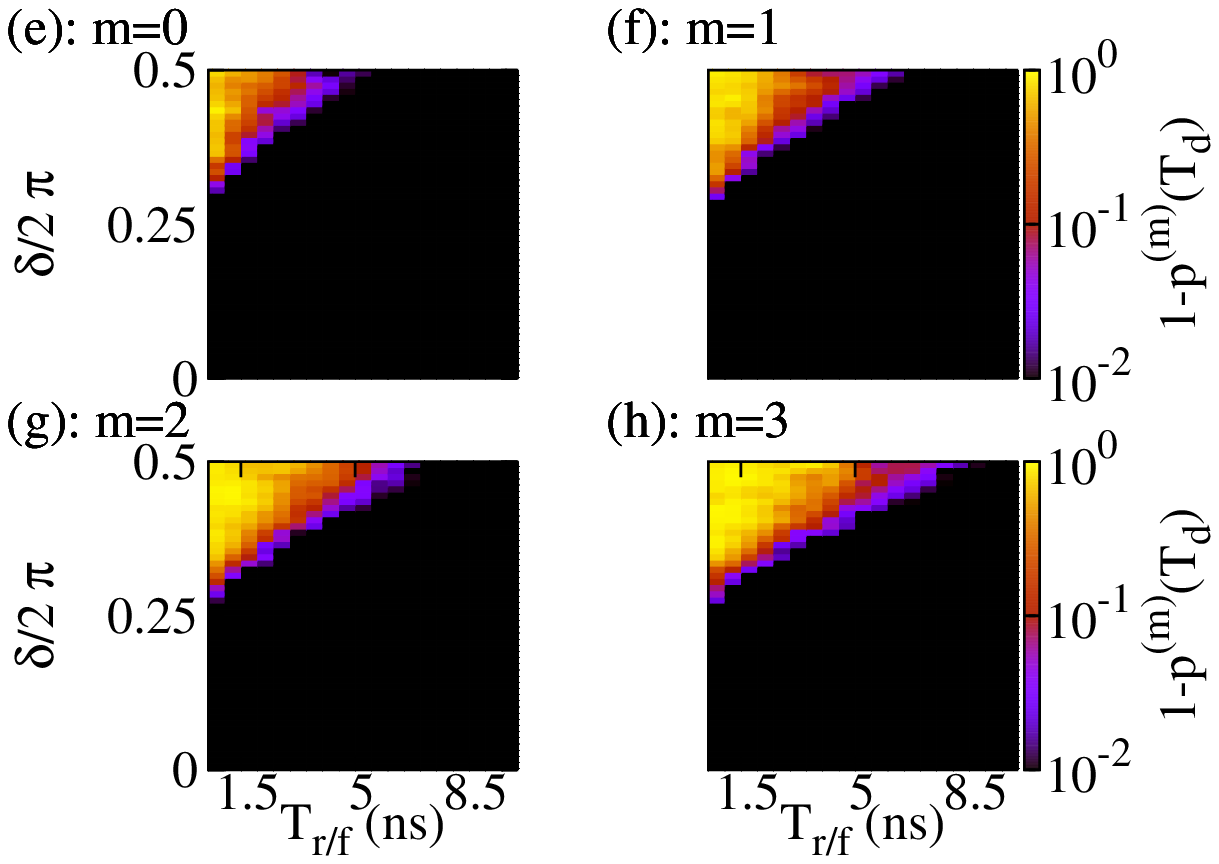} 
    \end{minipage}
    \caption{(Color online) Probabilities $1-p^{(m)}$ at time $T_{d}$ as functions of the rise and fall time $T_{\mathrm{r/f}}$ and the pulse amplitude $\delta$. We use the pulse given by \equref{eq:pulse} with $\omega^{D}=0$ and $T_{\mathrm{d}}=50$ ns, see \figref{fig:pulse_time_evo}(b) and the device parameters for a single flux-tunable transmon listed in \tabref{tab:device_parameter_flux_tunable_coupler_chip}, row $i=2$ to obtain the results. The results in (a-d) for $m=0$(a), $m=1$(b), $m=2$(c) and $m=3$(d) are obtained by solving the TDSE for the circuit Hamiltonian in \equref{eq:flux-tunable transmon}. Similarly, the results in (e-h) for $m=0$(e), $m=1$(f), $m=2$(g) and $m=3$(h) are obtained by solving the TDSE for the non-adiabatic effective Hamiltonian given by \equref{eq:HamTrafo_main}. At time $t=0$ the systems are initialised in the corresponding eigenstates $p^{(m)}(0)=1$. The simulations test whether or not we have left the pulse parameter regime where the adiabatic approximation is valid, \ie the bright areas indicate the parameters which lead to non-adiabatic transitions. Note that it is impossible to use the effective Hamiltonian in \equref{eq:flux-tunable transmon effective} to model such non-adiabatic transitions. Interestingly, for $m=0$, $m=1$ and $m=2$ the effective model given by the Hamiltonian in \equref{eq:HamTrafo_main} shows a qualitatively similar behaviour as the circuit model given by the Hamiltonian in \equref{eq:flux-tunable transmon}.}
    \label{fig:leakage}
\end{figure}
We now consider the second case, \ie non-adiabatic transitions driven by a unimodal pulse. Figures \ref{fig:leakage}(a-h) show the probabilities $1-p^{(m)}$ at time $T_{\mathrm{d}}$ as functions of the rise and fall time $T_{\mathrm{r/f}}$ and the pulse amplitude $\delta$. We use a unimodal pulse, see \figref{fig:pulse_time_evo}(b), with $\omega^{D}=0$ and $T_{\mathrm{d}}=50$ ns to obtain the results. In \figsref{fig:leakage}(a-d) we use the circuit Hamiltonian given by \equref{eq:flux-tunable transmon} to obtain the results for $m=0$(a), $m=1$(b), $m=2$(c) and $m=3$(d). Similarly, in \figsref{fig:leakage}(e-h) we use the effective Hamiltonian given by \equref{eq:HamTrafo_main} to obtain the results for $m=0$(e), $m=1$(f), $m=2$(g) and $m=3$(h). At time $t=0$ we initialise the system in the corresponding eigenstates, \ie $p^{(m)}(0)=1$. Therefore, the simulations test whether or not the pulse parameters are still in the regime where the adiabatic approximation, see \REFS\cite{Weinberg2015,Amin}, is valid. The bright areas correspond to pulse parameters which induce non-adiabatic transitions.

As one can see, the circuit model given by \equref{eq:flux-tunable transmon} and the effective model in \equref{eq:HamTrafo_main} yield qualitative similar results for $m=0$, $m=1$ and $m=2$. The results for $m=3$ deviate qualitatively and quantitatively.

As before, the time evolution of the effective Hamiltonian given by \equref{eq:flux-tunable transmon effective} for this scenario is trivial, \ie the system simply remains in its initial state such that $1-p^{(m)}(t)=0$ for all $m \in \{0,1,2,3\}$ and time $t$.

In summary, the effective flux-tunable Hamiltonian given by \equref{eq:flux-tunable transmon effective} cannot describe any of the transitions we can model with the Hamiltonians in \equaref{eq:flux-tunable transmon}{eq:HamTrafo_main}. Furthermore, we presented results which show that the effective flux-tunable Hamiltonian given by \equref{eq:HamTrafo_main} and the circuit Hamiltonian given by \equref{eq:flux-tunable transmon} generate qualitative and sometimes even quantitative similar pulse responses, see \figsref{fig:ResMapComp}(a-b) and \figsref{fig:TimeEvolComp}(a-b) for the case of resonant transitions and \figsref{fig:leakage}(a-h) for the case of non-adiabatic transitions. Some of the deviations we find, \eg small shifts in the transition frequency, might be explained by the fact that the spectrum of the effective model given by \equref{eq:HamTrafo_main} is not exactly the one of the circuit model given by \equref{eq:flux-tunable transmon}. Furthermore, additional deviations might be attributed to the fact that we truncate the cosine expansion up to a finite order, see the Hamiltonian in \equref{eq:flux-tunable transmon recast} and \secref{sec:FromCircuitToEffective Hamiltonians}. The full dynamic behaviour, with regard to the circuit model, might only be recovered if we include all terms.

\subsection{Simulations of suppressed transitions in the effective two-qubit model}\label{sec:Single-qubit operations effective}

In the previous section, we discussed the case of a single flux-tunable transmon. In this section we consider transitions in a two-qubit system which are suppressed in the effective model. Here we use the effective model Hamiltonian in \equref{eq:architecture I effective} and the parameters listed in \tabref{tab:device_parameter_flux_tunable_coupler_chip_effective} to obtain the results. The effective Hamiltonian describes a two-qubit system (two qubits and one coupler). We index the different states by using tuples of the form $z=(k_{0},m_{1},m_{0})$, where $k_{0}\in \{0,1,2,3\}$ is the coupler index, $m_{1}\in\{0,1,2,3\}$ is the index of the second qubit and $m_{0}\in\{0,1,2,3\}$ is the index of the first qubit. Previous work by the authors of \REFS\cite{McKay16,Roth19,Ganzhorn20} shows that at least the transitions $z=(0,0,1) \rightarrow z=(0,1,0)$ and $z=(0,1,1) \rightarrow z=(0,2,0)$ can be activated by modulating the coupler frequency given by \equref{eq:tunable frequency} with a microwave pulse, see also \secref{sec: two-qubit operations effective}.

Our aim is to model the following transitions $z=(0,0,0) \rightarrow z=(0,1,0)$ and $z=(0,0,0) \rightarrow z=(0,0,1)$ for a two-qubit system. We are able to model these transitions with the circuit Hamiltonian \equref{eq:architecture I} and the device parameters listed in \tabref{tab:device_parameter_flux_tunable_coupler_chip}, the pulse parameters are summarised in \tabref{tab:summary_circuit_hamiltonian_results}. However, we find that the effective model does not respond to pulses of the form \equref{eq:pulse}, with pulse parameters similar to the ones given in \tabref{tab:summary_circuit_hamiltonian_results}. Therefore, we search for the corresponding transitions in a more systematic way.

\renewcommand{\width}{1.0}
\begin{figure}[!tbp]
    \centering
    \begin{minipage}{0.45\textwidth}
        \centering
        \includegraphics[width=\width\textwidth]{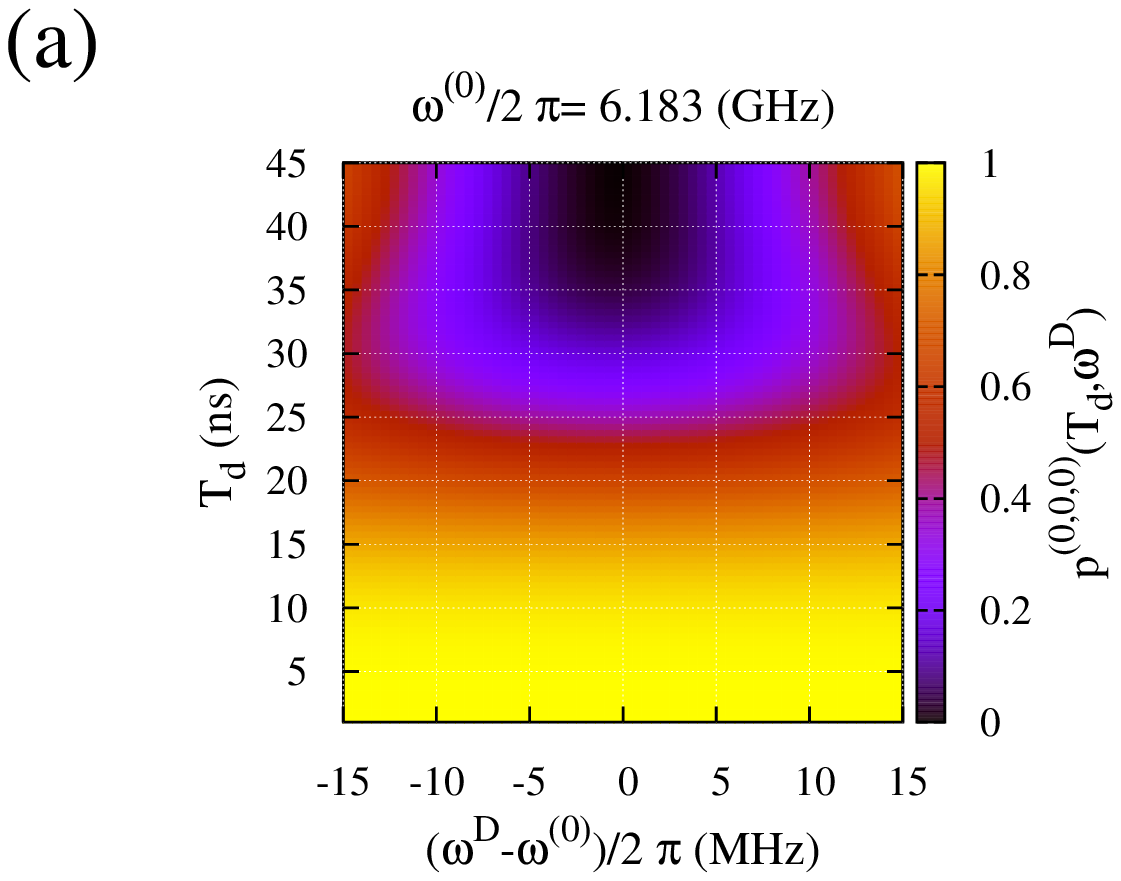} 
    \end{minipage}
    \begin{minipage}{0.45\textwidth}
        \centering
        \includegraphics[width=\width\textwidth]{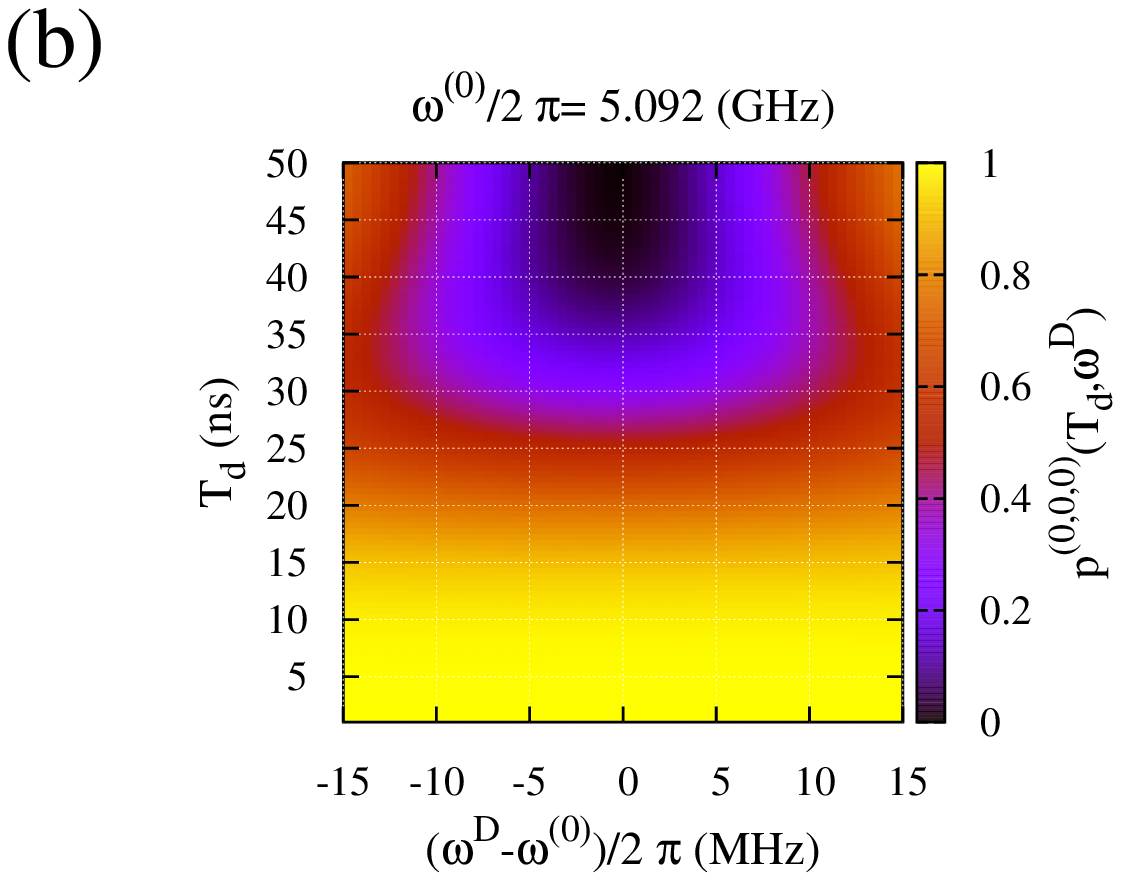} 
    \end{minipage}
    \caption{(Color online) Chevron pattern (a), for a pulse of the form \equref{eq:pulse} with the drive frequency $\omega^{D}=6.183$ GHz and the amplitude $\delta/2\pi=0.045$. Chevron pattern (b), for a pulse of the form \equref{eq:pulse} with the drive frequency $\omega^{D}=5.092$ GHz and the amplitude $\delta/2\pi=0.085$. We use the same rise and fall time $T_{\mathrm{r/f}}=T_{\mathrm{d}}/2$ for both cases. These patterns show how the circuit Hamiltonian \equref{eq:architecture I} (with the parameters listed in \tabref{tab:device_parameter_flux_tunable_coupler_chip}) reacts to two different pulses, characterised by the different pulse parameters. The color bar shows the probability $p^{(0,0,0)}$ as a function of the pulse duration time $T_{\mathrm{d}}$. The chevron patterns are used to calibrate control pulses, which are then used to obtain the results in \tabref{tab:summary_circuit_hamiltonian_results}.}
    \label{fig:chevron_patterns}
\end{figure}

We initialise the system in the state $z=(0,0,0)$ and compute the probability $p^{(0,0,0)}(\omega^{D},\delta,t)=|\braket{\phi^{(0,0,0)}|\Psi(\omega^{\mathrm{D}},\delta,t)}|^{2}$ for various control pulses, which are characterised by the drive frequency $\omega^{\mathrm{D}}$ and the amplitude $\delta$. This allows us to determine the value of the indicator
\begin{equation}\label{eq:cost function}
  \epsilon=1-\min_{(\omega^{D},\delta,t)\in \mathcal{G}}(p^{(0,0,0)}(\omega^{D},\delta,t)),
\end{equation}
where $\mathcal{G}\subseteq\mathbb{R}^{3}$ denotes a grid which ranges over a discrete set of pulse parameters and a discrete set of points in time.

Every row in \tabref{tab:grid_search} corresponds to a different search grid. In the first row we search for an excitation of the first qubit. This means we have to consider the frequency range $[4.90,5.30]$. Similarly, in the second row we search in the frequency range $[6.00,6.40]$. The last row serves as a reference. Here we simulate the free time evolution, i.e. we do not apply any external flux to the system. Since we do not want to activate transitions by accidentally creating an avoided crossing between different energies, we restrict the search range of the amplitude to $\delta/2\pi \in [0.000,0.110]$. The step parameters are set to $\Delta\omega/2\pi=10^{-5}$ GHz, $\Delta t=0.2$ ns and $\Delta\delta/2\pi=10^{-3}$.
\begin{table}[!tbp]
\caption{Results of the computation of the function $\epsilon=1-\min_{(\omega^{D},\delta,t)\in \mathcal{G}}(p^{(0,0,0)}(\omega^{D},\delta,t))$. Here $\omega^{D}$ denotes the drive frequency and $\delta$ is the pulse amplitude. The initial state of the system is $\ket{\phi^{(0,0,0)}}$ in all cases. The probability $p^{(0,0,0)}(\omega^{D},\delta,t)=|\braket{\phi^{(0,0,0)}|\Psi(\omega^{D},\delta,t)}|^{2}$ is determined for various pulses and points in times so that the minimum can be obtained. The first three columns show the search intervals for $\omega^{D}$, $\delta$ and $T_{d}$, which define the search grid $(\omega^{D},\delta,t) \in \mathcal{G}\subseteq\mathbb{R}^{3}$. The step parameters are set to $\Delta\omega/2\pi=10^{-5}$ GHz, $\Delta\delta/2\pi=10^{-3}$ and $\Delta t=0.2$ ns. The last column shows the result for $\epsilon$. The results are obtained with the system parameters listed in \tabref{tab:device_parameter_flux_tunable_coupler_chip_effective} and effective Hamiltonian \equref{eq:architecture I effective}.}
\label{tab:grid_search}
\begin{ruledtabular}
\begin{tabular}{cccccccc}
$\omega^{D}/2 \pi$ & $\delta/2 \pi$ & $T_{d}$ & $\epsilon$  \\
\hline
$[4.90,5.30]$ & $[0.000,0.110]$ & $[0,300]$ & $10^{-3}$  \\

$[6.00,6.40]$ & $[0.000,0.110]$ & $[0,300]$ & $10^{-3}$\\

$[0.00,0.00]$ & $[0.000,0.000]$ & $[0,300]$ & $10^{-3}$\\
\end{tabular}
\end{ruledtabular}
\end{table}
In all cases we find that $\epsilon \approx 0.001$. This means that the free time evolution yields the same result as the instances where we compute $\epsilon$ for cases where we apply pulses. The results suggest that the system reacts to these sets of pulses in the same way it does to no pulse at all, \ie the system remains mainly in its ground state.

Figures~\ref{fig:chevron_patterns}(a-b) show two chevron patterns obtained for the circuit Hamiltonian in \equref{eq:architecture I}. We used these figures to determine the pulse parameters for the results we presented in \tabref{tab:summary_circuit_hamiltonian_results}, see row three and four. The chevron patterns (a) and (b) are several MHz wide. Therefore, assuming that the effective Hamiltonian in \equref{eq:architecture I effective} allows us to model these operations, we would expect that $\epsilon \approx 1.000$. However, since this is not the case, we might conclude that we cannot model these transitions with the Hamiltonian in \equref{eq:architecture I effective}. Note that these results are in accordance with the single flux-tunable transmon case. Furthermore, there are other transitions, e.g.~$z=(0,0,0) \rightarrow z=(1,0,0)$, which seem to be suppressed. Therefore, our listing is not complete.

The deficit of the effective model Hamiltonian that it does not describe all the transitions might become relevant once we consider more and more qubits in one system, i.e.~if we consider the spectral crowding problem.

We also simulated the effective model given by \equref{eq:architecture I effective} with an additional non-adiabatic drive term given by \equref{eq:basis_trafo_main}, for the flux-tunable coupler. Here we find (data not shown) that one can model the transitions $z=(0,0,0) \rightarrow z=(0,1,0)$, $z=(0,0,0) \rightarrow z=(0,0,1)$ and others with the non-adiabatic effective model. The non-adiabatic effective model shows a similar response, see \figsref{fig:chevron_patterns}(a-b), as the circuit Hamiltonian given by \equref{eq:architecture I}.

\subsection{Simulation of unsuppressed transitions in the effective two-qubit model}\label{sec: two-qubit operations effective}

\renewcommand{\width}{1.0}
\begin{figure}[!tbp]
  \begin{minipage}{0.45\textwidth}
    \centering
    \includegraphics[width=\width\textwidth]{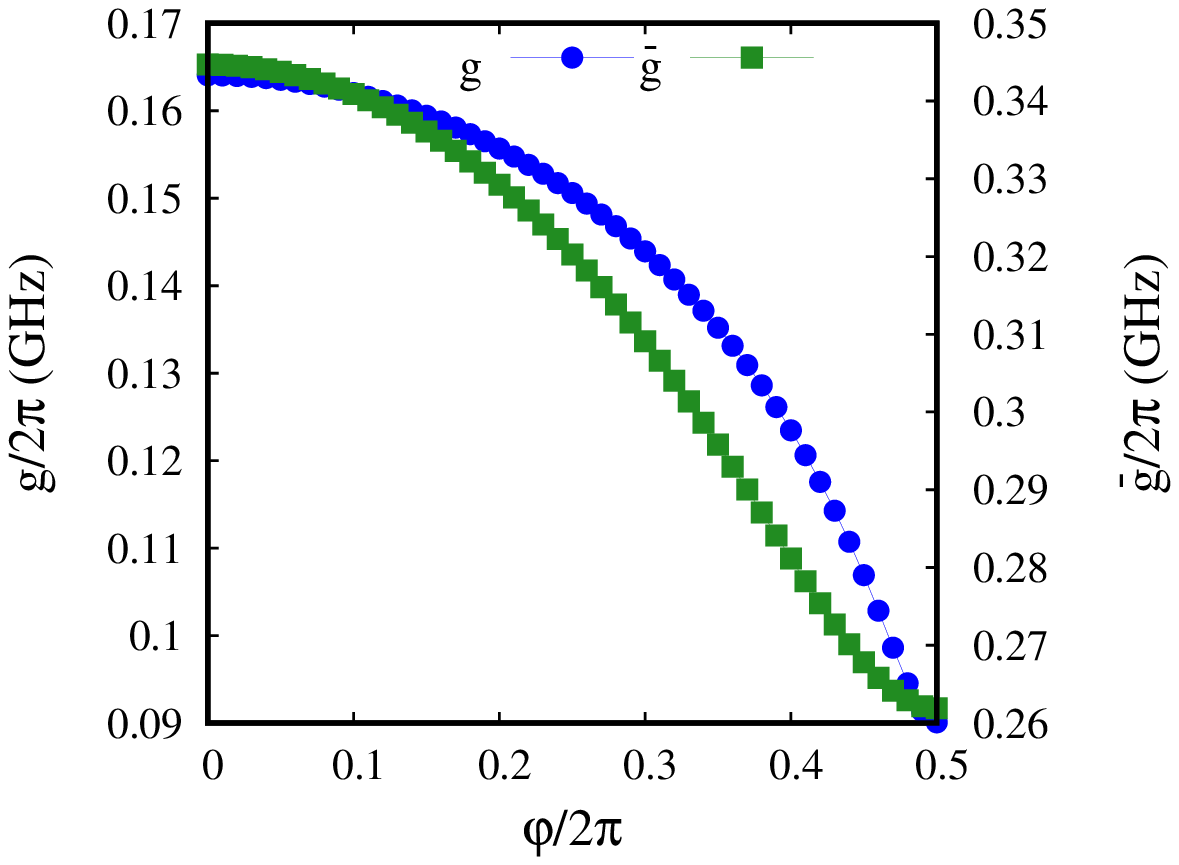}
  \end{minipage}
  \caption{(Color online) Effective interaction strengths $g$ (see \equref{eq:eff_int_I}) in blue on the left y-axis and $\bar{g}$ (see \equref{eq:eff_int_II}) in green on the right y-axis as functions of the external flux $\varphi$. We use the energies listed in \tabref{tab:device_parameter_flux_tunable_coupler_chip} ($i=2$) to obtain $g$. Similarly, for $\bar{g}$ we use the parameters listed in \tabref{tab:device_parameter_resonator_coupler_chip} ($i=1$).}
  \label{fig:effective interaction strength}
\end{figure}
\renewcommand{\width}{1.0}
\begin{figure}[!tbp]
  \begin{minipage}{0.45\textwidth}
    \centering
    \includegraphics[width=\width\textwidth]{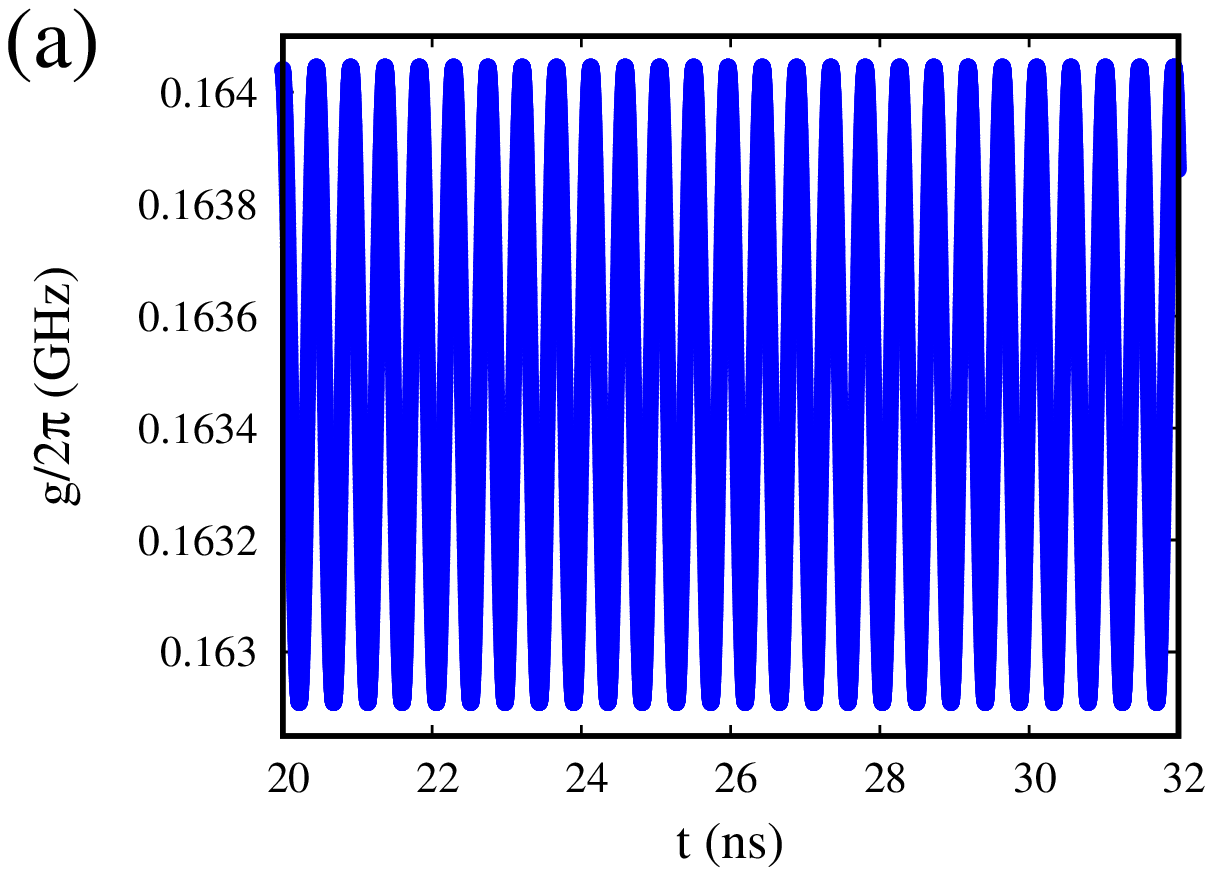}
  \end{minipage}
  \begin{minipage}{0.45\textwidth}
    \centering
    \includegraphics[width=\width\textwidth]{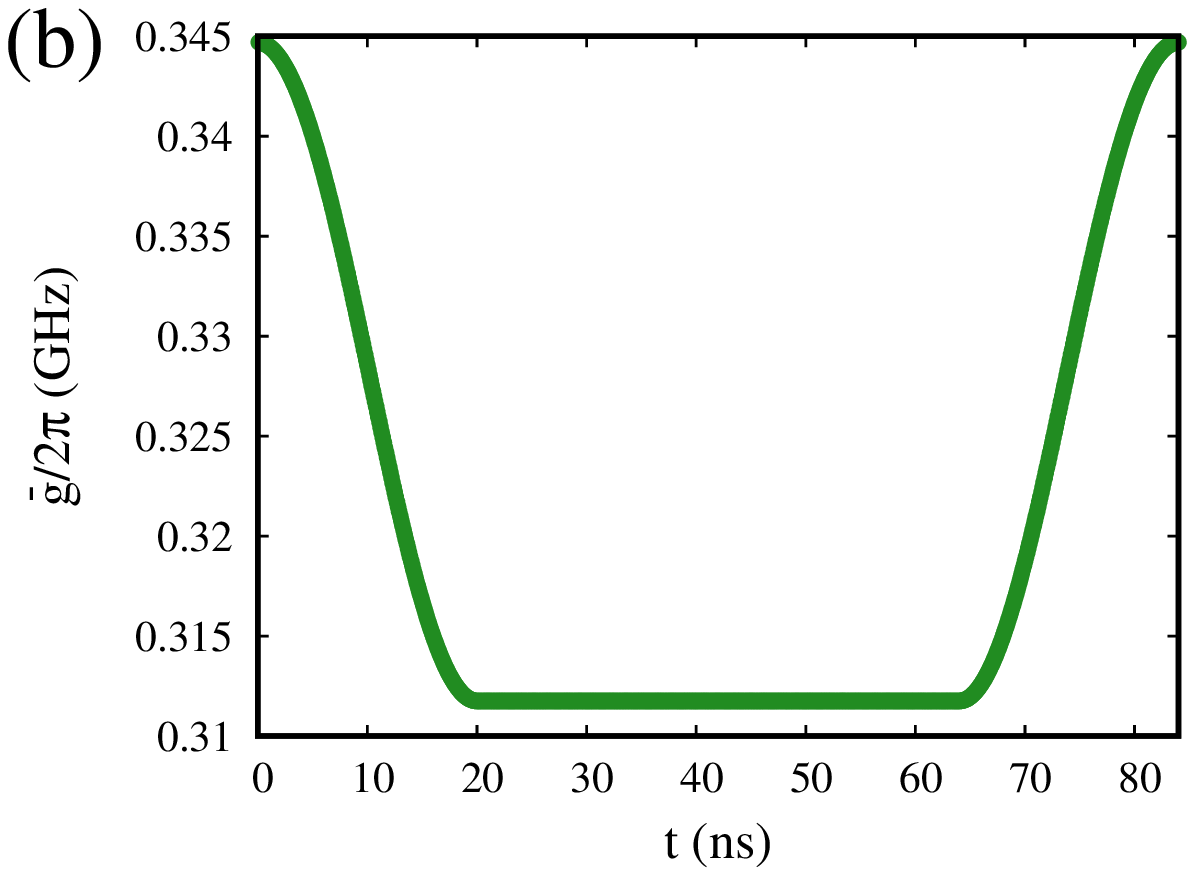}
  \end{minipage}
  \caption{(Color online) Effective interaction strength as a function of time for two different flux control pulses. Figure \ref{fig:eff_int_time_evo}(a): effective interaction strength $g(\varphi(t))$ (see \equref{eq:eff_int_I} and \equref{eq:pulse}) for architecture I. We use the energies listed in \tabref{tab:device_parameter_flux_tunable_coupler_chip} ($i=2$) to obtain $g$ and the same control pulse parameters as in \figref{fig:pulse_time_evo}(a). These control pulse parameters are also listed in \tabref{tab:summary_effective_hamiltonian_results} (row six). Figure \ref{fig:eff_int_time_evo}(b): effective interaction strength $\bar{g}(\varphi(t))$ (see \equref{eq:eff_int_II} and \equref{eq:pulse}) for architecture II. We use the energies listed \tabref{tab:device_parameter_resonator_coupler_chip} ($i=1$) to obtain $\bar{g}$ and the same control pulse parameters as in \figref{fig:pulse_time_evo}(b). These control pulse parameters are also listed in \tabref{tab:summary_effective_hamiltonian_results} (row nine).}
  \label{fig:eff_int_time_evo}
\end{figure}

It is common practice, see Refs.~\listcitetwo, that multi-qubit Hamiltonians are simplified by making assumptions about the effective parameters which influence the dynamics of the system. We begin this section with a discussion of one of these assumptions, namely that the effective interaction strength $g$ (see \equaref{eq:architecture I effective}{eq:architecture II effective}) between the different subsystems is time independent.

Figures~\ref{fig:effective interaction strength}(a-b) show the effective interaction strengths $g$ (in blue on the left y-axis) for architecture I and $\bar{g}$ (in green on the right y-axis) for architecture II as functions of the external flux $\varphi/2\pi$. The values for $g(\varphi)$ were determined with \equref{eq:eff_int_I} and the parameters listed in \tabref{tab:device_parameter_flux_tunable_coupler_chip}, row $i=2$. Similarly, the values for $\bar{g}(\varphi)$ are obtained with the parameters listed in \tabref{tab:device_parameter_resonator_coupler_chip}, row $i=1$ and \equref{eq:eff_int_II}. We can see that both effective interaction strengths show a similar qualitative and quantitative behaviour. As one can see, $g(\varphi)$ varies around $0.075$ GHz, over the interval $\varphi/2\pi \in [0,0.5]$, while $\bar{g}(\varphi)$ spans over a range of $0.08$ GHz.

Figures~\ref{fig:eff_int_time_evo}(a-b) shows the evolution of the effective interaction strength as a functions of time $t$. In \figref{fig:eff_int_time_evo}(a) we show the effective interaction strength $g(\varphi(t))$ (see \equref{eq:eff_int_I} and \equref{eq:pulse}) for architecture I. Here we use the same parameters as in \figref{fig:pulse_time_evo}(a) to model the control pulse $\varphi(t)$ and the energies listed in \tabref{tab:device_parameter_flux_tunable_coupler_chip}, row $i=2$, to obtain $g$. In this case we observe fast oscillating variations of $g$ at the order of $1$ MHz. Similarly, in \figref{fig:eff_int_time_evo}(b) we show the effective interaction strength $\bar{g}(\varphi(t))$ (see \equref{eq:eff_int_II} and \equref{eq:pulse}) for architecture II. Here we use the same control pulse parameters as in \figref{fig:pulse_time_evo}(b) and the energies listed in \tabref{tab:device_parameter_resonator_coupler_chip}, row $i=1$. As one can see, in this case we find that if the pulse has reached its plateau, the effective interaction strength has been reduced by about $31$ MHz.

Since architecture I is usually operated around a fixed flux offset $\varphi_{0}$, i.e.~we only use small pulse amplitudes $\delta$, we would expect that small variations of the effective interaction do not matter too much. The same reasoning would suggest that, in case of architecture II, the time-dependence of $\bar{g}(\varphi)$ is much more relevant since here we vary the external flux over a much larger interval. Furthermore, the unimodal pulse lowers the effective interaction strength temporarily, for about eighty percent of the total gate duration, and it does not oscillate. However, in the following section we show that this reasoning is not sound. We find that the time-dependent effective interaction strength affects architecture I much more than architecture II. We show this by performing all simulations twice, i.e.~we simulate the systems with and without a time-dependent interaction strength.

\begingroup
\begin{table*}[!tbp]
\caption{Summary of all model and pulse parameters used to perform simulations of the circuit Hamiltonians in \equref{eq:flux-tunable transmon}, \equref{eq:architecture I} and \equref{eq:architecture II}, see \appref{sec:CircuitHamiltonianSimulations}. The first column lists the model Hamiltonian and the system parameters (in form of references). The second column states which gate is modelled. The third column gives the states which are being controlled. The next columns show the following pulse parameters: the drive frequency $\omega^{D}/2 \pi$ in GHz, the amplitude $\delta/2 \pi$ in units of the flux quantum $\phi_{0}$, the rise and fall time $T_{\mathrm{r/f}}$ in ns and the gate duration $T_{\mathrm{d}}$ in ns. The last column shows the number of basis states $N_{m}$ which are needed to obtain an accurate solution.}
\label{tab:summary_circuit_hamiltonian_results}
\begin{ruledtabular}
\begin{tabular}{cccccccc}
Hamiltonian and parameters  &Gate & States $z$  &$\omega^{D}/2 \pi$ & $\delta/2 \pi$ & $T_{\mathrm{r/f}}$ & $T_{\mathrm{d}}$ & $N_{m}$  \\
\hline
\equref{eq:flux-tunable transmon} and \tabref{tab:device_parameter_flux_tunable_coupler_chip}  &$X$ & $\{(0),(1)\}$ & $7.636$ & $0.001$ & $10$ & $20$ & $3$  \\

\equref{eq:flux-tunable transmon} and \tabref{tab:device_parameter_flux_tunable_coupler_chip}  &$X$ & $\{(0),(1)\}$ & $7.636$ & $0.01$ & $100$ & $200$ & $3$ \\

\equref{eq:architecture I} and \tabref{tab:device_parameter_flux_tunable_coupler_chip}  & $X$ & $\{(0,0,0),(0,1,0)\}$ &$6.183$ & $0.045$  & $22.5$ & $45$ & $3$\\

\equref{eq:architecture I} and \tabref{tab:device_parameter_flux_tunable_coupler_chip}  & $X$ & $\{(0,0,0),(0,0,1)\}$ &$5.092$ & $0.085$  & $25$ & $50$ & $3$\\

\equref{eq:architecture I} and \tabref{tab:device_parameter_flux_tunable_coupler_chip}  & $\text{Iswap}$ &  $\{(0,1,0),(0,0,1)\}$ &$1.089$ & $0.075$ & $13$ & $209.40$  & $6$  \\

\equref{eq:architecture I} and \tabref{tab:device_parameter_flux_tunable_coupler_chip} &$\text{Cz}$ & $\{(0,1,1),(0,2,0)\}$ & $0.809$ & $0.085$ & $13$ & $297.55$ & $8$\\

\equref{eq:architecture II} and \tabref{tab:device_parameter_resonator_coupler_chip}    &$\text{Iswap}$ & $\{(0,1,0),(0,0,1)\}$ & $0$ & $0.289$ & $20$ & $100$ & $14$\\

\equref{eq:architecture II} and \tabref{tab:device_parameter_resonator_coupler_chip}  &$\text{Cz}$ & $\{(0,1,1),(0,0,2)\}$ & $0$ & $0.3335$ & $20$ & $125$ & $16$\\
\end{tabular}
\end{ruledtabular}

\caption{Summary of all pulse parameters we use to perform the simulations of the effective models \equref{eq:HamTrafo_main}, \equref{eq:flux-tunable transmon effective}, \equaref{eq:architecture I effective}{eq:architecture II effective}. The first column lists the model Hamiltonian and the system parameters (in form of references). The second column shows which case we simulate. In case A we use a static interaction strength and a non-adjusted spectrum to model the system. In case B we use a time-dependent interaction and a non-adjusted spectrum to obtain the results. Similarly, in case C we use a time-dependent interaction strength and an adjusted spectrum. The third column displays the figure which contains the results. The fourth column states which gate we model. The fifth column shows the states which are being controlled. The next columns show the following pulse parameters: the drive frequency $\omega^{D}/2 \pi$ in GHz, the amplitude $\delta/2\pi$ in units of the flux quantum $\phi_{0}$, the rise and fall time $T_{\mathrm{r/f}}$ in ns and the gate duration $T_{\mathrm{d}}$ in ns. In the last column we state whether or not is was possible to model the gate (see \secref{sec:Single-qubit operations effective} for more details). If it is not possible to model a transition, we label the corresponding parameters with not applicable (n/a).}
\label{tab:summary_effective_hamiltonian_results}
\begin{ruledtabular}
\begin{tabular}{cccccccccc}
Hamiltonian and parameters & Case &  Fig. &Gate & States $z$ &$\omega^{D}/2 \pi$ & $\delta/2 \pi$ & $T_{\mathrm{r/f}}$ & $T_{\mathrm{d}}$ & Can be modelled?   \\
\hline
\equref{eq:HamTrafo_main} and \tabref{tab:device_parameter_flux_tunable_coupler_chip} & n/a & n/a &$X$ & $\{(0),(1)\}$ & $7.643$ & $0.01$ & $10$ & $20$ & Yes \\

\equref{eq:HamTrafo_main} and \tabref{tab:device_parameter_flux_tunable_coupler_chip} & n/a & \figref{fig:TimeEvolComp}(b) & $X$ & $\{(0),(1)\}$ & $7.643$ & $0.001$ & $100$ & $200$ & Yes \\

\equref{eq:flux-tunable transmon effective} and \tabref{tab:device_parameter_flux_tunable_coupler_chip_effective} & n/a & n/a &$X$ & $\{(0),(1)\}$ & n/a & n/a & n/a & n/a & No  \\

\equref{eq:flux-tunable transmon effective} and \tabref{tab:device_parameter_flux_tunable_coupler_chip_effective} & n/a & n/a &$X$ & $\{(0),(1)\}$ & n/a & n/a & n/a & n/a & No \\

\equref{eq:architecture I effective} and \tabref{tab:device_parameter_flux_tunable_coupler_chip_effective} & n/a & n/a & $X$ & $\{(0,0,0),(0,1,0)\}$ & n/a & n/a  & n/a & n/a & No\\

\equref{eq:architecture I effective} and \tabref{tab:device_parameter_flux_tunable_coupler_chip_effective} & n/a & n/a & $X$ & $\{(0,0,0),(0,0,1)\}$ & n/a & n/a  & n/a & n/a & No\\

\equref{eq:architecture I effective} and \tabref{tab:device_parameter_flux_tunable_coupler_chip_effective} & A &\figref{fig:two_qubit_gates_tunable_coupler_eff}(a) & $\text{Iswap}$ &  $\{(0,1,0),(0,0,1)\}$ &$1.088$ & $0.075$ & $13$ & $139.6$  & Yes  \\

\equref{eq:architecture I effective} and \tabref{tab:device_parameter_flux_tunable_coupler_chip} & B &\figref{fig:two_qubit_gates_tunable_coupler_eff}(b) & $\text{Iswap}$ &  $\{(0,1,0),(0,0,1)\}$ &$1.089$ & $0.075$ & $13$ & $205.4$  & Yes  \\

\equref{eq:architecture I effective} and \tabref{tab:device_parameter_flux_tunable_coupler_chip_effective} & A &\figref{fig:two_qubit_gates_tunable_coupler_eff}(c) &$\text{Cz}$ & $\{(0,1,1),(0,2,0)\}$ & $0.807$ & $0.085$ & $13$ & $196.5$ & Yes\\

\equref{eq:architecture I effective} and \tabref{tab:device_parameter_flux_tunable_coupler_chip} & B &\figref{fig:two_qubit_gates_tunable_coupler_eff}(d) &$\text{Cz}$ & $\{(0,1,1),(0,2,0)\}$ & $0.807$ & $0.085$ & $13$ & $272.00$ & Yes\\

\equref{eq:architecture II effective} and \tabref{tab:device_parameter_resonator_coupler_chip_effective}   & A &\figref{fig:two_qubit_gates_resonator_coupler_eff}(a) &$\text{Iswap}$ & $\{(0,1,0),(0,0,1)\}$ & $0$ & $0.297$ & $20$ & $84$ & Yes\\

\equref{eq:architecture II effective} and \tabref{tab:device_parameter_resonator_coupler_chip}   & C &\figref{fig:two_qubit_gates_resonator_coupler_eff}(b) &$\text{Iswap}$ & $\{(0,1,0),(0,0,1)\}$ & $0$ & $0.289$ & $20$ & $96$ & Yes\\

\equref{eq:architecture II effective} and \tabref{tab:device_parameter_resonator_coupler_chip_effective} & A &\figref{fig:two_qubit_gates_resonator_coupler_eff}(c) &$\text{Cz}$ & $\{(0,1,1),(0,0,2)\}$ & $0$ & $0.343$ & $20$ & $105$ & Yes\\

\equref{eq:architecture II effective} and \tabref{tab:device_parameter_resonator_coupler_chip} & C &\figref{fig:two_qubit_gates_resonator_coupler_eff}(d) &$\text{Cz}$ & $\{(0,1,1),(0,0,2)\}$ & $0$ & $0.334$ & $20$ & $121$ & Yes\\
\end{tabular}
\end{ruledtabular}
\end{table*}
\endgroup

In \appref{sec:CircuitHamiltonianSimulations}, we study transitions between states of the circuit Hamiltonian models which can be used to implement Iswap and Cz gates on different circuit architectures, see circuit Hamiltonians in \equaref{eq:architecture I}{eq:architecture II}. In case of architecture I we applied a harmonic control pulse of the form \equref{eq:pulse} to the tunable coupler. On architecture II we activated transitions between different states by means of a unimodal pulse, i.e.~in \equref{eq:pulse} we set $\omega^{D}=0$. Here we create avoided crossings between different energy levels. In \secaref{sec:two-qubitArchitectureI}{sec:two-qubitArchitectureII} we repeat this analysis with the effective model Hamiltonians \equaref{eq:architecture I effective}{eq:architecture II effective} and compare the results with the ones for the circuit Hamiltonian models which can be found in \tabref{tab:summary_circuit_hamiltonian_results}. A summary of all results for the effective models can be found in \tabref{tab:summary_effective_hamiltonian_results}.

\subsubsection{Architecture I}\label{sec:two-qubitArchitectureI}
We consider the model Hamiltonian \equref{eq:architecture I effective}. The simulation parameters are listed in \tabsref{tab:device_parameter_flux_tunable_coupler_chip}{tab:device_parameter_flux_tunable_coupler_chip_effective}. Note that we need the capacitive and Josephson energies if we model the time-dependent effective interaction strength with \equref{eq:eff_int_I}. We first discuss the two different Iswap transitions (see \figsref{fig:two_qubit_gates_tunable_coupler_eff}(a,b)) and then the Cz transitions (see \figsref{fig:two_qubit_gates_tunable_coupler_eff}(c,d)). Afterwards, we further investigate the transitioning from a model with a static effective interaction strength $g$ to a model with a time-dependent effective interaction strength $g(t)$ (see \figref{fig:sketch_int_strength_sim} and \figsref{fig:eff_int_scaling}(a-b)).
\renewcommand{\width}{0.45}
\begin{figure}[!tbp]
    \centering
        \begin{minipage}{0.5\textwidth}
        \centering
        \includegraphics[width=\width\textwidth]{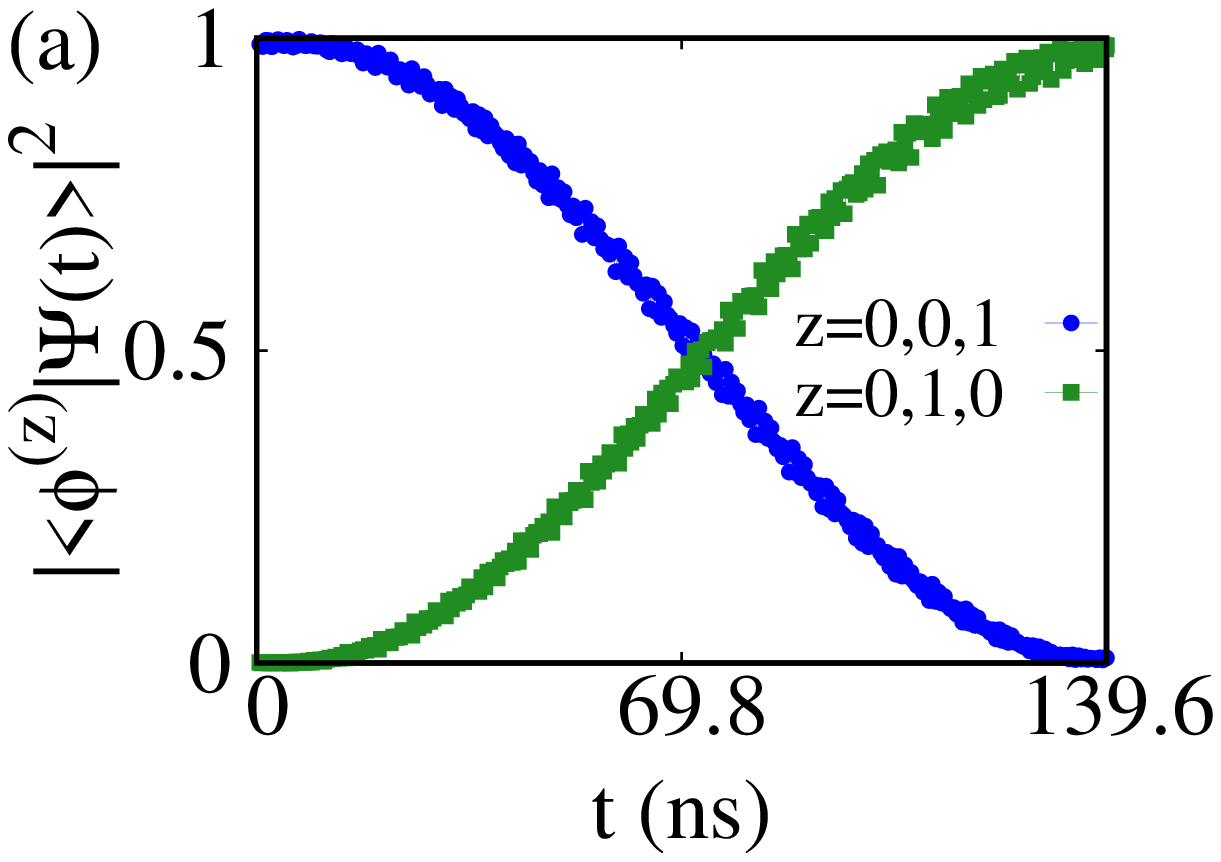} 
        \includegraphics[width=\width\textwidth]{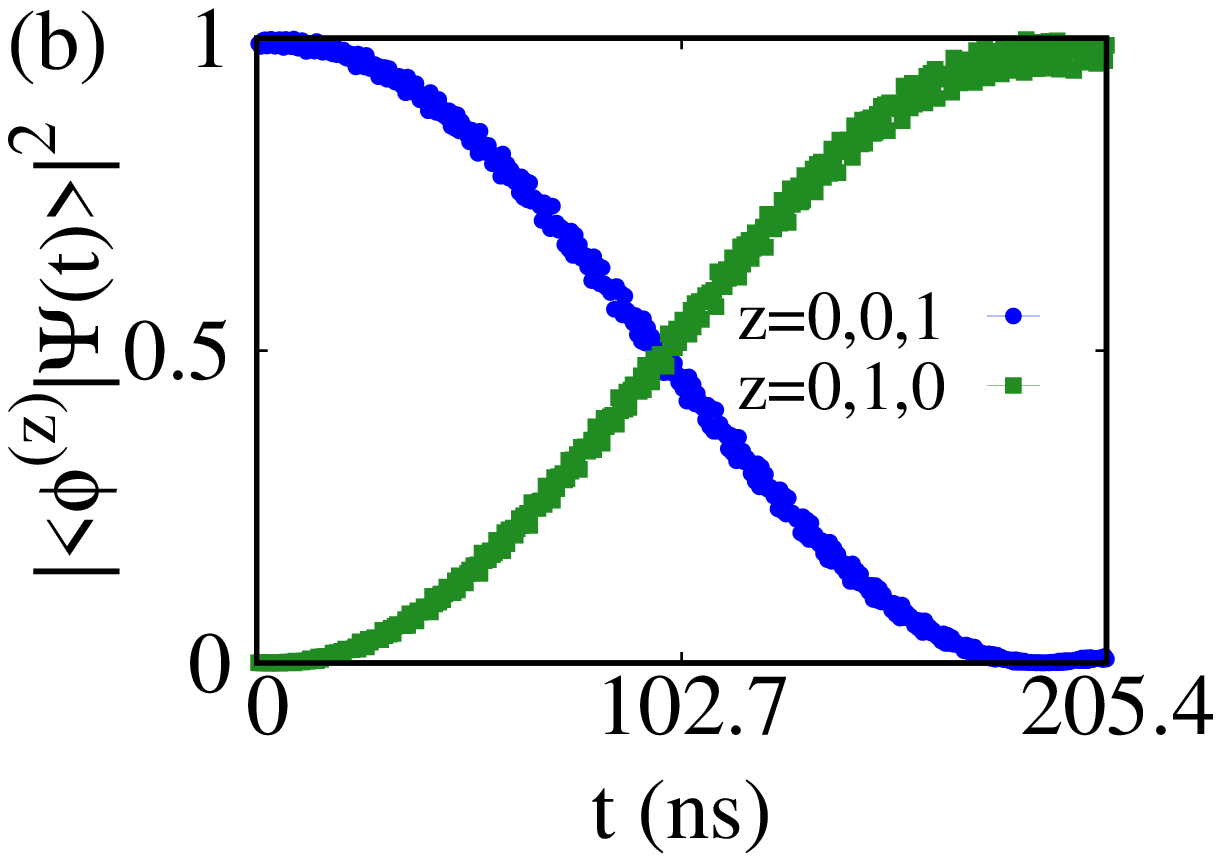} 
        \includegraphics[width=\width\textwidth]{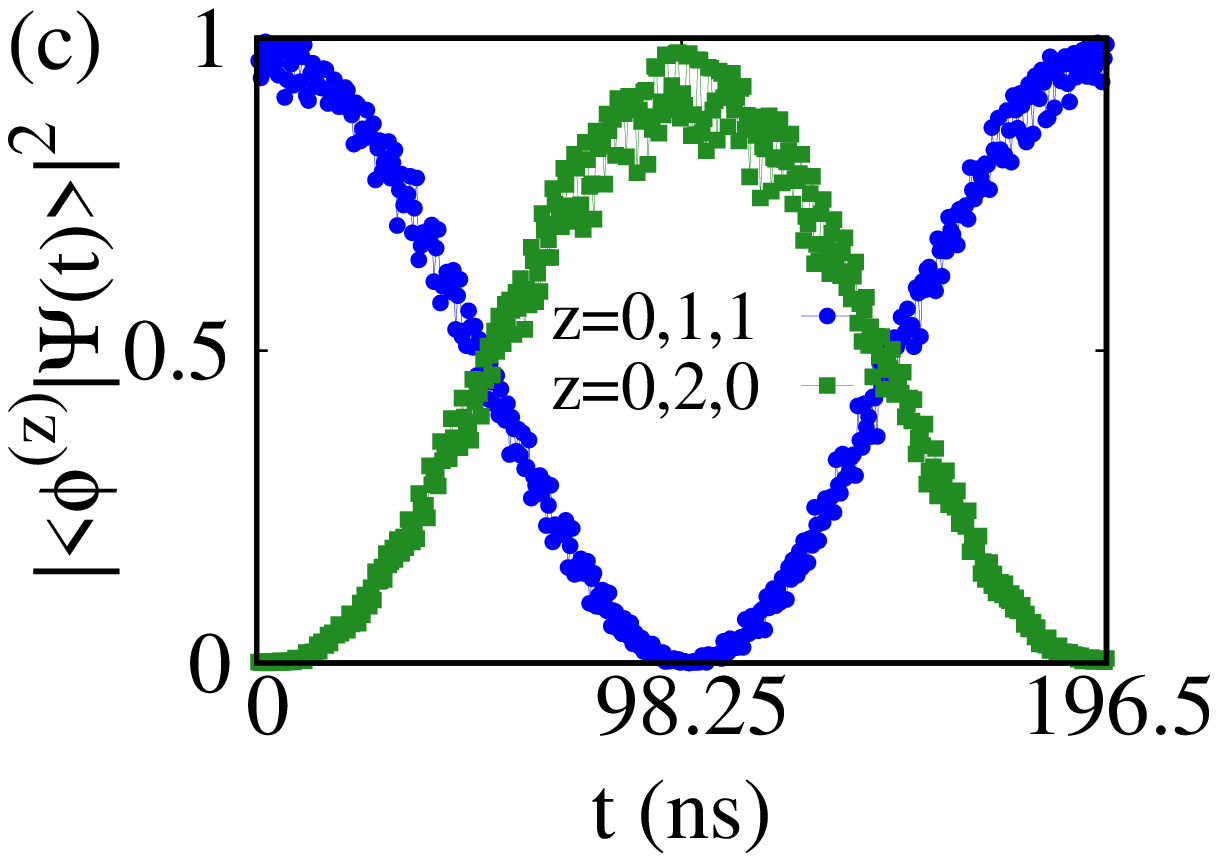} 
        \includegraphics[width=\width\textwidth]{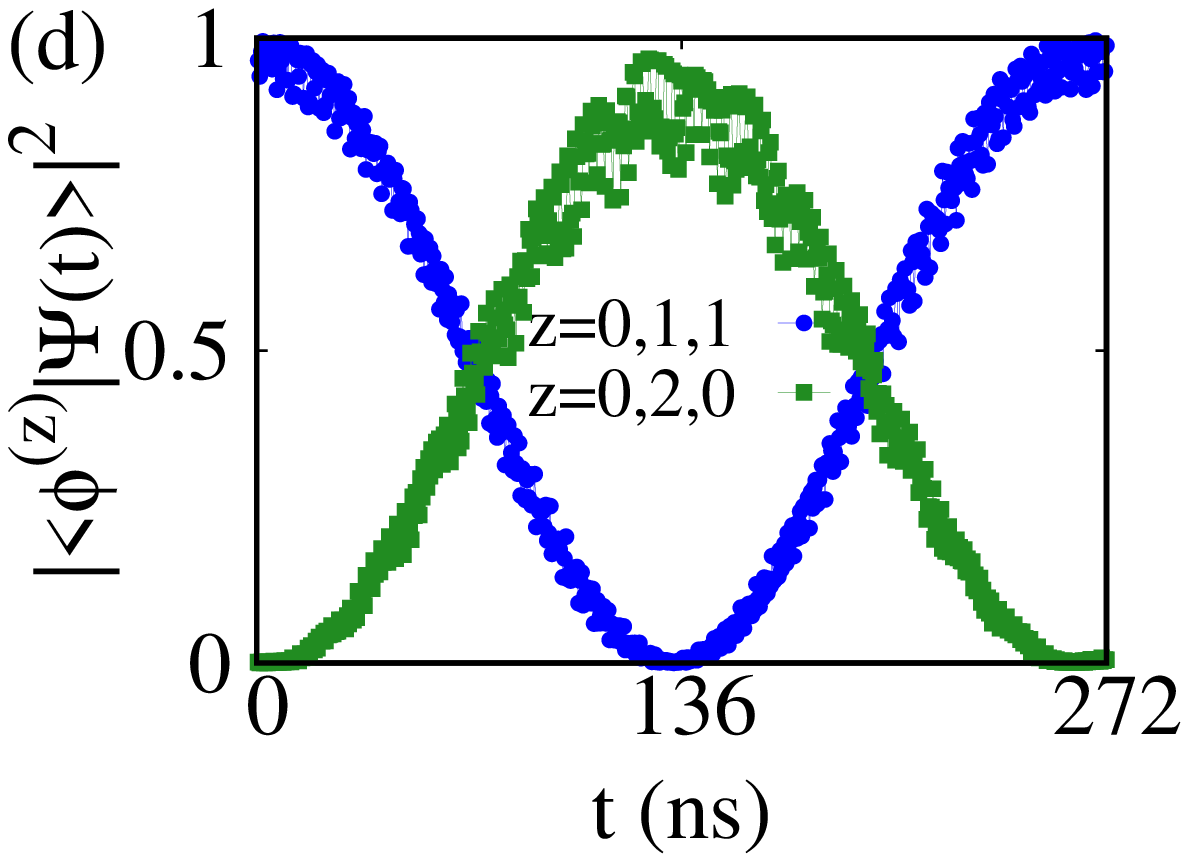} 
        \end{minipage}
        \caption{(Color online) Panels (a) and (b): Probabilities $p^{(0,0,1)}(t)$ and $p^{(0,1,0)}(t)$ as functions of time $t$. Panels (c) and (d): Probabilities $p^{(0,1,1)}(t)$ and $p^{(0,2,0)}(t)$ as functions of time $t$. In panel (a,c) we model the system without a time-dependent effective interaction strength, see \equref{eq:eff_int_I}. In panel (b,d) we include the time dependence. In all cases we use the Hamiltonian in \equref{eq:architecture I effective}, the parameters listed in \tabsref{tab:device_parameter_flux_tunable_coupler_chip}{tab:device_parameter_flux_tunable_coupler_chip_effective} and a pulse of the form \equref{eq:pulse} to obtain the results. The pulse parameters are discussed in the main text. The $z=(0,1,1) \rightarrow z=(0,2,0)$ transitions are usually used to implement Cz operations and the $z=(0,0,1) \rightarrow z=(0,1,0)$ transitions are often used to realise Iswap operations, see Refs.\cite{Ganzhorn20,Bengtsson2020}. Interestingly, we observe a large shift in the pulse duration $T_{\mathrm{d}}$ if we model the system with a time-dependent effective interaction strength, see \figref{fig:eff_int_time_evo}(a).}
        \label{fig:two_qubit_gates_tunable_coupler_eff}
\end{figure}

Figure~\ref{fig:two_qubit_gates_tunable_coupler_eff}(a) shows the probabilities $p^{(0,0,1)}(t)$ and $p^{(0,1,0)}(t)$ as functions of time $t$. We use a static effective interaction strength $g$ to model the system, i.e.~we use the effective interaction strength which is determined by the flux offset $\varphi_{0}/2\pi=0.15$. We find a resonance frequency or optimal drive frequency of $\omega^{D}=1.088$ GHz. This frequency deviates only $2$ MHz from the one we found for the corresponding circuit Hamiltonian model, see \tabref{tab:summary_circuit_hamiltonian_results}. The drive amplitude which is $\delta/2\pi=0.075$, is the same amplitude we use in \tabref{tab:summary_circuit_hamiltonian_results}. However, with these pulse parameters we find a gate duration of $139.6$ ns. This means we can implement this gate around 70 ns faster than in the case of the circuit Hamiltonian \equref{eq:architecture I}, see \tabref{tab:summary_circuit_hamiltonian_results}. This is a rather strong difference.

Figure~\ref{fig:two_qubit_gates_tunable_coupler_eff}(b) shows the probabilities $p^{(0,0,1)}(t)$ and $p^{(0,1,0)}(t)$ as functions of time $t$. We use a time-dependent effective interaction strength to model the dynamics of the system.

Note that the effective interaction strengths $g(\varphi)$ for an external flux of $\varphi/2\pi=0.075$ and $\varphi/2\pi=0.15$ deviate from one another by roughly 3 MHz. Apart from the effective interaction strength, we only adjusted the drive frequency slightly. Here we find an optimal drive frequency of $\omega^{D}=1.089$ GHz. As one can see, the gate duration in this case is 205.4 ns. Therefore, we find that the deviations between the gate durations, for both models \equaref{eq:architecture I}{eq:architecture I effective}, decrease to 4 ns if we model the system with a time-dependent interaction strength.

Figures~\ref{fig:two_qubit_gates_tunable_coupler_eff}(c,d) show the same scenarios for the Cz operation, i.e. we display the time evolution of $p^{(0,1,1)}(t)$ and $p^{(0,2,0)}(t)$ for two different models. In \figref{fig:two_qubit_gates_tunable_coupler_eff}(c) we model the system with a time-independent effective interaction strength and in \figref{fig:two_qubit_gates_tunable_coupler_eff}(d) we include the time dependence. In both cases we find the optimal drive frequency $\omega^{D}=0.807$ GHz. If we compare this drive frequency with the one we obtained for the circuit Hamiltonian, see \tabref{tab:summary_circuit_hamiltonian_results}, we see that there is a shift of $2$ MHz. Additionally, both control pulses are calibrated with an amplitude of $\delta/2\pi=0.085$.

We observe that if we model the system with a time-independent effective interaction strength, we find a gate duration of 196.5 ns. Including the time dependence leads to a gate duration of 272 ns. A comparison between theses results and the ones given in \appref{sec:CircuitHamiltonianSimulations} leads to a deviation of around 25 ns if we include the time-dependent effective interaction strength.

In order to better understand the behaviour of the transitioning from a model with a static effective interaction strength to a model with a time-dependent effective interaction strength, we performed more simulations. The results are displayed in \figsref{fig:eff_int_scaling}(a-b). Additionally, in \figref{fig:sketch_int_strength_sim} we show a functional sketch of the control pulses we use to obtain the results presented in \figsref{fig:eff_int_scaling}(a-b).

Figure~\ref{fig:sketch_int_strength_sim} shows that we use the control pulse \equref{eq:pulse} to model the tunable coupler frequency given by \equref{eq:tunable frequency} with a pulse amplitude $\delta/2\pi=\text{const.}$ and the effective interacting strength given by \equref{eq:eff_int_I} with pulse amplitudes $\delta^{*}/2\pi \in [0,0.125]$. All the remaining pulse parameters are the same for both pulses. We use $\delta/2\pi=0.075$ in \figref{fig:eff_int_scaling}(a) for the two-qubit gate Iswap transitions $z=(0,0,1) \rightarrow z=(0,1,0)$ and $\delta/2\pi=0.085$ in \figref{fig:eff_int_scaling}(b) for the two-qubit gate Cz transitions $z=(0,1,1) \rightarrow z=(0,2,0)$. If we use $\delta^{*}=0$ to model the  static effective interaction strength, we model the scenarios we presented in \figref{fig:two_qubit_gates_tunable_coupler_eff}(a,c). Similarly, if we use $\delta^{*}=\delta$, we model the scenario we presented in \figref{fig:two_qubit_gates_tunable_coupler_eff}(b,d). The values in between $\delta^{*}\in (0,\delta)$ show the transition from one case to the other. Additionally, we added some more amplitudes $\delta^{*}\in (\delta, 2\pi 0.125]$ to have some additional data which might shine some light on the effect. Note that we use the pulse duration $T_{\mathrm{d}}=300$ ns and the rise and fall time $T_{\mathrm{r/f}}=13$ ns for all simulations.

Figures \ref{fig:eff_int_scaling}(a-b) show the probabilities $p^{(z)}(t)$ for $z=(0,0,1)$(a) and $z=(0,1,1)$(b) as functions of time $t$ for different pulse amplitudes $\delta^{*}$ as explained above, see \figref{fig:sketch_int_strength_sim}. We use the Hamiltonian given by \equref{eq:architecture I effective} and the parameters listed in \tabref{tab:device_parameter_flux_tunable_coupler_chip} to model the dynamics of a system of type architecture I. The control pulses and all pulse parameters except the drive frequency $\omega^{D}$ are discussed in the preceding paragraph.

In \figref{fig:eff_int_scaling}(a) we model the Iswap transitions $z=(0,0,1) \rightarrow z=(0,1,0)$. Here we use $\omega^{D}/2\pi=1.088$ GHz, blue lines and unfilled markers and $\omega^{D}/2\pi=1.089$ GHz, green lines and filled markers, to model the dynamics of $p^{(0,0,1)}(t)$. Note that the drive frequency which leads to full population exchange between the two states involved only shifts by one MHz over the range $\delta^{*}/2\pi \in [0,0.125]$. As one can see, at first for $\delta^{*}/2\pi \in [0,0.010]$ the qualitative and quantitative behaviour of the overall transition $z=(0,0,1) \rightarrow z=(0,1,0)$ is barely affected by the time-dependent effective interaction strength. Then for $\delta^{*}/2\pi \in [0.050,0.125]$ every increase in the pulse amplitude leads to a shift of the first minimum of $p^{(0,0,1)}(t)$ of more than $25$ ns.

In \figref{fig:eff_int_scaling}(b) we use $\omega^{D}/2\pi=0.807$ GHz, blue lines and unfilled markers and $\omega^{D}/2\pi=0.808$ GHz, green lines and filled markers, to model the dynamics of $p^{(z)}(t)$. Here we find a similar qualitative behaviour as in \figref{fig:eff_int_scaling}(a). At first, the overall behaviour of the transition $z=(0,1,1) \rightarrow z=(0,2,0)$ is not much affected by $g(t)$. Then we can observe how the first minimum of $p^{(0,1,1)}(t)$ moves roughly in steps of $25$ ns to the right of the x-axis.

\renewcommand{\width}{1}
\begin{figure}[!tbp]
  \begin{minipage}{0.45\textwidth}
    \centering
    \includegraphics[width=\width\textwidth]{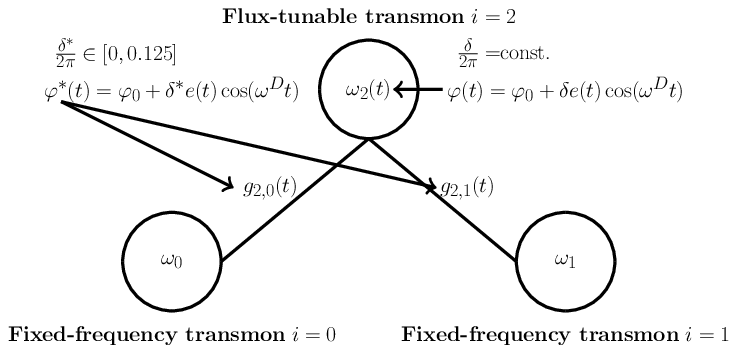} %
  \end{minipage}
  \caption{Functional sketch of the control pulses, see \equref{eq:pulse}, we use to determine the results in \figsref{fig:eff_int_scaling}(a-b). We use the Hamiltonian given by \equref{eq:architecture I effective} and the device parameters listed in \tabref{tab:device_parameter_flux_tunable_coupler_chip} to model the dynamics of a system of type architecture I. The intention is to investigate the transition from a model with a static effective interaction strength $g_{j,i}(t)$ given by \equref{eq:eff_int_I} with $\delta^{*}/2\pi=0$ to a model where the effective interaction strength oscillates with $\delta^{*}/2\pi \in (0,0.125]$, see also \figref{fig:eff_int_time_evo}(a). Therefore we keep the pulse amplitude $\delta$ for the tunable coupler frequency given by \equref{eq:tunable frequency} constant. We use $\delta/2\pi=0.075$ in \figref{fig:eff_int_scaling}(a) to model the two-qubit Iswap transitions and $\delta/2\pi=0.085$ in \figref{fig:eff_int_scaling}(b) to model the two-qubit Cz transitions. Note that these are the same pulse amplitudes we use in \figsref{fig:two_qubit_gates_tunable_coupler_eff}(a,b) and \figsref{fig:two_qubit_gates_tunable_coupler_eff}(c,d), respectively. Furthermore, if $\delta^{*}=0$ we simulate the scenarios we show in \figsref{fig:two_qubit_gates_tunable_coupler_eff}(a,c) and if $\delta^{*}=\delta$ we simulate the scenarios we show in \figsref{fig:two_qubit_gates_tunable_coupler_eff}(b,d). However, in \figsref{fig:eff_int_scaling}(a-b) the pulse duration $T_{\mathrm{d}}$ is set to 300 ns for all cases.\label{fig:sketch_int_strength_sim}}
\end{figure}

\renewcommand{\width}{1.0}
\begin{figure}[!tbp]
  \begin{minipage}{0.45\textwidth}
    \centering
    \includegraphics[width=\width\textwidth]{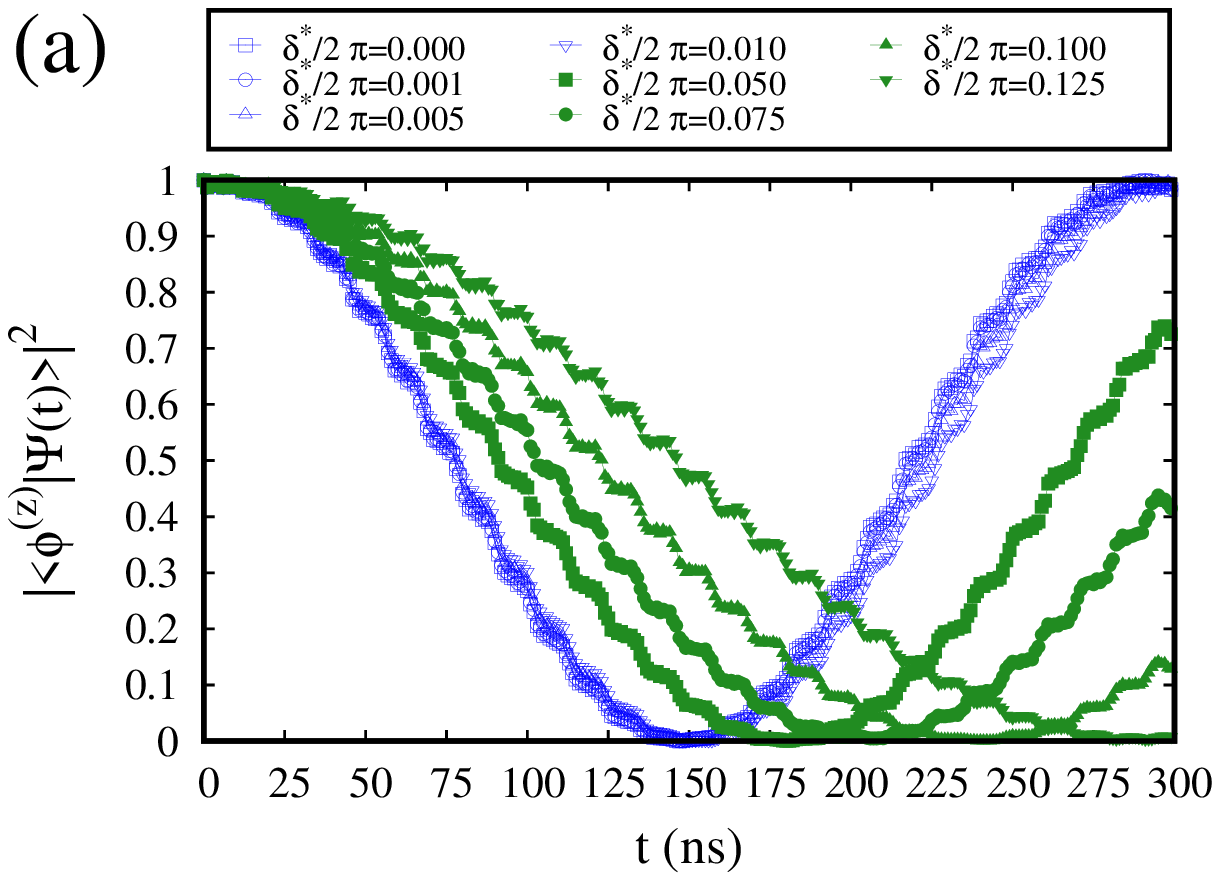}
  \end{minipage}
  \begin{minipage}{0.45\textwidth}
    \centering
    \includegraphics[width=\width\textwidth]{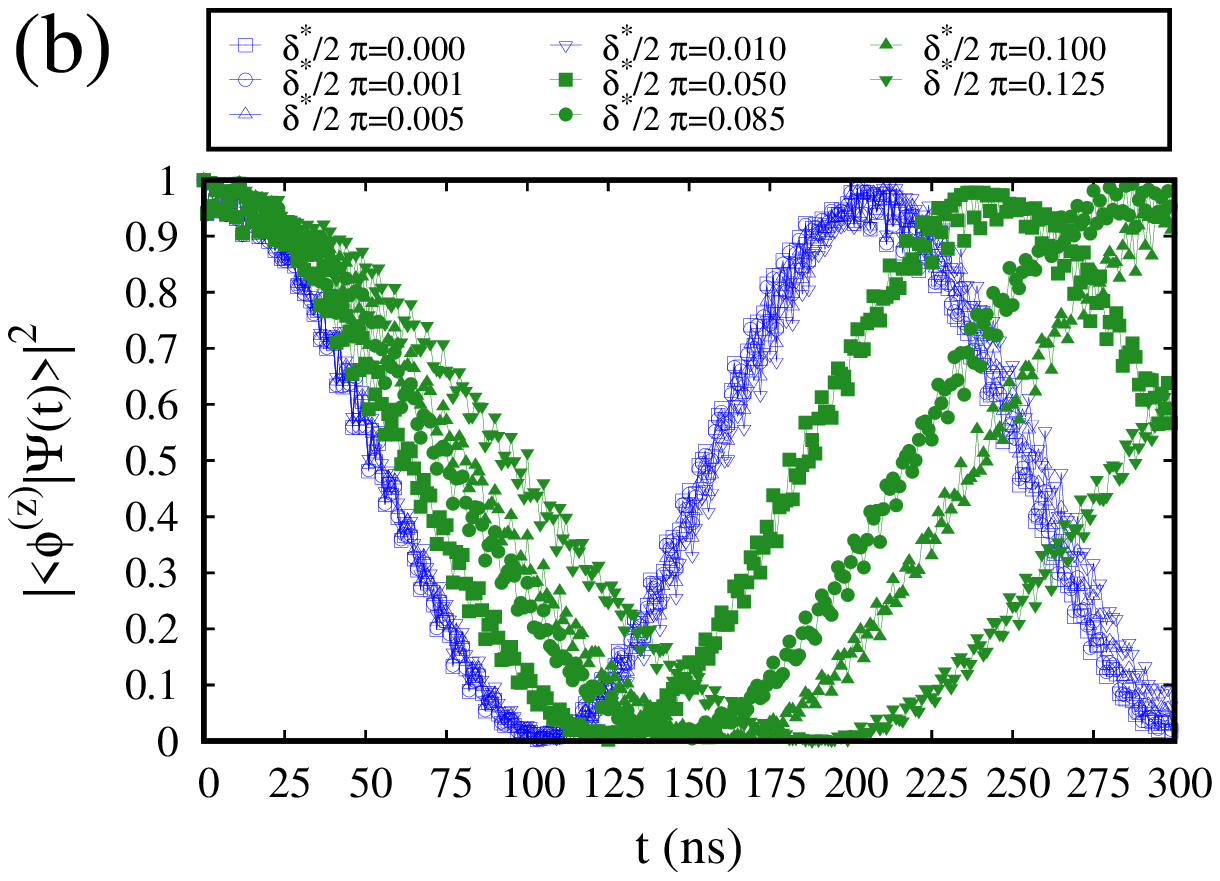}
  \end{minipage}
  \caption{(Color online) Probabilities $p^{(z)}(t)$ as functions of time $t$ for $z=(0,0,1)$(a) and $z=(0,1,1)$(b). In (a) we model transitions which might be used to implement Iswap gates. Similarly, in (b) we model transitions which might be used to implement Cz gates. Here we use the effective Hamiltonian given by \equref{eq:architecture I effective}, the device parameters listed in \tabref{tab:device_parameter_flux_tunable_coupler_chip} and the pulse \equref{eq:pulse} to obtain the results. The system is modelled with a time-dependent interaction strength $g(t)$ given by \equref{eq:eff_int_I}. Panel(a-b) show the route from the model where we use a static interaction strength, \ie with pulse amplitude $\delta^{*}=0$, to the model where the interaction strength is dynamic, \ie with pulse amplitude $\delta^{*} \neq0$. Here $\delta^{*}$ denotes the amplitude we use to model the time-dependent $g(t)$ given by \equref{eq:eff_int_I}, see also \figref{fig:eff_int_time_evo}(a). The procedure is graphically illustrated in \figref{fig:sketch_int_strength_sim}. In order to better understand how a time-dependent $g(t)$ affects the dynamics of the system, we turn on the dynamic interaction strength $\delta^{*}/2\pi \in [0,0.125]$ while keeping the amplitude $\delta$ for the tunable frequency given by \equref{eq:tunable frequency} fixed. We use $\delta/2\pi=0.075$(a) to model the Iswap transition and $\delta/2\pi=0.085$(b) to model the Cz transition. Note that these are the same amplitudes $\delta$ we use to obtain the results in \figref{fig:two_qubit_gates_tunable_coupler_eff}. In this scenario, we need to slightly adjust the drive frequencies $\omega^{D}$ as we increase $\delta^{*}$. We use $\omega^{D}/2\pi=1.088$ GHz, blue lines and unfilled markers and $\omega^{D}/2\pi=1.089$ GHz, green lines and filled markers, to model the Iswap transitions in (a). Similarly, we use $\omega^{D}/2\pi=0.807$ GHz, blue lines and unfilled markers and $\omega^{D}2\pi=0.808$ GHz, green lines and filled markers, to model the Cz transitions in (b). All results are obtained with the rise and fall time $T_{\text{r/f}}=13$ ns and the gate duration $T_{\mathrm{d}}=300$ ns.\label{fig:eff_int_scaling}}
\end{figure}

The results presented in \figsref{fig:two_qubit_gates_tunable_coupler_eff}(a-d) and \figsref{fig:eff_int_scaling}(a-b) lead to the question why the oscillations of the effective interaction strength $g(t)$ are so relevant. However, even after performing more simulations, we were not able to find a conclusive theoretical explanation for this effect. Here we simulated the time evolution of the spectrum and the relevant probabilities while turning on and off various time dependencies in the model.We leave this problem for future research.

Additionally, we also simulated the effective model given by \equref{eq:architecture I effective} with an additional non-adiabatic drive term given by \equref{eq:basis_trafo_main}, for the flux-tunable coupler. Here we find (data not shown) that the Iswap $z=(0,0,1) \rightarrow z=(0,1,0)$ and Cz $z=(0,1,1) \rightarrow z=(0,2,0)$ transitions are barely affected by the non-adiabatic drive term. Note that we tested this only for the pulse parameters listed in \tabref{tab:summary_effective_hamiltonian_results} in row seven to ten.

The remaining deviations between the effective and circuit model might be attributed to additional approximations made. For instance, we model the interaction between the different subsystems with an operator which is the result of a perturbative analysis, see \REF\cite{Koch}. Second, \REF\cite{WillschDennis2020Phd} shows that such approximations can lead to deviations which increase with time; in this case a free time evolution was considered.

In general, we find that if we consider short timescales of around 250 ns, both Hamiltonians in \equref{eq:architecture I} and \equref{eq:architecture I effective}, predict similar outcomes for only marginally different control pulses if we model the system with a time-dependent interaction strength.

\subsubsection{Architecture II}\label{sec:two-qubitArchitectureII}

\renewcommand{\width}{0.45}
\begin{figure}[!tbp]
    \centering
        \begin{minipage}{0.5\textwidth}
        \includegraphics[width=\width\textwidth]{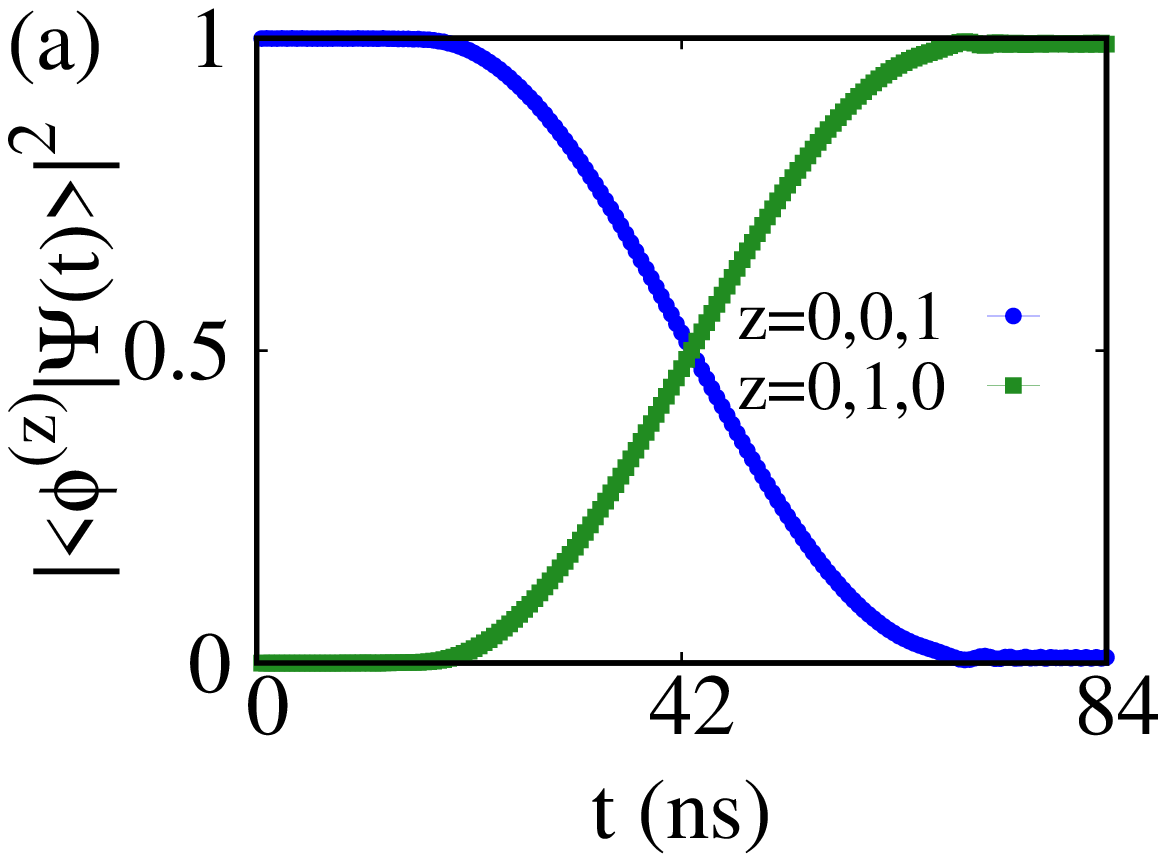}
        \includegraphics[width=\width\textwidth]{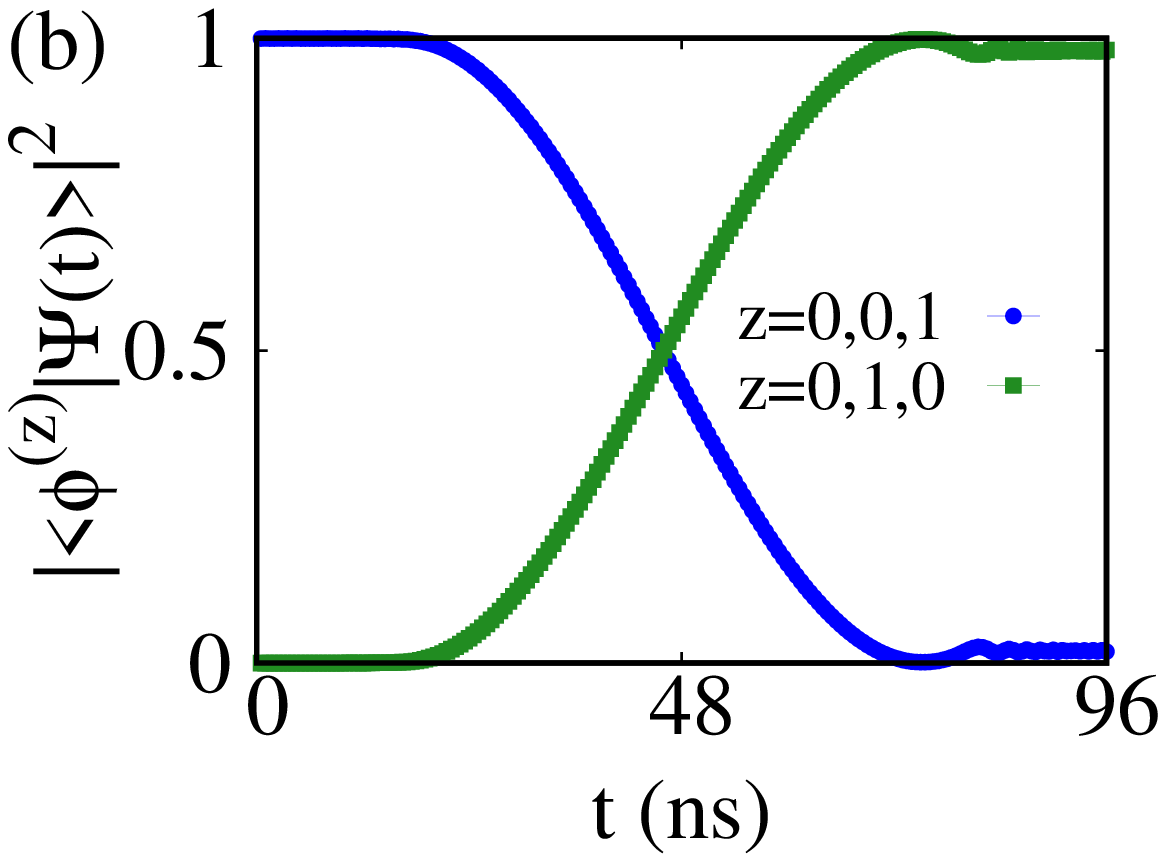}
        \includegraphics[width=\width\textwidth]{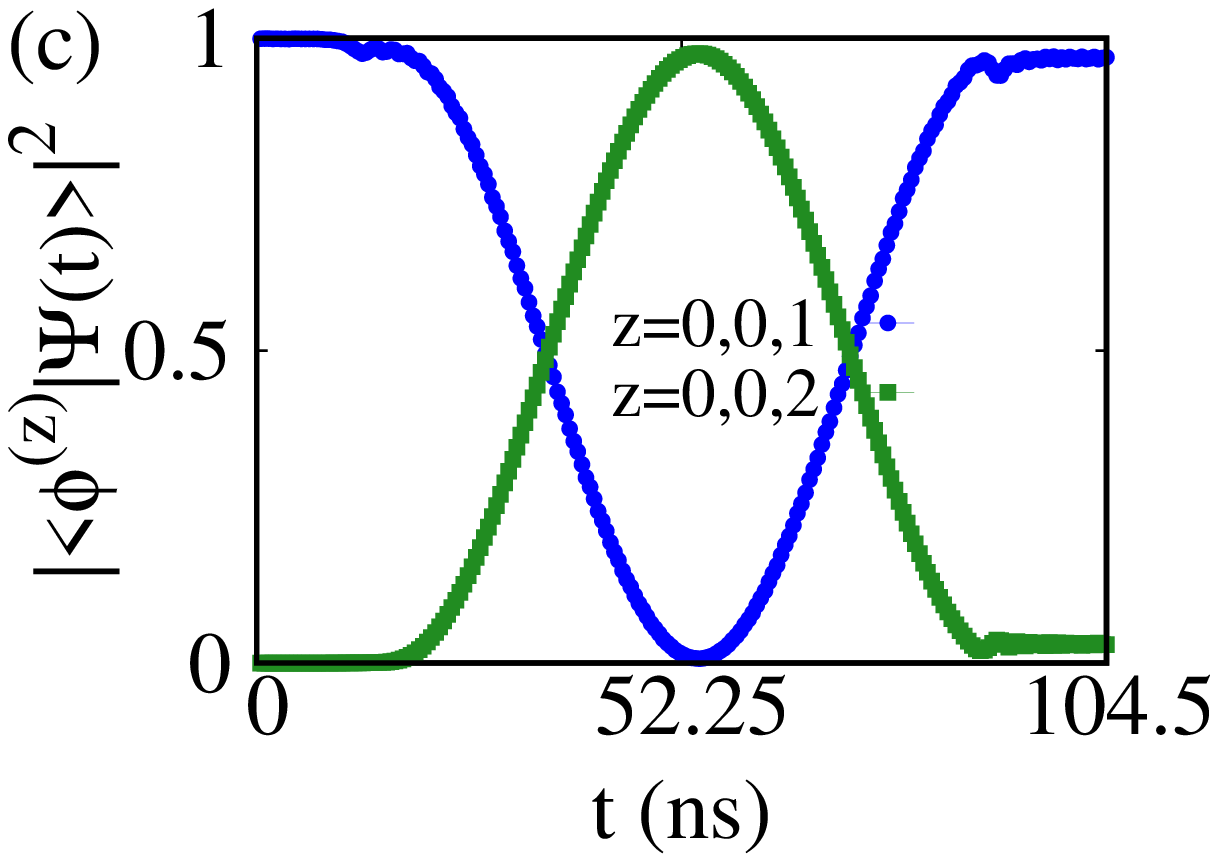}
        \includegraphics[width=\width\textwidth]{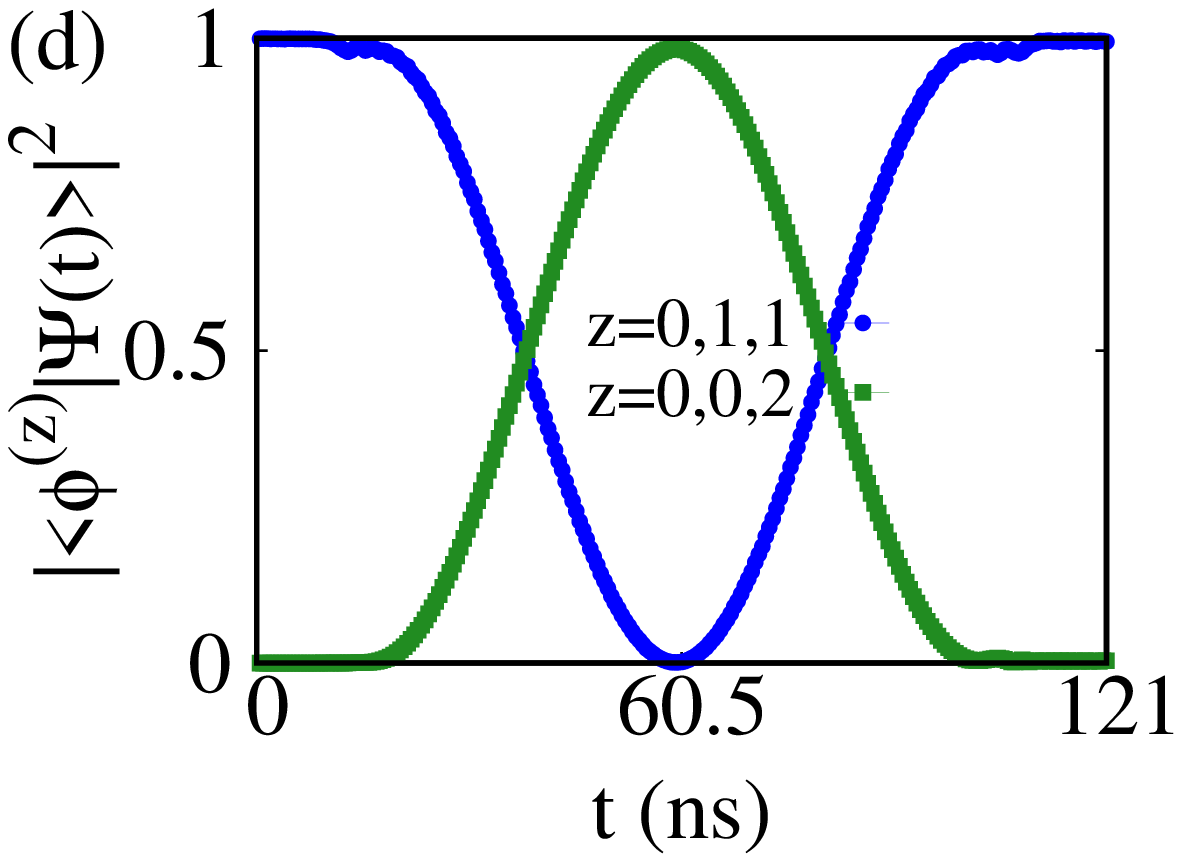}
        \end{minipage}
        \caption{(Color online) Panels (a) and (b): Probabilities $p^{(0,0,1)}(t)$ and $p^{(0,1,0)}(t)$ as functions of time $t$. Panels (c) and (d): Probabilities $p^{(0,1,1)}(t)$ and $p^{(0,0,2)}(t)$ as functions of time $t$. In panel (a,c) we model the system without a time-dependent effective interaction strength, see \equref{eq:eff_int_II}. In panel (b,d) we include the time dependence. In all cases we use the Hamiltonian in \equref{eq:architecture II effective}, the parameters listed in \tabsref{tab:device_parameter_resonator_coupler_chip}{tab:device_parameter_resonator_coupler_chip_effective} and a pulse of the form \equref{eq:pulse} to obtain the results. The pulse parameters are discussed in the main text. The $z=(0,1,1) \rightarrow z=(0,2,0)$ transitions are usually used to implement Cz operations and the $z=(0,0,1) \rightarrow z=(0,1,0)$ transitions are often used to realise Iswap operations. We observe a modest shift in the pulse duration $T_{\mathrm{d}}$ if we model the system with a time-dependent effective interaction strength, see \figref{fig:eff_int_time_evo}(b).}
        \label{fig:two_qubit_gates_resonator_coupler_eff}
\end{figure}

In the following, we compare the results of the second circuit Hamiltonian \equref{eq:architecture II} with the ones we obtain for Hamiltonian given by \equref{eq:architecture II effective}. Here we use the parameters listed in \tabsref{tab:device_parameter_resonator_coupler_chip}{tab:device_parameter_resonator_coupler_chip_effective} to obtain the results. Note that we need the parameters in \tabref{tab:device_parameter_resonator_coupler_chip} if we model the system with a time-dependent interaction strength and an adjusted spectrum, see \equaref{eq: expansion frequency}{eq: expansion anharmonicity}. Furthermore, we use a pulse of the form \equref{eq:pulse} with $\omega^{\mathrm{D}}=0$ and $T_{\mathrm{r/f}}=20$ ns in all cases. As before, we first discuss the Iswap gate (see \figsref{fig:two_qubit_gates_resonator_coupler_eff}(a,b)) and then the Cz gate (see \figsref{fig:two_qubit_gates_resonator_coupler_eff}(c,d)).

Figure~\ref{fig:two_qubit_gates_resonator_coupler_eff}(a) shows the probabilities $p^{(0,0,1)}(t)$ and $p^{(0,1,0)}(t)$ as functions of time $t$. We use a time-independent effective interaction strength to model the dynamics of the system. We find the optimal drive amplitude $\delta/2 \pi=0.297$ and a gate duration of $T_{\mathrm{d}}=84$ ns. Consequently, we observe a 16 ns discrepancy if we compare these results with the one we obtained for the circuit Hamiltonian model, see \tabref{tab:summary_circuit_hamiltonian_results}. Furthermore, the pulse amplitude has shifted. This can be explained by the fact that the flux-tunable frequency of the effective model $\omega(\varphi)$ as well as the corresponding anharmonicity $\alpha$ start to deviate from the numerically exact spectrum for large external fluxes $\varphi$, see \appref{sec: Effective Hamiltonians appendix} and \figsref{fig:approximation}(a-b).

We can correct the spectrum by using more accurate expressions (see \equaref{eq: expansion frequency}{eq: expansion anharmonicity}) for the qubit frequency and the anharmonicity. Figure~\ref{fig:two_qubit_gates_resonator_coupler_eff}(b) shows the probabilities $p^{(0,0,1)}(t)$ and $p^{(0,1,0)}(t)$ as functions of time $t$. Here we model the system with a time-dependent effective interaction strength $\bar{g}(t)$ (see \equref{eq:eff_int_II}). Furthermore, we also adjust the spectrum. We find the optimal pulse amplitude $\delta/2\pi=0.289$. This is the same amplitude we determined for the circuit Hamiltonian \equref{eq:architecture II}, see \tabref{tab:summary_circuit_hamiltonian_results}. We find a gate duration of 96 ns. Therefore, the discrepancies between the different gate duration times have decreased to 4 ns. Note that this is the same deviation we found for the other system, when we modelled the Iswap operation.

We also simulated the case (data not shown) where only the spectrum is adjusted and the effective interaction strength is constant. As before, we compute the tunable qubit frequency and anharmonicity with the series expansions in \equaref{eq: expansion frequency}{eq: expansion anharmonicity}. Here we also find an optimal pulse amplitude $\delta/2\pi=0.289$. Therefore, we conclude that this is purely a consequence of the deviations in the qubit frequency and anharmonicity, see \appref{sec: Effective Hamiltonians appendix} and \figsref{fig:approximation}(a-b).

Figures~\ref{fig:two_qubit_gates_resonator_coupler_eff}(c-d) show the probabilities $p^{(0,1,1)}(t)$ and $p^{(0,0,2)}(t)$ as functions of time $t$. Here we model the Cz gate with two different model Hamiltonians, \ie with and without the time-independent effective interaction strength and an adjusted spectrum. Figures~\ref{fig:two_qubit_gates_resonator_coupler_eff}(c,d) show the same characteristics as \figsref{fig:two_qubit_gates_resonator_coupler_eff}(a,b). We find that if we do not use an adjusted spectrum, the optimal control pulse amplitude $\delta$ requires adjustment. Furthermore, if we assume that the effective interaction strength is constant, we find a gate duration which is about $20$ ns shorter. If we adjust the spectrum, we find that the shift of the optimal drive amplitude disappears. Similarly, if we include the time-dependent effective interaction strength, we see that the gate duration increases to 121 ns. This means the differences between the effective and the circuit Hamiltonian model decrease to 4 ns. Therefore we might conclude that the time-dependence of the effective interaction strength is not negligible if the aim is to approximate the time evolution of the corresponding circuit Hamiltonian.

Finally, we also simulated the effective model given by \equref{eq:architecture II effective} with additional non-adiabatic drive terms given by \equref{eq:basis_trafo_main}, for the flux-tunable transmon qubits. Here we find (data not shown) that the Iswap $z=(0,0,1) \rightarrow z=(0,1,0)$ and Cz $z=(0,1,1) \rightarrow z=(0,0,2)$ transitions are barely affected by the non-adiabatic drive terms which we add to the model. Note that we tested this only for the pulse parameters listed in \tabref{tab:summary_effective_hamiltonian_results} in row eleven to fourteen.

In summary, we observe that if we adjust the spectrum of the effective model and include the time dependence of the effective interaction strength, the effective Hamiltonian \equref{eq:architecture II effective} and the circuit Hamiltonian \equref{eq:architecture II} predict similar outcomes. However, we also found that unless the model is adjusted properly, the outcomes can deviate quite strongly. Note that the deviations are already observable for the rather small time scales considered here, and such deviations typically tend to grow with time.


\section{Summary and Conclusions}\label{sec:SummaryAndConclusions}

We have implemented two simulators to solve the TDSE for two different but related generic models of a superconducting quantum processor. The first model is a lumped-element model, i.e.~a circuit Hamiltonian. The second model is an approximation of the first one, i.e.~an effective model Hamiltonian. Both models aim to describe a set of interacting transmon qubits (fixed-frequency and/or flux-tunable) and transmission line resonators. The interaction between the different subsystems is always of the dipole-dipole type.

The first simulation code, for the circuit Hamiltonian model, enables us to simulate the model without making any approximations. The second simulation code, for the effective Hamiltonian model, allows us to simulate the system with various approximations being turned on or off. A basic version of the simulation code for the effective model is available at  Ref.~\cite{JugitJUSQUACE}. This simulation framework provides us with the tools to study the validity of different approximations, which are often made to make analytical calculations feasible.

For our study we consider three different systems. The first system is a single flux-tunable transmon. The second system\change{, architecture I,} consists of two fixed-frequency transmons, coupled to a flux-tunable transmon. \change{The flux-tunable transmon works as a coupler, see \figref{fig:arch_sketch}(a)}. The third system\change{, architecture II,} is made up of two flux-tunable transmons, coupled to a transmission line resonator. \change{Here the resonator functions only as a coupler element, see \figref{fig:arch_sketch}(b).}

We found that the effective model Hamiltonian given by \equref{eq:HamTrafo_main} allows us to approximate the dynamic behaviour of the circuit Hamiltonian \equref{eq:flux-tunable transmon} quite well. However, for some transition scenarios some deviations still remain, see \figref{fig:leakage}(d, h). Furthermore, as can be expected, the adiabatic effective Hamiltonian in \equref{eq:flux-tunable transmon effective} cannot describe any dynamic transitioning behaviour. The results are discussed in \secref{sec: Revised derivation of the effective Hamiltonian for flux-tunable transmons}.

In addition, it seems that if we use the adiabatic effective Hamiltonian \equref{eq:flux-tunable transmon effective} to model flux-tunable transmons in multi-qubit systems, see the effective model Hamiltonian given by \equref{eq:architecture I effective}, we suppress additional resonant transitions. A summary of these results is provided in \tabref{tab:summary_effective_hamiltonian_results}\change{, see rightmost column.} However, we can recover these resonant transitions by adding the non-adiabatic drive term in \equref{eq:basis_trafo_main} to every flux-tunable transmon in the effective model Hamiltonian. The results are discussed in \secref{sec:Single-qubit operations effective}. Once larger superconducting processors are built, with more than a few transmon qubits, we face the problem of spectral crowding. However, if we base our analysis of this problem only on the transition frequencies which are relevant for the effective model, we might overlook frequencies which are crucial for this issue.

Our analysis shows that assuming the effective interaction strength to be time independent can affect the gate durations of some two-qubit gates quite strongly. Here we consider the difference between two effective models\change{, with and without a time-dependent interaction strength} and the difference with respect to the circuit Hamiltonian model. A summary of these results can be found in \tabref{tab:summary_effective_hamiltonian_results}\change{, see the second-last column.} For example, if we model two-qubit Cz gate interactions in architecture I, see \figref{fig:arch_sketch}(a), with and without a time-dependent interaction strength and the effective Hamiltonian given by \equref{eq:architecture I effective}, we find that the gate duration deviates up to about 75 ns. The deviations with respect to the circuit Hamiltonian model for the same transitions are about 100 ns if we do not include the time dependence into the effective model. These deviations seem too large to be neglected. The time-dependent effective interaction strength affects the gate durations of architecture II, see \figref{fig:arch_sketch}(b), to a lesser extent. Additionally, we found that for the pulses we model in this work, the non-adiabatic drive term in \equref{eq:basis_trafo_main} barely affects the two-qubit gate transitions in architecture I and II. The results are discussed in \secref{sec: two-qubit operations effective}.

The focus of our analysis has been put on the dynamics of the very basic state-transition mechanism. For future work, it might be interesting to see whether or not the different models generate different error signatures, once complete quantum circuits are simulated, see \REF\cite{Willsch2017GateErrorAnalysis}. It seems plausible that these errors are very sensitive to changes to the model. The challenge here is to make a fair comparison between two different models that are parameterised in terms of the pulse parameters.

\begin{acknowledgments}
The authors gratefully acknowledge the Gauss Centre for Supercomputing e.V.~(www.gauss-centre.eu) for funding this project by providing computing time on the GCS Supercomputer JUWELS \cite{JUWELS} at Jülich Supercomputing Centre (JSC). H.L.~acknowledges support from the project OpenSuperQ (820363) of the EU Quantum Flagship.
D.W.'s work was partially supported by the Q(AI)$^{2}$ project.
D.W. and M.W. acknowledge support from the project J\"ulich UNified Infrastructure for Quantum computing (JUNIQ) that has received funding from the German Federal Ministry of Education and Research (BMBF) and the Ministry of Culture and Science of the State of North Rhine-Westphalia.
\end{acknowledgments}


\appendix

\section{Derivation of the effective Hamiltonian for a flux-tunable transmon by means of a cosine expansion}\label{app:DiscussionOfCosineExpansionArgument}

The goal of the main text was to present a comparison between the descriptions of the full circuit Hamiltonian in \equref{eq:flux-tunable transmon} and the effective Hamiltonian given by \equref{eq:flux-tunable transmon effective}. In this appendix, we outline the steps that are often implicitly made to derive the effective Hamiltonian. \change{Note that throughout this work we use $\hbar=1$.}

We derive the effective Hamiltonian given by \equref{eq:flux-tunable transmon effective} in a step-wise manner. We start from the circuit Hamiltonian
\begin{equation}\label{eq:FFTHamiltonianAgain}
 \hat{H}_{\text{Tun}} = E_{C} \hat{n}^2 -  E_{J,\text{eff}}(t) \cos(\hat{\varphi}-\varphi_{\text{eff}}(t)),
\end{equation}
given by \equref{eq:flux-tunable transmon recast} in the main text and perform an expansion of the cosine to second order. The corresponding second-order expansion reads
\begin{equation}
  \hat{H}=E_{C} \hat{n}^{2} + \frac{E_{J,\text{eff}}(t)}{2} \left(\hat{\varphi}-\varphi_{\text{eff}}(t)\right)^{2},
\end{equation}
where we neglect the $-E_{J,\text{eff}}(t)$ term which only contributes a non-measurable phase to the dynamics of the system. We obtain the instantaneous eigenstates in $\varphi$-space for this Hamiltonian,
\begin{equation}\label{eq:tdHB}
  \psi^{(m)}(x(t))=\frac{1}{\sqrt{2^{m} m!}} \left(\frac{\xi(t)}{\pi}\right)^{\frac{1}{4}} e^{-\frac{x^{2}(t)}{2}} \mathcal{H}_{m}(x(t)),
\end{equation}
where $\xi(t)=(E_{J,\text{eff}}(t)/2 E_{C})^{1/2}$, $x(t)=\sqrt{\xi(t)}(\varphi-\varphi_{\text{eff}}(t))$ and $\mathcal{H}_{m}$ denotes the Hermite polynomial of order $m$. The corresponding eigenvalues
\begin{equation}
   E^{(m)}(t)=\omega(t) m + 1/2,
\end{equation}
where $\omega(t)=\sqrt{2 E_{C} E_{J,\text{eff}}(t)}$ can be determined analytically.

We intend to model the system in the time-dependent basis
\begin{equation}
  \mathcal{B}(t)=\{\ket{\psi^{(m)}(t)}\}_{m \in \mathbb{N}},
\end{equation}
such that the transformed state vector reads
\begin{equation}
  \ket{\Psi^{*}(t)}=\hat{\mathcal{W}}(t)\ket{\Psi(t)},
\end{equation}
where $\hat{\mathcal{W}}(t)$ denotes the unitary transformation which maps the basis states $\mathcal{B}(0)$ to the basis states $\mathcal{B}(t)$. This requires that we transform the Hamiltonian operator
\begin{equation}\label{eq:HamTrafo}
  \hat{H}^{*}(t)=\hat{\mathcal{W}}(t)\hat{H}(t)\hat{\mathcal{W}}^{\dagger}(t) -i \hat{\mathcal{W}}(t)\partial_{t}\hat{\mathcal{W}}^{\dagger}(t),
\end{equation}
such that TDSE for the state $\ket{\Psi^{*}(t)}$ retains its original form, see \REFS\cite{WillschMadita2020PhD,Weinberg2015}.

The drive term
\begin{equation}
  \hat{\mathcal{D}}(t)=-i\hat{\mathcal{W}}(t)\partial_{t}\hat{\mathcal{W}}^{\dagger}(t),
\end{equation}
in \equref{eq:HamTrafo} can be expressed as
\begin{equation}\label{eq:basis_trafo}
    \hat{\mathcal{D}}(t)= -i \sqrt{\frac{\xi(t)}{2}} \dot{\varphi_{\text{eff}}}(t)  \left(\hat{b}^{\dagger}-\hat{b}\right)+ \frac{i}{4}\frac{\dot{\xi}(t)}{\xi(t)}\left(\hat{b}^{\dagger}\hat{b}^{\dagger}-\hat{b}\hat{b}\right)
\end{equation}
where we assume that $\xi(t)\neq0$ for all times $t$. Here we adjusted a derivation which can be found in \REF\cite[Section 5.1.2]{WillschMadita2020PhD}. We also find
\begin{equation}
  \dot{\varphi_{\text{eff}}}(t)=\dot{\varphi}(t)\frac{d}{2 \left(\cos\left(\frac{\varphi(t)}{2}\right)^{2}+d^{2} \sin\left(\frac{\varphi(t)}{2}\right)^{2}\right)}
\end{equation}
and
\begin{equation}
  \frac{\dot{\xi}(t)}{\xi(t)}=\dot{\varphi}(t)\frac{(d^{2}-1) \sin(\varphi(t))}{8 \left(\cos\left(\frac{\varphi(t)}{2}\right)^{2}+d^{2} \sin\left(\frac{\varphi(t)}{2}\right)^{2}\right)}.
\end{equation}
so that the first (second) drive term in \equref{eq:basis_trafo} disappears if $d=0$ ($d=1$).

The Hamiltonian in the time-dependent harmonic basis reads
\begin{equation}
\begin{split}
    \hat{H}_{1}^{*}&= \omega(t) \hat{b}^{\dagger}\hat{b}\\
    &+-i \sqrt{\frac{\xi(t)}{2}} \dot{\varphi_{\text{eff}}}(t) (\hat{b}^{\dagger}-\hat{b})\\
    &+\frac{i}{4}\frac{\dot{\xi}(t)}{\xi(t)}\left(\hat{b}^{\dagger}\hat{b}^{\dagger}-\hat{b}\hat{b}\right).
\end{split}
\end{equation}
\change{Here we made use of the definitions
\begin{equation}\label{eq:charge_op_def_HB}
  \hat{n}=-\sqrt{\frac{\xi(t)}{2}} \left(\hat{b}^{\dagger}+\hat{b}\right)
\end{equation}
and
\begin{equation}\label{eq:flux_op_def_HB}
  \left(\hat{\varphi}-\varphi_{\text{eff}}(t)\hat{I}\right)=\frac{-i}{\sqrt{2\xi(t)}}\left(\hat{b}^{\dagger}-\hat{b}\right),
\end{equation}
for the charge and the shifted flux operator, respectively.}

If one models the system with two basis states only, one can express the Hamiltonian in terms of the Pauli $\hat{\sigma}^{(z)}$ and $\hat{\sigma}^{(y)}$ operators. The result reads
\begin{equation}
  \hat{H}_{1,I}^{*}= - \frac{\omega(t)}{2} \hat{\sigma}^{(z)} + -\sqrt{\frac{\xi(t)}{2}} \dot{\varphi_{\text{eff}}}(t)\hat{\sigma}^{(y)}.
\end{equation}
The term $-(\omega(t)/2) \hat{\sigma}^{(z)}$ is sometimes used to model flux-tunable transmons as two-level systems, see \REFS\cite{Roth19,McKay16,Yan18}. Obviously, in such a case, one neglects the contribution of the higher-order terms in the cosine expansion. Furthermore, one neglects all contributions of the drive term $\hat{\mathcal{D}}(t)$ which originated from the fact that we use a time-dependent basis to describe the dynamics.

We now expand the cosine to the quartic order and neglect all terms which only contribute a non-measurable phase. The corresponding effective Hamiltonian reads
\begin{equation}
  \hat{H}_{2}=E_{C} \hat{n}^{2} + \frac{E_{J,\text{eff}}(t)}{2} \left(\hat{\varphi}-\varphi_{\text{eff}}(t)\right)^{2} - \frac{E_{J,\text{eff}}(t)}{24} \left(\hat{\varphi}-\varphi_{\text{eff}}(t)\right)^{4}.
\end{equation}
If we model the system in the basis $\mathcal{B}(t)$, we find the Hamiltonian
\begin{equation}\label{eq:2ndTD}
  \hat{H}_{2}^{*}=\omega(t) \hat{b}^{\dagger} \hat{b} - \frac{E_{C}}{48} \left( \hat{b}^{\dagger}-\hat{b}\right)^{4} + \hat{\mathcal{D}}(t).
\end{equation}
One can split the operator
\begin{equation}\label{eq:PowerFour}
  \left( \hat{b}^{\dagger}-\hat{b}\right)^{4}=\hat{D}+\hat{V},
\end{equation}
into a diagonal $\hat{D}$ and a \change{off-diagonal} part $\hat{V}$. We make use of this decomposition and define another effective Hamiltonian
\begin{equation}\label{eq:2ndDiagonal}
  \hat{H}_{2,I}^{*}=\omega^{\prime}(t) \hat{b}^{\dagger} \hat{b}+ \frac{\alpha}{2} \hat{b}^{\dagger} \hat{b} \left(\hat{b}^{\dagger} \hat{b}-\hat{I}\right)  + \hat{\mathcal{D}}(t),
\end{equation}
where we only keep the diagonal contributions $\hat{D}$ of the operator given by \equref{eq:PowerFour}. Here $\omega^{\prime}(t)=\omega(t)+\alpha $.
Additionally, we define the effective Hamiltonian
\begin{equation}\label{eq:2ndDiagonalDrive}
  \hat{H}_{2,II}^{*}=\omega^{\prime}(t) \hat{b}^{\dagger} \hat{b}+ \frac{\alpha}{2} \hat{b}^{\dagger} \hat{b} \left(\hat{b}^{\dagger}\hat{b}-\hat{I}\right),
\end{equation}
where we only take into account the diagonal part $\hat{D}$ but neglect the operator $\hat{V}$ and the drive term $\hat{\mathcal{D}}(t)$. We emphasise that this Hamiltonian is often used, see \REFS\listciteone, to describe flux-tunable transmons and the subject of the main text.

It should be obvious that neglecting the drive term $\hat{\mathcal{D}}(t)$ in Hamiltonian \equref{eq:2ndDiagonalDrive} prevents us from modelling transitions between the different basis states in $\mathcal{B}(t)$, \ie dropping the drive term makes the Hamiltonian \equref{eq:2ndDiagonalDrive} diagonal in the basis $\mathcal{B}(t)$. Note that here we consider the route from the model given by \equref{eq:2ndTD} to the model given by \equref{eq:2ndDiagonalDrive}. The drive term in \equref{eq:2ndDiagonal} still allows us to model transitions between the different basis states of the system.

In principle, if one defines an effective Hamiltonian $\hat{H}_{\text{E}}(t)$ by dropping one or more terms in a given model Hamiltonian $\hat{H}_{\text{M}}(t)$, one has to consider how the time-evolution operators
\begin{equation}
  \hat{\mathcal{U}}_{\text{E}}(t,t_{0})=\mathcal{T} \exp\left( -i \int_{t_{0}}^{t} \hat{H}_{\text{E}}(t^{\prime}) dt^{\prime} \right)
\end{equation}
and
\begin{equation}
  \hat{\mathcal{U}}_{\text{M}}(t,t_{0})=\mathcal{T} \exp\left( -i \int_{t_{0}}^{t} \hat{H}_{\text{M}}(t^{\prime}) dt^{\prime} \right)
\end{equation}
deviate from one another, and not only the Hamiltonians themselves. Consequently, one has to consider an appropriate operator norm. A general discussion of this subject, with explicit examples, is provided by \REF\cite{Burgarth21}.

This makes approximating time-dependent Hamiltonians a rather complex subject. For example, dropping the drive term $\hat{\mathcal{D}}(t)$ only constitutes to a kind of adiabatic approximation for a single flux-tunable transmon, see \REF\cite{Weinberg2015}. However, the adiabatic approximation is formulated in terms of the instantaneous eigenstates of a system. Therefore, once we describe a collection of interacting transmons, we have to reconsider how the corresponding time-evolution operators for the effective model $\hat{\mathcal{U}}_{\text{E}}$ and the original model $\hat{\mathcal{U}}_{\text{M}}$ deviate, \ie in this case we have to reconsider the error which is caused by defining the effective Hamiltonian. Additionally, in general one cannot predict how dropping different terms, see \equaref{eq:2ndDiagonal}{eq:2ndDiagonalDrive}, changes the deviations with respect to the original model, see \equref{eq:2ndTD} or \equref{eq:FFTHamiltonianAgain}. Therefore, we decided to simulate both models independently, and to compare their predictions as shown in the main text, see \secref{sec: Revised derivation of the effective Hamiltonian for flux-tunable transmons}.

\section{Series expansion of the qubit frequency and anharmonicity}\label{sec: Effective Hamiltonians appendix}

In the main text, see \secref{sec:Results}, we model various single-qubit and two-qubit transitions with effective and circuit Hamiltonian models. If we compare the pulse parameters for some of these transitions, see \secref{sec: Revised derivation of the effective Hamiltonian for flux-tunable transmons} and \tabsref{tab:summary_circuit_hamiltonian_results}{tab:summary_effective_hamiltonian_results}, we find that some of these parameters which one can associate with the energy of a flux-tunable transmon deviate. Consequently, these differences might be attributed to the fact that if we model the energies of flux-tunable transmons with the expression
\begin{equation}\label{eq:spec}
  \left(E^{(m)}(\varphi(t))-E^{(0)}(\varphi)\right) = \left(m \omega(\varphi)+\frac{\alpha(\varphi)}{2}m(m-1)\right),
\end{equation}
the results are not accurate for some choices of the external fluxes $\varphi$. Here $\omega(\varphi)$ denotes the tunable frequency given by \equref{eq:tunable frequency} and $\alpha(\varphi)=\text{const.}$ is the anharmonicity of the flux-tunable transmon. \change{Note that throughout this work we use $\hbar=1$}. Furthermore, we removed the explicit time dependence $\varphi(t)\rightarrow \varphi$ since the spectrum exhibits symmetries with respect to the variable $\varphi$, see Hamiltonian \equref{eq:flux-tunable transmon}.

\renewcommand{\width}{1.0}
\begin{figure}[!tbp]
  \begin{minipage}{0.45\textwidth}
    \centering
    \includegraphics[width=\width\textwidth]{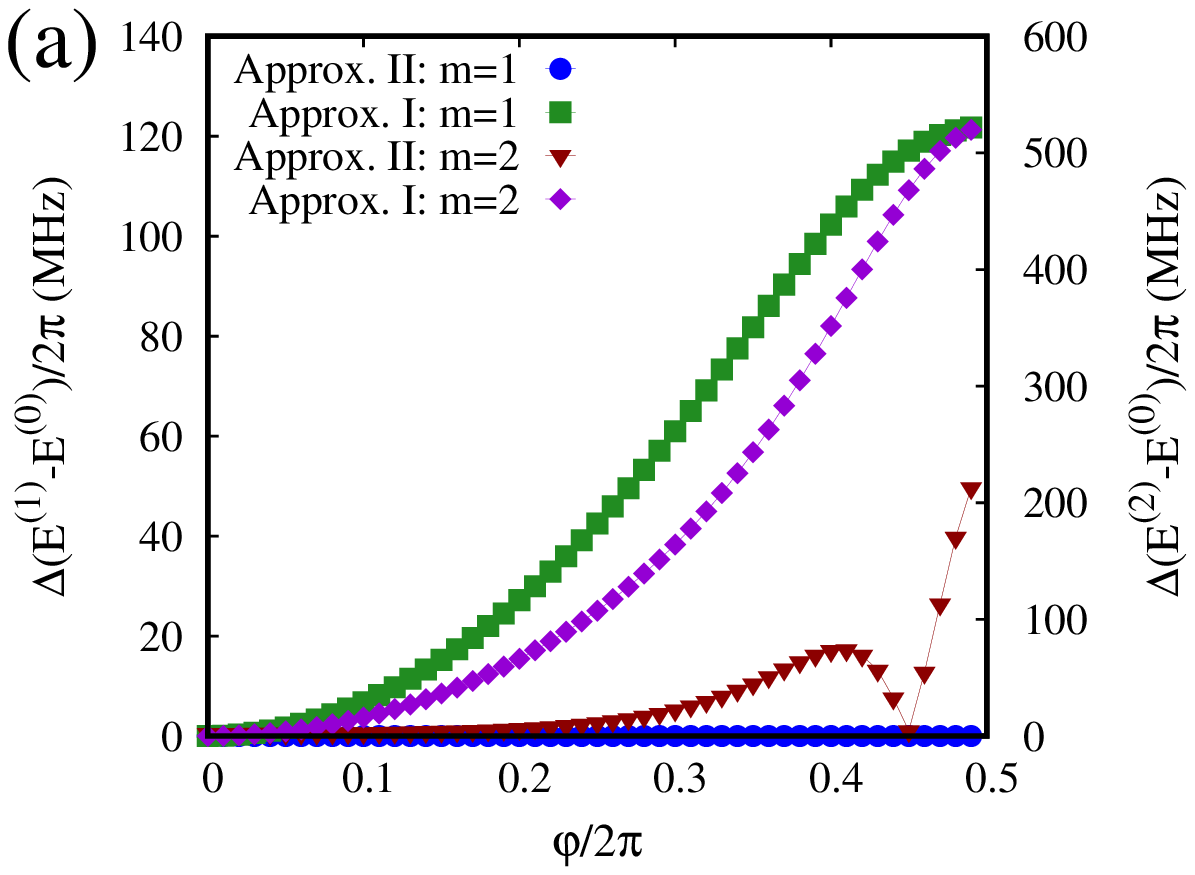}
  \end{minipage}
  \begin{minipage}{0.45\textwidth}
    \centering
    \includegraphics[width=\width\textwidth]{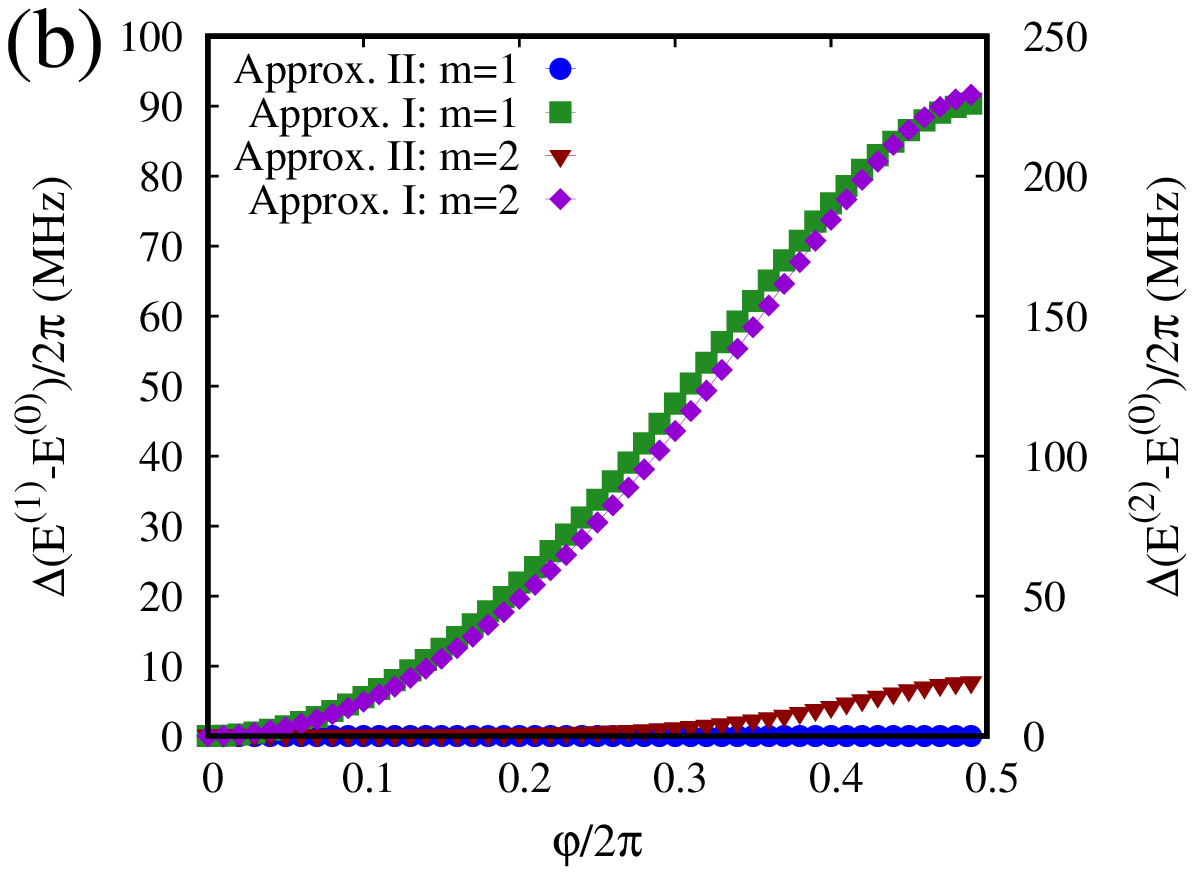}
  \end{minipage}
  \caption{(Color online) Deviations between the numerically exact spectrum and two different approximations of this spectrum as a functions of the external flux $\varphi$ for an asymmetry factor $d=0.33$(a) and $d=0.5$(b). First, we compute the numerically exact spectrum with fifty charge states. Then we use two different sets of expressions for the qubit frequency and anharmonicity to determine the approximated spectrum with \equref{eq:spec}. Approximation I: $\omega(\varphi)$ is given by \equref{eq:tunable frequency} and $\alpha(\varphi)=\text{const.}$. Approximation II: $\tilde{\omega}(\varphi)$ is given by \equref{eq: expansion frequency} and $\tilde{\alpha}(\varphi)$ is given by \equref{eq: expansion anharmonicity}. In the end, we use \equref{eq:spec_deviation} to determine the deviations between the numerically exact energies and the different approximants, see approximation I and II. We use the capacitive and Josephson energies as well as the qubit frequency and anharmonicity listed in \tabref{tab:device_parameter_resonator_coupler_chip}, row $i=0$(a) and $i=1$(b).}
  \label{fig:approximation}
\end{figure}

In this appendix, we compare the spectra of the circuit Hamiltonian given by \equref{eq:flux-tunable transmon} with the one of the effective flux-tunable transmon given by \equref{eq:flux-tunable transmon effective}. Furthermore, it is possible to make use of alternative expressions which allow us to approximate the spectrum with higher precision. Two such expressions were given in \REF\cite{Didier}. The corresponding flux-tunable transmon qubit frequency is of the form
\begin{equation}\label{eq: expansion frequency}
  \tilde{\omega}(\varphi)=\sqrt{2 E_{C} E_{J_{\text{eff}}}(\varphi)} - \frac{E_{C}}{4} \sum_{n=0}^{24} a_{n} \xi(\varphi)^{n}.
\end{equation}
Similarly, the flux-dependent qubit anharmonicity can be expressed as
\begin{equation}\label{eq: expansion anharmonicity}
  \tilde{\alpha}(\varphi)= - \frac{E_{C}}{4} \sum_{n=0}^{24} b_{n} \xi(\varphi)^{n},
\end{equation}
where $a_{n}$ and $b_{n}$ are real coefficients and the function $\xi(\varphi)$ can be expressed as
\begin{equation}
  \xi(\varphi)=\sqrt{\frac{E_{C}}{2 E_{J_{\text{eff}}}(\varphi)}}.
\end{equation}
We emphasise that the parameters $a_{n}$ and $b_{n}$ can be of order $10^{6}$ for large $n$. Furthermore, for some system parameters, we found this to be the case for a system with an asymmetry factor $d=0$, we find that $\xi(\varphi) \rightarrow 1$ if $\varphi/2\pi \rightarrow 0.5$. Here the approximation can break down.

In the following, the flux-tunable frequencies $\omega(\varphi)$ and $\tilde{\omega}(\varphi)$ and the anharmonicities $\alpha(\varphi)$ and $\tilde{\alpha}(\varphi)$ are only given by the functions we specify, \ie we do not include further corrections. Figures~\ref{fig:approximation}(a,b) show the deviations
\begin{align}\label{eq:spec_deviation}
\begin{split}
\Delta \left(E^{(m)}(\varphi)-E^{(0)}(\varphi)\right)&=\bigg|\left(E_{\text{exact.}}^{(m)}(\varphi)-E_\text{exact.}^{(0)}(\varphi)\right)\\
-&\left(m \omega(\varphi)+\frac{\alpha(\varphi)}{2}m(m-1)\right)\bigg|,
\end{split}
\end{align}
for $m=1$ (on the left y-axis in green and blue) and $m=2$ (on the right y-axis in red and violet) between the numerically exact spectrum of the Hamiltonian in \equref{eq:flux-tunable transmon} and two different sets of expressions for the qubit frequency and anharmonicity in \equref{eq:spec_deviation} as a function of the external flux $\varphi$. First, we use the parameters listed in \tabref{tab:device_parameter_resonator_coupler_chip}, row $i=0$(a) and  $i=1$(b), to compute the numerically exact values for two different asymmetry factors $d=0.33$(a) and $d=0.5$(b). Then we compute the approximated spectrum by means of \equref{eq:spec}. Here we consider two different approximations.

Approximation I: We use \equref{eq:tunable frequency} for $\omega(\varphi)$, $\alpha(\varphi)=\text{const.}$ and \equref{eq:spec} to compute the energies.

Approximation II: We use the series expansions $\tilde{\omega}(\varphi)$, $\tilde{\alpha}(\varphi)$ (see \equaref{eq: expansion frequency}{eq: expansion anharmonicity} respectively) and \equref{eq:spec} to do the same. Both, \equaref{eq: expansion frequency}{eq: expansion anharmonicity} were taken from \REF\cite{Didier}. Note that for $m=1$, see \equaref{eq:spec}{eq:spec_deviation}, the deviations between the different spectra become independent of $\alpha(\varphi)$.

As one can see, approximation I, \ie the first set of expressions \equref{eq:tunable frequency} and $\alpha(t)=\text{const.}$, deviates more from the exact solution, than approximation II, \ie \equaref{eq: expansion frequency}{eq: expansion anharmonicity}. In both cases, the deviations grow as the external flux $\varphi$ approaches the value $0.5$. Furthermore, the asymmetry factor $d$ seems to influence how well the spectrum is approximated. If we compare \figsref{fig:approximation}(a,b), we find that in \figref{fig:approximation}(b) the deviations can be smaller, e.g.~by a factor of ten (compare right y-axis of \figsref{fig:approximation}(a,b)).

The deviations in the spectrum can change the behaviour of the system once a flux pulse is applied. In particular, if we implement non-adiabatic two-qubit gates, see \REFS\cite{Foxen20,DiCarlo09}, the spectrum determines whether or not transitions occur. This becomes even more important if we consider several flux-tunable transmon qubits in one system. Here the errors, in terms of the spectrum, might add up and enhance or suppress different transitions between states. Therefore, an accurate modelling of the spectrum is important.

\section{\change{Simulation algorithm}}\label{sec:Methodology}
In this \change{appendix}, we discuss how we obtain the numerical results presented in \secref{sec:Results}.

The formal solution of the TDSE (with $\hbar=1$)
\begin{equation}
  i\partial_{t}\ket{\Psi(t)}=\hat{H}(t)\ket{\Psi(t)},
\end{equation}
for an arbitrary time-dependent Hamiltonian $\hat{H}(t)$, reads
\begin{equation}
  \hat{\mathcal{U}}(t,t_{0}) = \mathcal{T} \exp\left( -i \int_{t_{0}}^{t} \hat{H}(t^{\prime}) dt^{\prime} \right),
\end{equation}
where $\mathcal{T}$ is the time-ordering symbol. Numerical calculations require that this expression is discretised, with steps of length $\tau$.
The corresponding time-evolution operator,
\begin{equation}\label{eq:TimeEvolutionOperator}
  \hat{U}(t+\tau,t) = \exp\left( -i \tau \hat{H}(t+\frac{\tau}{2})\right),
\end{equation}
can then be implemented for every time step (using the mid-point rule \cite{Suzuki1993GeneralDecompositionTheoryOrderedExponentials}).

In this work we use the so-called product-formula algorithm, see \REFS\cite{deraedt1987productformula,huyghebaert1990productFormulaTimeDependentErrorBounds}, to solve the TDSE. This algorithm is explicit, inherently unitary, and unconditionally stable by construction. Here the time step parameter $\tau$ needs to be chosen small enough, with respect to the energy scales and the other relevant time scales of $\hat{H}(t)$, such that the exact mathematical solution of the TDSE is obtained up to some fixed numerical precision. Practically, this means that we decrease $\tau$ until it is small enough such that the relevant decimals do not change anymore. This procedure has to be repeated every time we make changes to the system, \ie if we change the system parameters or the control pulse parameters.

Furthermore, to compute e.g.~the spectrum of a Hamiltonian, we use a standard diagonalisation algorithm to obtain the eigenvalues and eigenstates of a Hamiltonian $\hat{H}(t)$.

The simulations of resonant transitions in the effective single flux-tunable transmon model, see \equref{eq:HamTrafo_main}, in \secref{sec: Revised derivation of the effective Hamiltonian for flux-tunable transmons} require at least four instantaneous basis states. Furthermore, the simulations of non-adiabatic transitions in \secref{sec: Revised derivation of the effective Hamiltonian for flux-tunable transmons} are performed with twenty instantaneous basis states.

For the simulations of the effective two-qubit models, see \equaref{eq:architecture I effective}{eq:architecture II effective}, in \secaref{sec:Single-qubit operations effective}{sec: two-qubit operations effective} we use four basis states for all fixed-frequency transmons, flux-tunable transmons and also for the resonators. The simulation basis here consists of the bare harmonic basis states.

The simulations of the circuit models are performed in the bare transmon basis, for more details see \appref{sec:CircuitHamiltonianSimulations}. Here we use as many states as necessary, \ie we increase the number of basis states $N_{m}$ for all transitions we model until the numerical values of the observables converge to some fixed numerical precision. This allows us to obtain an approximation free, numerical solution of the TDSE for the circuit Hamiltonian.

\section{Circuit Hamiltonian simulations}\label{sec:CircuitHamiltonianSimulations}
In this appendix we discuss the results of the circuit Hamiltonian simulations. A summary of the relevant results can be found in \tabref{tab:summary_circuit_hamiltonian_results}. We begin with a discussion of the simulation details in \secref{sec:SimulationOfCircuitHamiltoniansInTheTransmonBasis}. Then, in \secref{sec:SimulationsOfTheSuppressedTransitions}, we discuss the transitions which are suppressed in the effective model, see \secref{sec:Single-qubit operations effective}. In the end, in \secref{sec:SimulationsOfTheUnsuppressedTransitions}, we discuss the transitions which are unsuppressed in the effective model, see \secref{sec: two-qubit operations effective}.

\subsection{Simulation of circuit Hamiltonians in the transmon basis}\label{sec:SimulationOfCircuitHamiltoniansInTheTransmonBasis}
If we intend to simulate the circuit Hamiltonians given in \equref{eq:flux-tunable transmon}, \equaref{eq:architecture I}{eq:architecture II} without performing any approximations, we can perform the simulations in the transmon bare basis
\begin{equation}
  \ket{\phi^{(z)}}=\mathop{\otimes}\limits_{j=0}^{J-1} \ket{\phi^{(m_{j})}},
\end{equation}
where $z=m_{0}, ...,m_{J-1}$ is a placeholder for the different subsystem indices $m_{j}$. We form this basis by means of the bare basis states
\begin{equation}
  \ket{\phi^{(m_{j})}},
\end{equation}
of the corresponding subsystems. These states are the eigenstates of the Hamiltonians given in \equref{eq:fixed-frequency transmon}, \equref{eq:flux-tunable transmon} and \equref{eq: transmission line resonators} at time $t=0$. For simplicity, we call this basis the transmon basis. We need to be able to change the number of basis states $N_{m}$, to allow us to extend the basis up to the point where the relevant decimals of the observables do not change anymore. The numerical error which stems from the discretisation of the time domain can be controlled by decreasing the time grid parameter $\tau$ up to a point where convergence has been reached. Obviously, both parameters $N_{m}$ and $\tau$ have to be changed together.

We are satisfied with the accuracy if the probabilities
\begin{equation}
  p^{(z)}(t)=|\braket{ \phi^{(z)}|\Psi(t)}|^2,
\end{equation}
we are interested in agree to the third decimal. Here $\ket{\Psi(t)}$ denotes the solution of the TDSE. Note that we use at least three basis states for the transmons in the system. If not stated otherwise, transmission line resonators are modelled with four states.

\subsection{Circuit Hamiltonian simulations of transitions that are suppressed in the effective model}\label{sec:SimulationsOfTheSuppressedTransitions}
We start our discussion with a single, isolated flux-tunable transmon. The system itself is defined by the parameters in \tabref{tab:device_parameter_flux_tunable_coupler_chip} and we model the system with circuit Hamiltonian \equref{eq:flux-tunable transmon}. Here we consider the flux-tunable transmon with label $i=2$.
\renewcommand{\width}{0.45}
\begin{figure}[!tbp]
  \centering
  \begin{minipage}{0.5\textwidth}
    \centering
    \includegraphics[width=\width\textwidth]{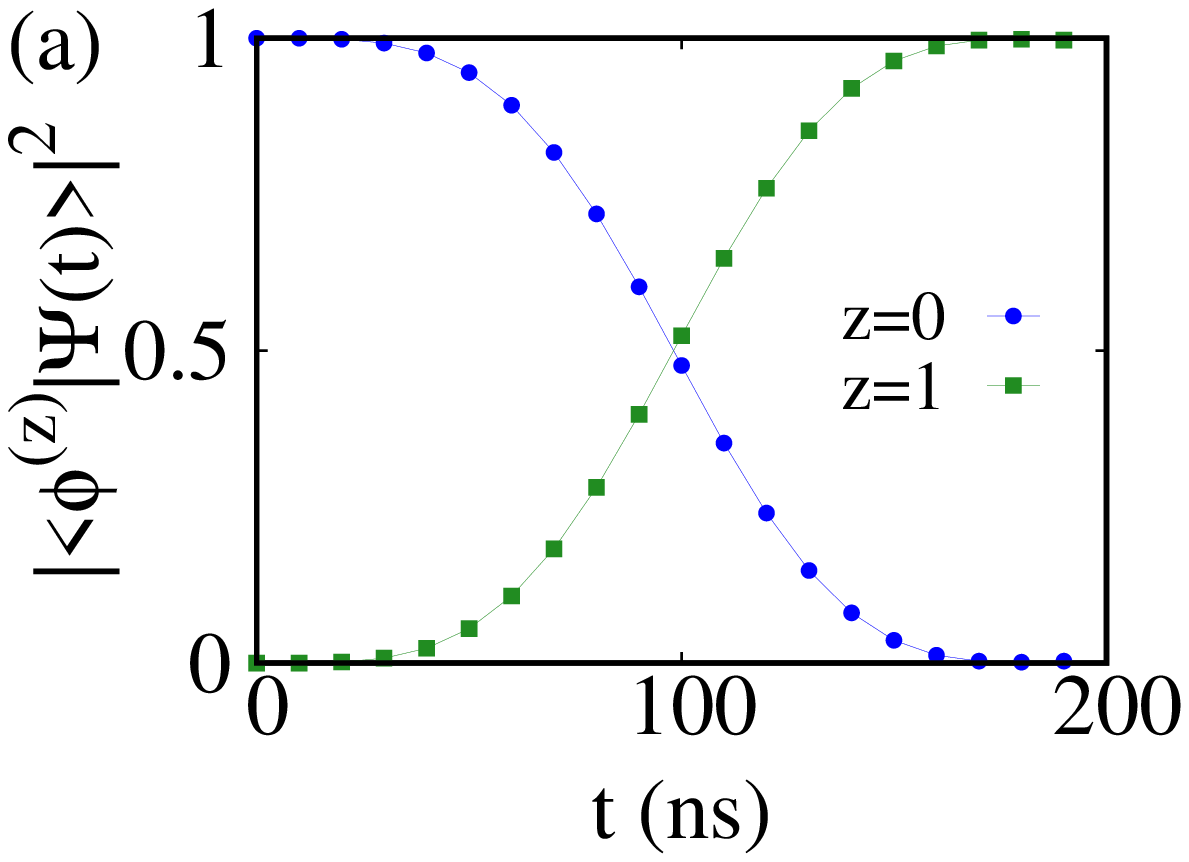}
    \includegraphics[width=\width\textwidth]{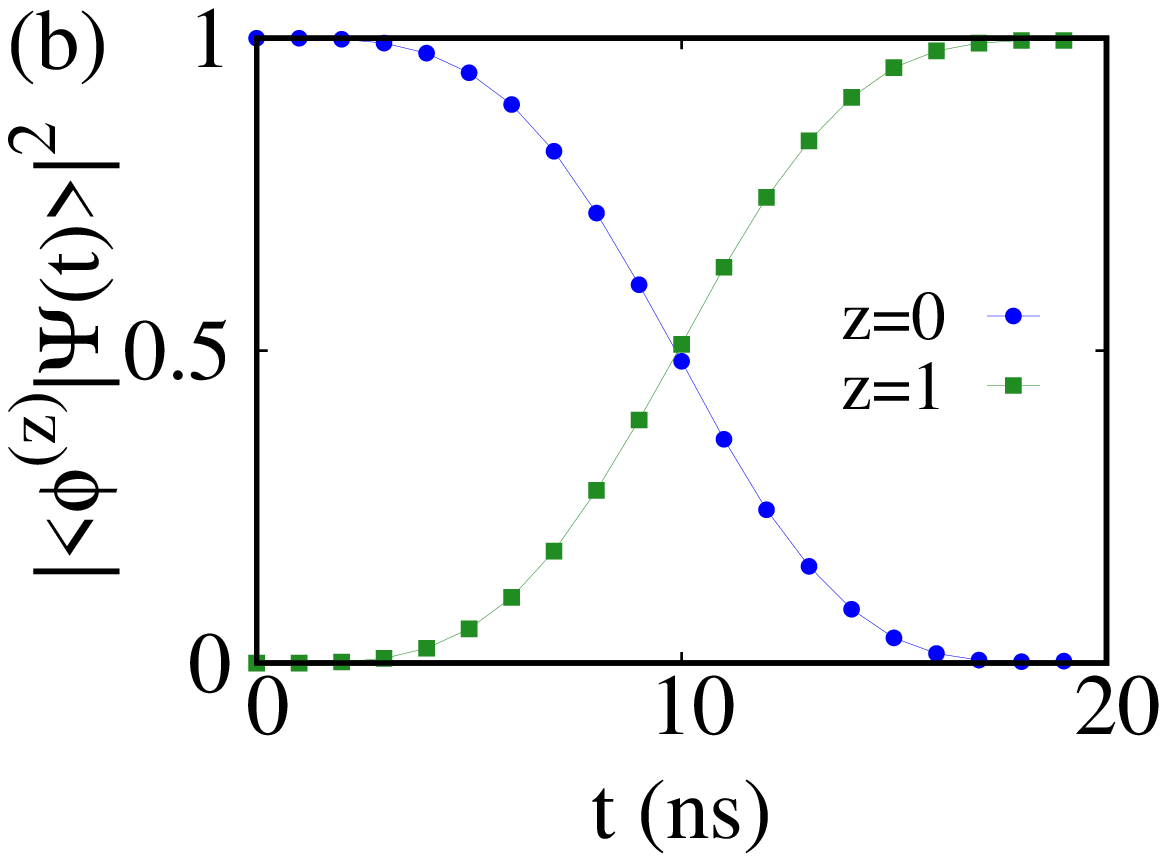}
    \caption{(Color online) Probabilities $p^{(0)}(t)$ and $p^{(1)}(t)$ as functions of time $t$. We use three transmon basis states $N_{m}=3$ to model the dynamics of the system, a control pulse of the form \equref{eq:pulse} and a drive frequency $\omega^{D}$ equal to the qubit frequency $\omega$ (see \tabref{tab:device_parameter_flux_tunable_coupler_chip}, row $i=2$). The rise and fall time $T_{\mathrm{r/f}}$ is set to half the duration time $T_{\mathrm{d}}$. The system is initialised in the state $\ket{\psi^{(0)}}$. The pulse amplitude $\delta/2\pi$ is set to (a) $\delta/2\pi = 0.001$ and (b) $\delta/2\pi = 0.01$. We can observe that an increase in the pulse amplitude $\delta$ by a factor of ten, leads to a decrease of the pulse duration $T_{\mathrm{d}}$ by a factor of ten (roughly). Note that these transitions cannot be modelled with the effective Hamiltonian \equref{eq:flux-tunable transmon effective}.}
    \label{fig:single_qubit_swap_amp}
  \end{minipage}
\end{figure}
\renewcommand{\width}{0.45}
\begin{figure}[!tbp]
  \centering
  \begin{minipage}{0.5\textwidth}
    \centering
    \includegraphics[width=\width\textwidth]{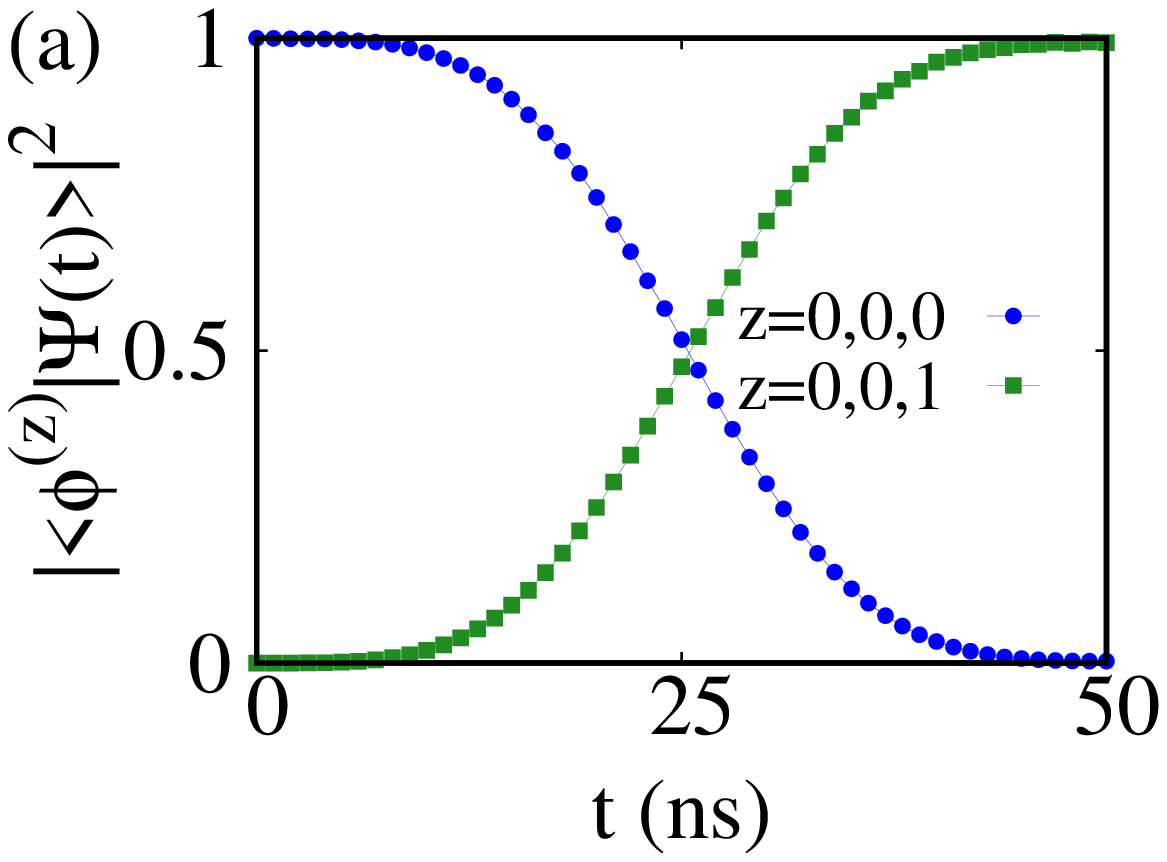}
    \includegraphics[width=\width\textwidth]{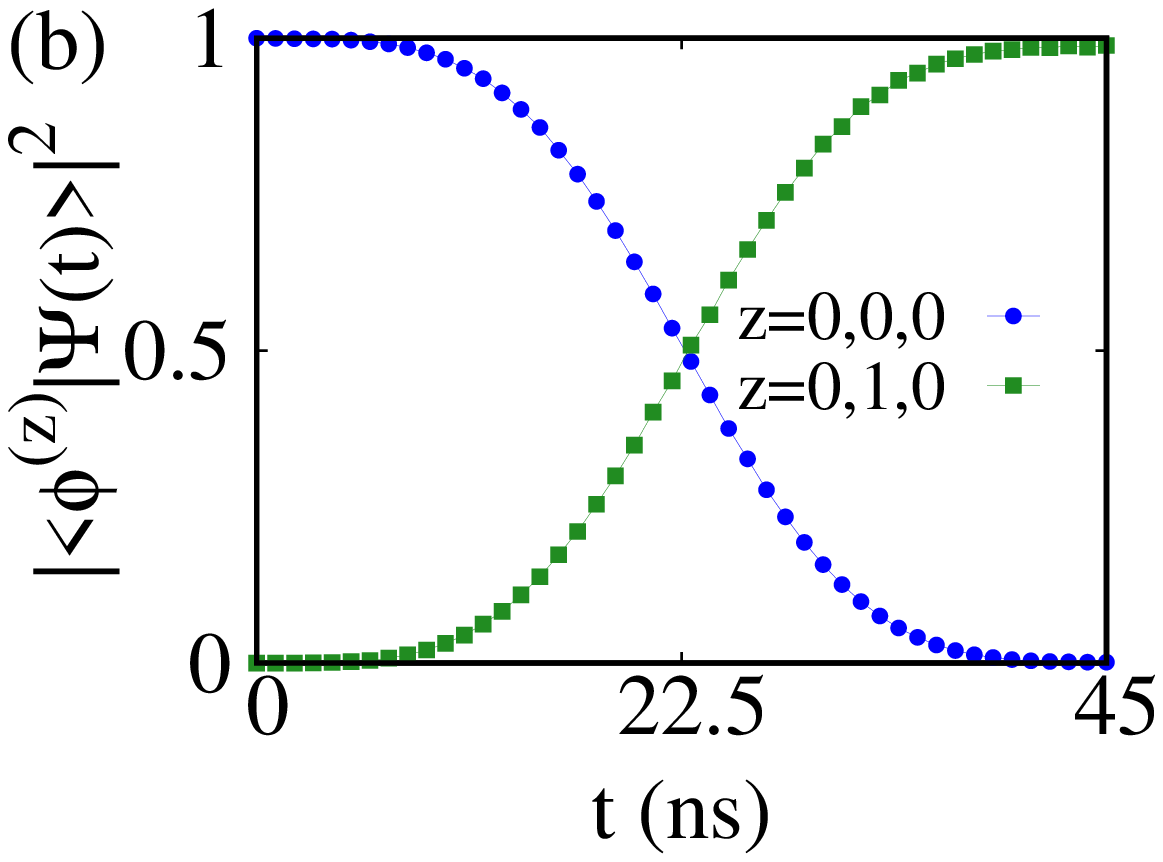}
    \caption{(Color online) Probabilities $p^{(z)}(t)$ for (a) $z = (0,0,0)$ and $z = (0,0,1)$ and (b) $z = (0,0,0)$ and $z = (0,1,0)$ as a function of time $t$. In both cases we use three basis states $N_{m}=3$ to model the dynamics of the system, a control pulse of the form \equref{eq:pulse} and a rise and fall time $T_{\mathrm{r/f}}$ set to half the duration time $T_{\mathrm{d}}$. (a) We use the drive frequency $\omega^{D}=5.092$ GHz, the pulse amplitude $\delta/2 \pi=0.085$. (b) We use the drive frequency $\omega^{D}=6.183$ GHz and the pulse amplitude $\delta/2 \pi=0.045$.  The initial state of the system is always $\ket{\psi^{(0,0,0)}}$. Note that we were not able to activate these transitions in the effective model of architecture I, see Hamiltonian \equref{eq:architecture I effective}.}
    \label{fig:two_qubits_with_coupler_qubit_single_qubit_op}
  \end{minipage}
\end{figure}

Figure~\ref{fig:single_qubit_swap_amp}(a) shows the time evolution of the probabilities $p^{(z)}(t)$, for the two lowest eigenstates $z\in \{0,1\}$. We use a control pulse of the form \equref{eq:pulse}, see \figref{fig:pulse_time_evo}(a), where we set $\omega^{\mathrm{D}}$ equal to the qubit frequency $\omega$. The rise and fall time $T_{\mathrm{r/f}}$ is set to half of the pulse duration $T_{\mathrm{d}}$. The x-axis displays the duration time. The pulse amplitude in this case is set to $\delta/2\pi=0.001$. The system is initially in the state $\ket{\phi^{(0)}}$ and we are able to implement a smooth transition between the states $\ket{\phi^{(0)}}$ and $\ket{\phi^{(1)}}$.

Figure~\ref{fig:single_qubit_swap_amp}(b) shows the results for a similar scenario. Here we increase the amplitude by one order of magnitude, i.e.~we use $\delta/2 \pi=0.01$. The time evolution shows that the duration $T_{\mathrm{d}}$ has decreased roughly by a factor of ten. Note that the transitions between the states $\ket{\phi^{0}}$ and $\ket{\phi^{1}}$ cannot be modelled with the effective Hamiltonian \equref{eq:flux-tunable transmon effective}.

In both cases it is sufficient to use three basis states to model the dynamics of the system, i.e.~increasing the number of basis states above three has no real impact on the probabilities we are interested in.

While it is possible to generate similar results (data not shown) for amplitudes in the range $\delta/2 \pi \in [0.001,0.01]$ we find that for amplitudes $\delta/2\pi \gg  0.01$ it is not possible to implement a smooth transition between both states. Application of the pulse does not conserve the probability in the subspace $\{\ket{\phi^{(0)}},\ket{\phi^{(1)}}\}$.

Next we study a system which consists of three transmons. We add two fixed-frequency transmons to the flux-tunable transmon. This means the corresponding circuit Hamiltonian is of the form \equref{eq:architecture I}. \tabref{tab:device_parameter_flux_tunable_coupler_chip} shows the corresponding system parameters. These parameters are motivated by a series of experiments reported in \REF\cite{Ganzhorn20}. Figures~\ref{fig:two_qubits_with_coupler_qubit_single_qubit_op}(a,b) show the system's response to a harmonic pulse of the form \equref{eq:pulse}, see \figref{fig:pulse_time_evo}(a).

In \figref{fig:two_qubits_with_coupler_qubit_single_qubit_op}(a) we use the drive frequency $\omega^{D}=6.183$ and the amplitude $\delta/2 \pi=0.045$. Here the figure shows the probabilities $p^{(z)}(t)$, for $z=(0,0,0)$ and $z=(0,1,0)$, as a function of time $t$. In this case the intention is to drive the $z=(0,0,0) \rightarrow z=(0,1,0)$ transition.

Figure~\ref{fig:two_qubits_with_coupler_qubit_single_qubit_op}(b) shows a similar case. Here we use the drive frequency $\omega^{D}=5.092$ and the amplitude $\delta/2 \pi=0.085$. Since we intend to drive the $z=(0,0,0) \rightarrow z=(0,0,1)$ transition, we display the corresponding probabilities $p^{(z)}(t)$ as a function of time $t$.

In both cases the initial state is set to $\ket{\phi^{(0,0,0)}}$ and we find a duration time $T_{\mathrm{d}}$ of around 50 ns.

Figures~\ref{fig:two_qubits_with_coupler_qubit_single_qubit_op}(a,b) show that we are able to implement transitions between the state pairs $z=(0,0,0)$ and $z=(0,1,0)$ as well as $z=(0,0,0)$ and $z=(0,0,1)$. In addition, it is also possible (data not shown) to drive transitions of the form $z=(0,0,1) \rightarrow z=(0,1,1)$ and  $z=(0,1,0) \rightarrow z=(0,1,1)$, simply by changing the initial state of the system and leaving all other parameters. Note that we were not able to activate these transitions in the effective model of architecture I, see Hamiltonian \equref{eq:architecture I effective}. Here we do not consider the transmon $i=2$ (see \tabref{tab:device_parameter_flux_tunable_coupler_chip}) since it is considered to be a coupler and not an actual qubit. However, it is possible to drive the transition $z=(0,0,0) \rightarrow z=(1,0,0)$.

For both cases we find that it is sufficient to use three transmon basis states to model the dynamics of the system.

\subsection{Circuit Hamiltonian simulations of the unsuppressed transitions in the effective two-qubit models}\label{sec:SimulationsOfTheUnsuppressedTransitions}

We investigate the transitions which are unsuppressed in the effective model. Here we differentiate between two cases. We first discuss transitions which are used to implement two-qubit gates by means of harmonic microwave pulses, see \REFS\cite{Roth19,McKay16,Ganzhorn20}. In this case we simulate circuit Hamiltonian \equref{eq:architecture I}, with the parameters listed in \tabref{tab:device_parameter_flux_tunable_coupler_chip}. As a second case, we study transitions which are activated by unimodal pulses, i.e.~gates which are implemented by means of adiabatic passage techniques, see \REFS\cite{Vitanov2001,DiCarlo09}. In this case we simulate circuit Hamiltonian \equref{eq:architecture II}. The corresponding system parameters can be found in \tabref{tab:device_parameter_resonator_coupler_chip}.

\subsubsection{Architecture I}

Figures~\ref{fig:iswap_tunable_coupler}(a-d) show the time evolution of the probabilities $p^{(0,0,1)}(t)$ and $p^{(0,1,0)}(t)$ as a function of time $t$. We use $N_{m}=3$ (a), $N_{m}=4$ (b), $N_{m}=6$ (c) and $N_{m}=15$ (d) basis states to model the dynamics of the system. The transition we model here is often used to implement an Iswap gate. The drive frequency is $\omega^{\mathrm{D}}=1.089$ GHz, which corresponds roughly to the frequency difference $\Delta\omega=1.100$ GHz between the individual transmon qubits $i=1$ and $i=0$. The frequency shift stems from the fact that the states $\{ \ket{\phi^{(z)}} \}$ are not exact eigenstates of the full circuit Hamiltonian. The drive amplitude is set to $\delta/2\pi = 0.075$ and the initial state of the system is $ \ket{ \phi^{(0,0,1)}}$.

The time evolutions in \figsref{fig:iswap_tunable_coupler}(a-d) clearly show that three or four basis states are not sufficient to describe this operation, i.e.~if we compare the solutions (a) and (b) with the reference solution (c)/(d) we find substantial qualitative and quantitative differences. We find that we need at least six transmon basis states to model the system. Note that we simulated the same system as before when studying the single-qubit operations. We conclude that the number of states which is needed to model different types can vary, i.e., it is not a system property but it depends on the type of transition we simulate.

\renewcommand{\width}{0.45}
\begin{figure}[!tbp]
    \centering
    \begin{minipage}{0.5\textwidth}
        \centering
        \includegraphics[width=\width\textwidth]{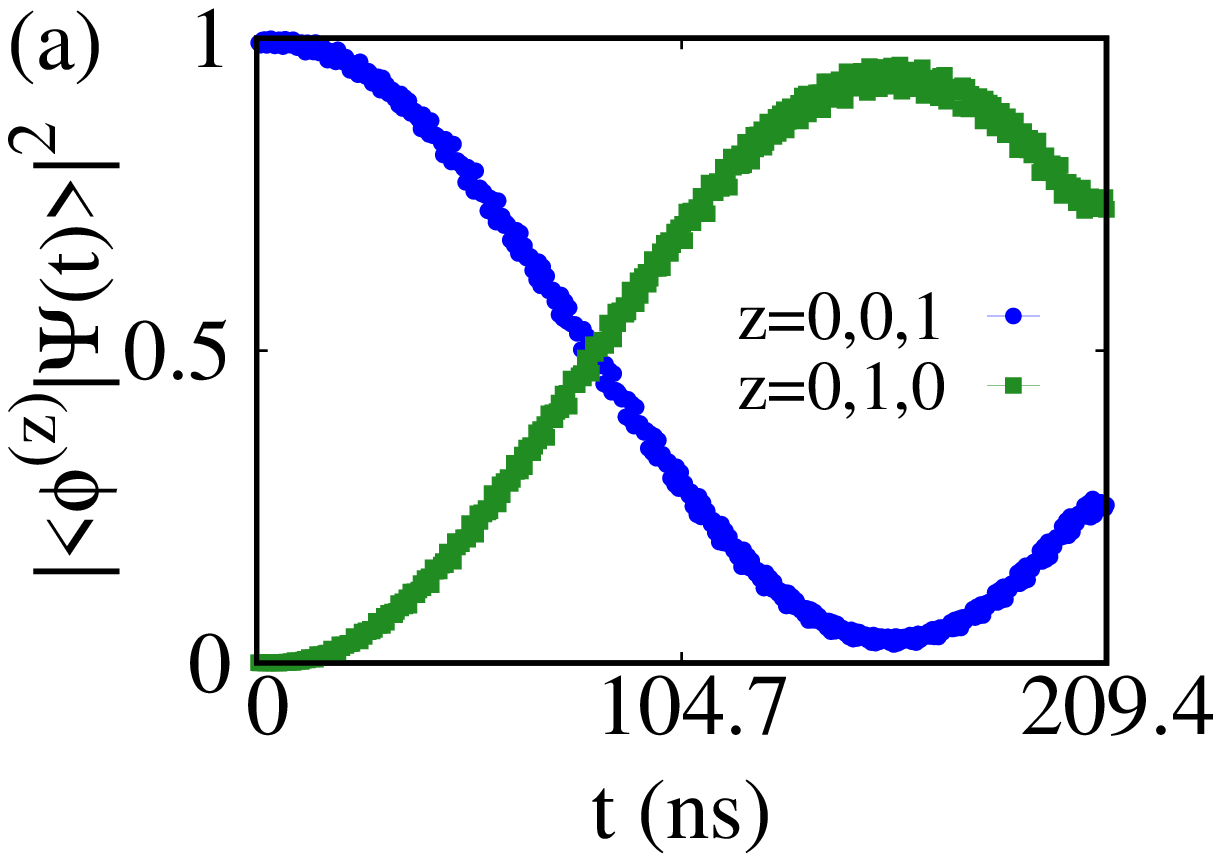}
        \includegraphics[width=\width\textwidth]{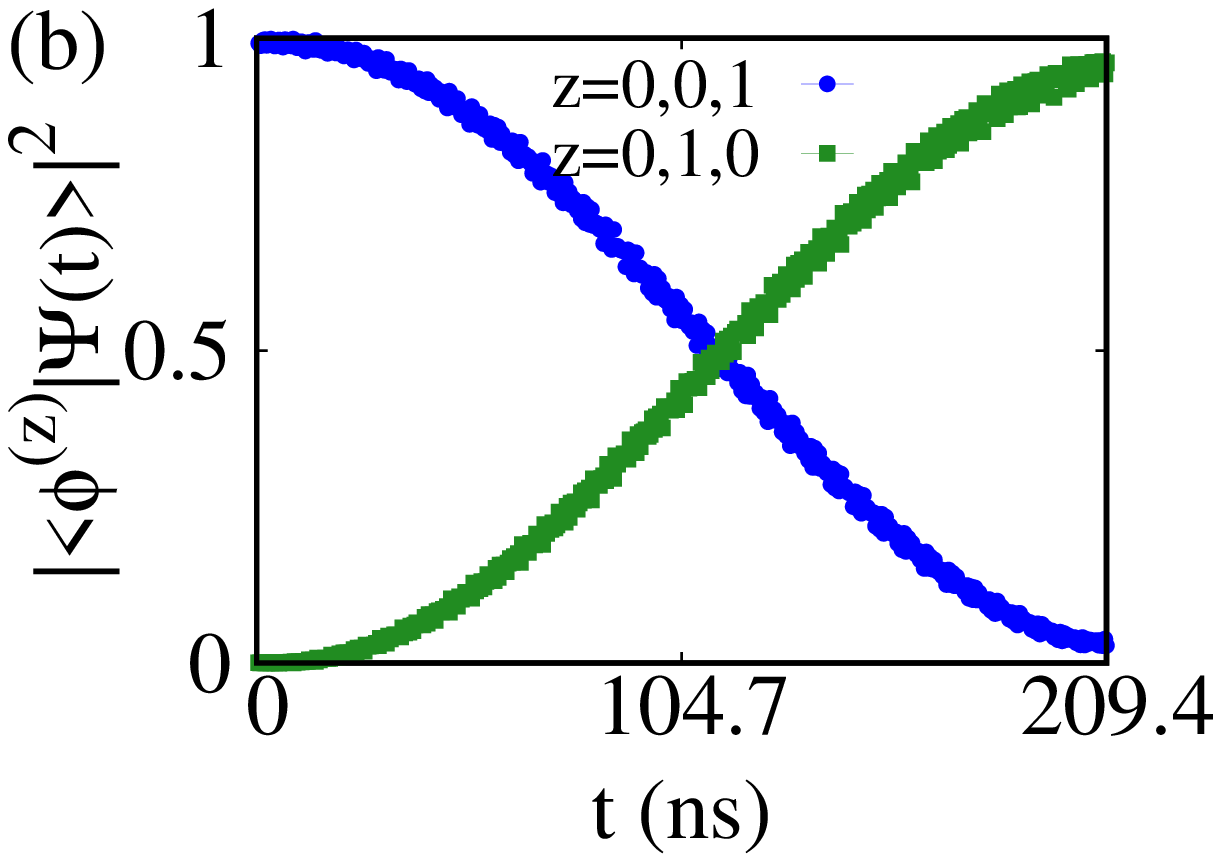}
        \includegraphics[width=\width\textwidth]{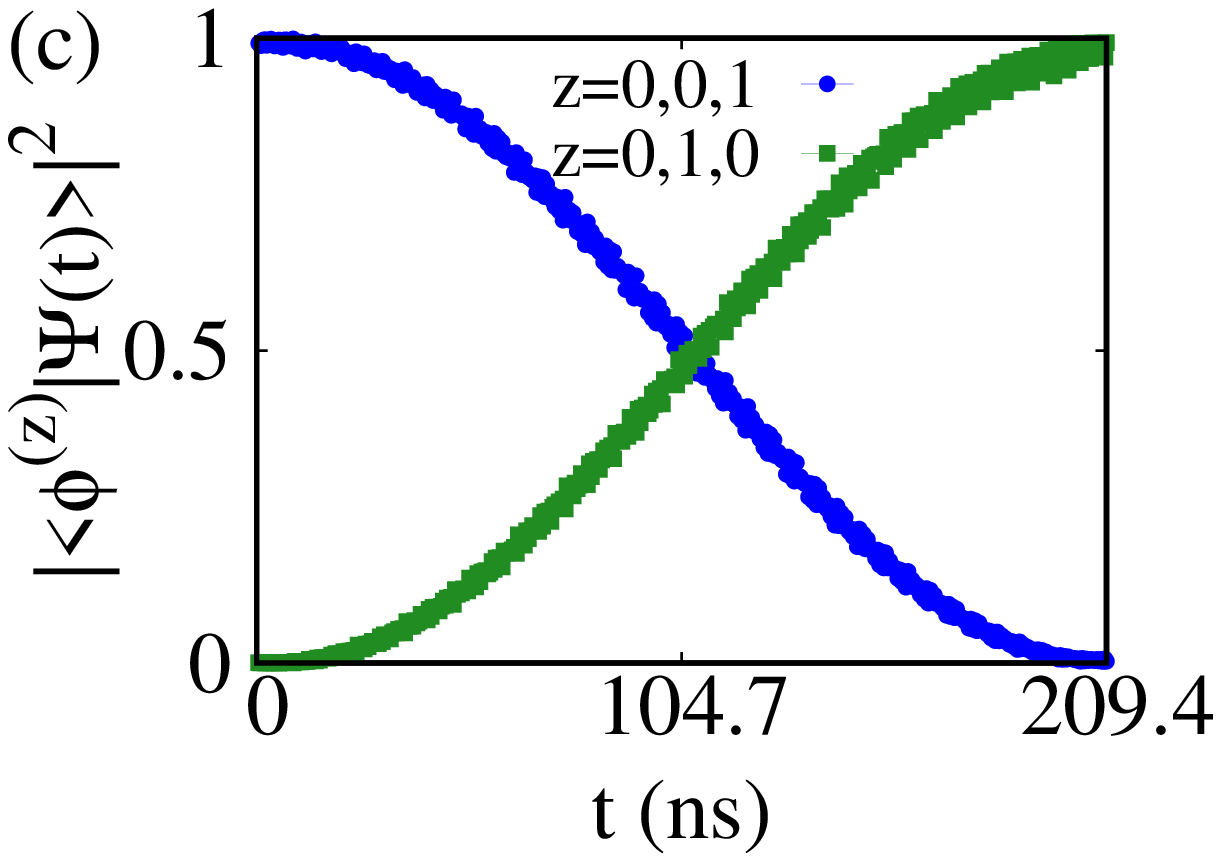}
        \includegraphics[width=\width\textwidth]{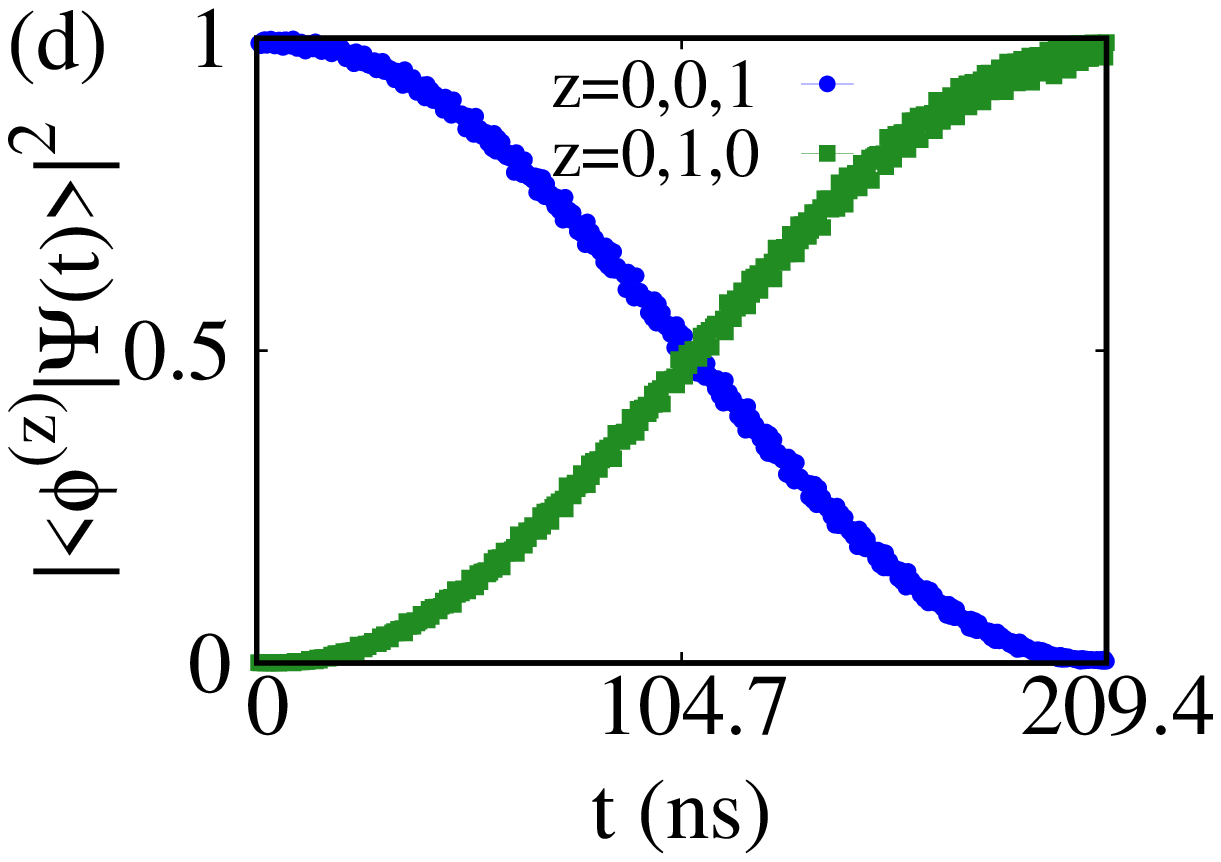}
        \caption{(Color online) Probabilities $p^{(0,0,1)}(t)$ and $p^{(0,1,0)}(t)$ as functions of time $t$. We use $N_{m}=3$ (a), $N_{m}=4$ (b), $N_{m}=6$ (c) and $N_{m}=15$ (d) basis states to model the system. We use a control pulse of the form \equref{eq:pulse}, with the pulse parameters $\omega^{D}=1.089$ GHz, $T_{\mathrm{r/f}}=13$ ns and $\delta/2\pi=0.075$. The pulse duration is $T_{\mathrm{d}}=209.40$ ns. The system we simulate is defined by \equref{eq:architecture I} and \tabref{tab:device_parameter_flux_tunable_coupler_chip}. The $z=(0,0,1)\rightarrow z=(0,1,0)$ transition is often used to implement Iswap operations, see \REF\cite{Ganzhorn20}. We find that numerical accurate modelling of the dynamic behaviour of the system seems to require at least $N_{m}=6$ transmon basis states.}
        \label{fig:iswap_tunable_coupler}
    \end{minipage}
\end{figure}
\renewcommand{\width}{0.45}
\begin{figure}[!tbp]
    \centering
    \begin{minipage}{0.5\textwidth}
        \centering
        \includegraphics[width=\width\textwidth]{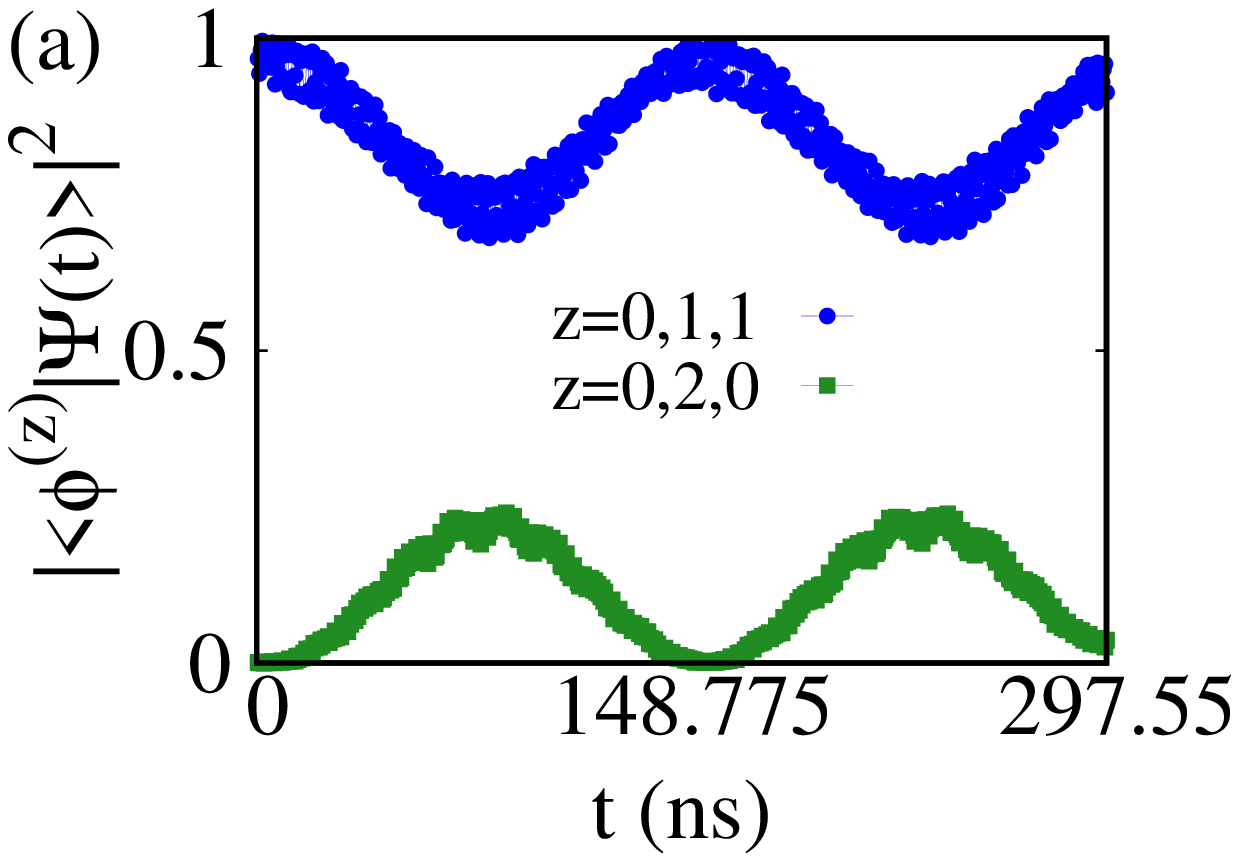}
        \includegraphics[width=\width\textwidth]{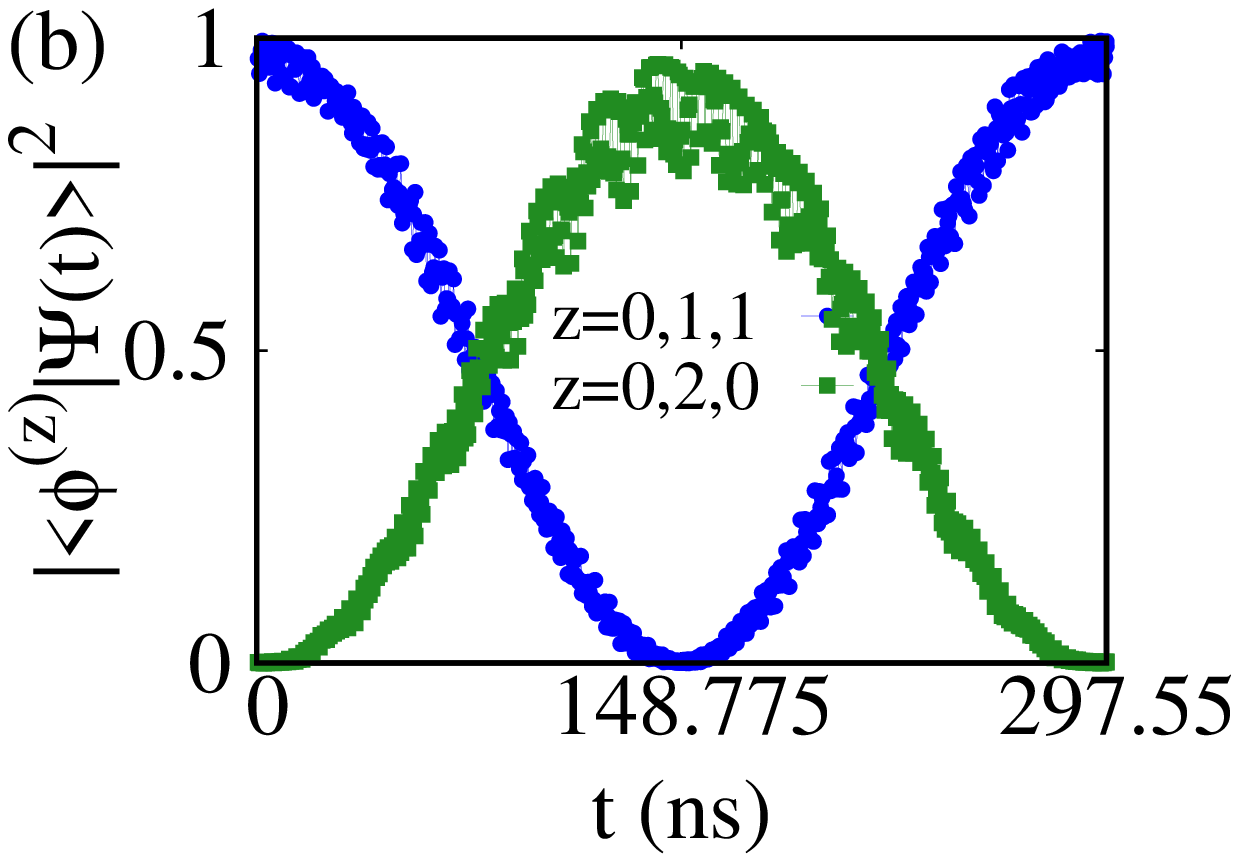}
        \includegraphics[width=\width\textwidth]{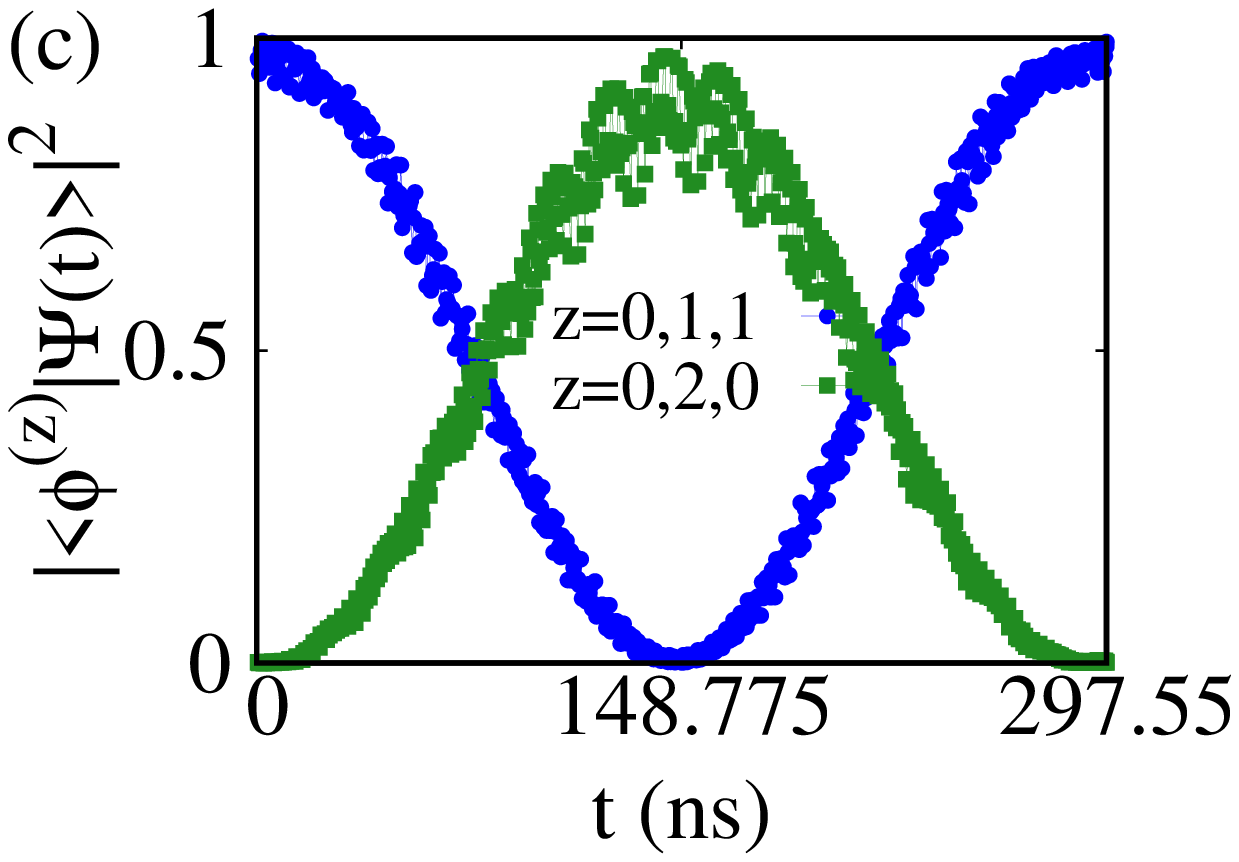}
        \includegraphics[width=\width\textwidth]{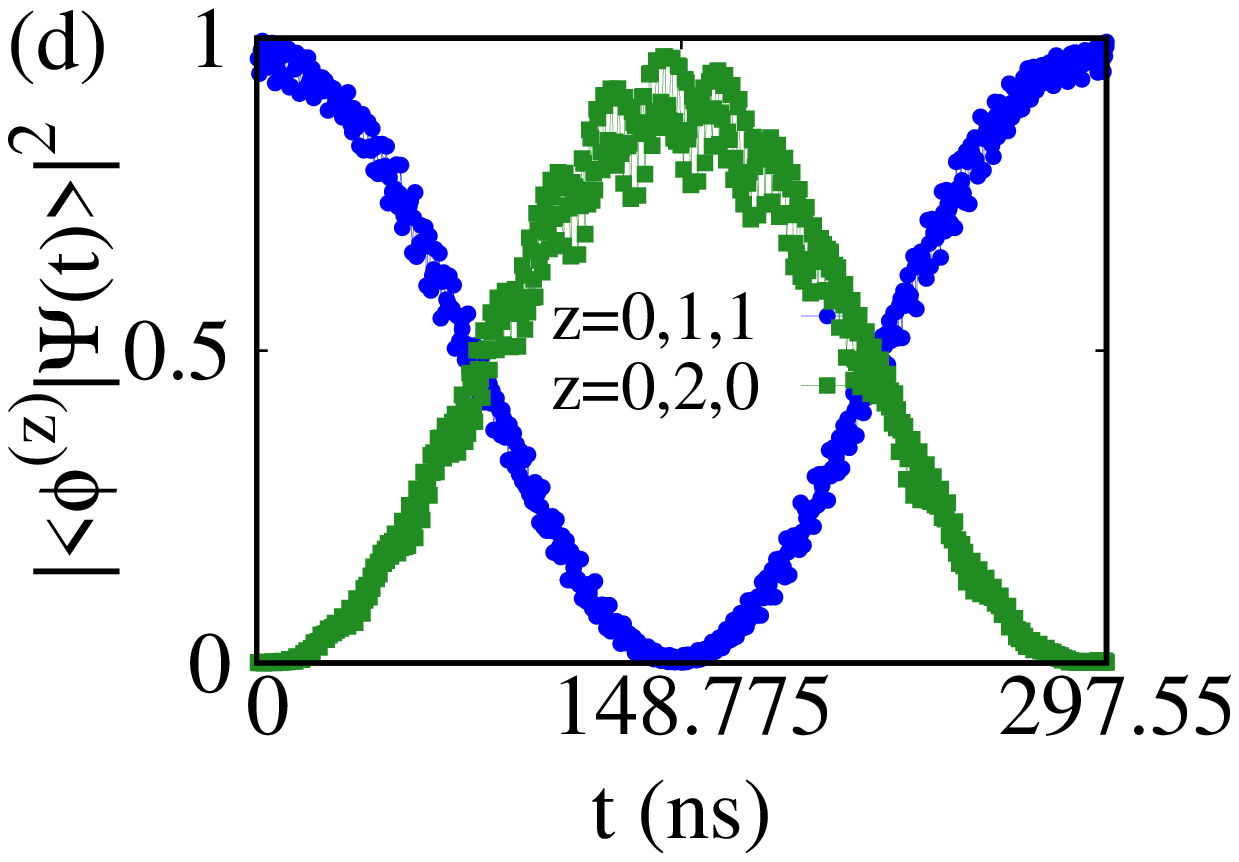}
        \caption{(Color online) Probabilities $p^{(0,1,1)}(t)$ and $p^{(0,2,0)}(t)$ as functions of time $t$. We use $N_{m}=3$ (a), $N_{m}=4$ (b), $N_{m}=8$ (c) and $N_{m}=15$ (d) basis states to model the system. We use a control pulse of the form \equref{eq:pulse}, with the pulse parameters $\omega^{\mathrm{D}}=0.809$ GHz, $T_{\mathrm{r/f}}=13$ ns and $\delta/2\pi=0.085$. The pulse duration is $T_{\mathrm{d}}=297.55$ ns. The system we simulate is defined by \equref{eq:architecture I} and \tabref{tab:device_parameter_flux_tunable_coupler_chip}. The $z=(0,1,1)\rightarrow z=(0,2,0)$ transition is usually used to implement Cz operations, see \REFS\cite{Ganzhorn20,Bengtsson2020}. We find that numerical accurate modelling of the dynamic behaviour of the system seems to require at least $N_{m}=8$ transmon basis states.}
        \label{fig:cz_tunable_coupler}
    \end{minipage}
\end{figure}

Figures~\ref{fig:cz_tunable_coupler}(a-d) show the time evolution of the probabilities $p^{(0,1,1)}(t)$ and $p^{(0,2,0)}(t)$ as a function of time $t$. We use $N_{m}=3$ (a), $N_{m}=4$ (b), $N_{m}=8$ (c) and $N_{m}=15$ (d) transmon basis states to model the system. This transition is often used to implement a Cz operation, see \REF\cite{Bengtsson2020}. The corresponding drive frequency is $\omega^{\mathrm{D}}=0.809$ GHz, which corresponds roughly to the energy difference, in GHz, of the two states involved. The pulse amplitude is $\delta/2\pi = 0.085$.

We observe that if we model this particular Cz operation, we find severe qualitative and quantitative deviations between the solutions (a) and (b) and (c)/(d) . Here we should use eight basis states to accurately model the dynamics of the system.

The Iswap and Cz operations we studied here are implemented with gate durations $T_{\mathrm{d}}$ between $200$ and $300$ ns. It is possible to implement shorter gate durations, by increasing the amplitude (data not shown). However, this almost always means we have to increase the number of basis states $N_{m}$ to obtain an accurate solution.

Furthermore, we repeated the same analysis for two additional devices. The corresponding device parameters were motivated by experiments carried out by the authors of \REFS\cite{Roth19,Roth20,Bengtsson2020}. Here we found similar results (data not shown), namely that we need at least six or eight basis states to describe Iswap and Cz operations, with similar gate durations.

The results we obtained for the Iswap and Cz gates indicate that the influence of the higher levels $\{\ket{\phi^{m>2}}\}$ on the subspace $\{\ket{\phi^{m\leq2}}\}$ is not negligible when it comes to modelling these operations. It seems to be the case that higher levels are instrumental in providing enough interaction strength, between the different subsystems, so that we can actually implement the operations (see \figsref{fig:iswap_tunable_coupler}(a-b) and \figsref{fig:cz_tunable_coupler}(a-b) in particular). Additionally, we can observe the trend that larger amplitudes seem to require more basis states $N_{m}$. Of course, all previous statements have to be restricted to the specific circuit Hamiltonian we studied here.

\subsubsection{Architecture II}

The second system we consider is defined by means of the circuit Hamiltonian \equref{eq:architecture II} and the parameters listed in \tabref{tab:device_parameter_resonator_coupler_chip}. Here we use a unimodal pulse (we set $\omega^{D}=0$) of the form \equref{eq:pulse} to implement two-qubit operations. Note that we apply the control pulse to the second flux-tunable transmon (see \tabref{tab:device_parameter_resonator_coupler_chip} row $i=1$).

\renewcommand{\width}{0.45}
\begin{figure}[!tbp]
    \centering
    \begin{minipage}{0.5\textwidth}
        \centering
        \includegraphics[width=\width\textwidth]{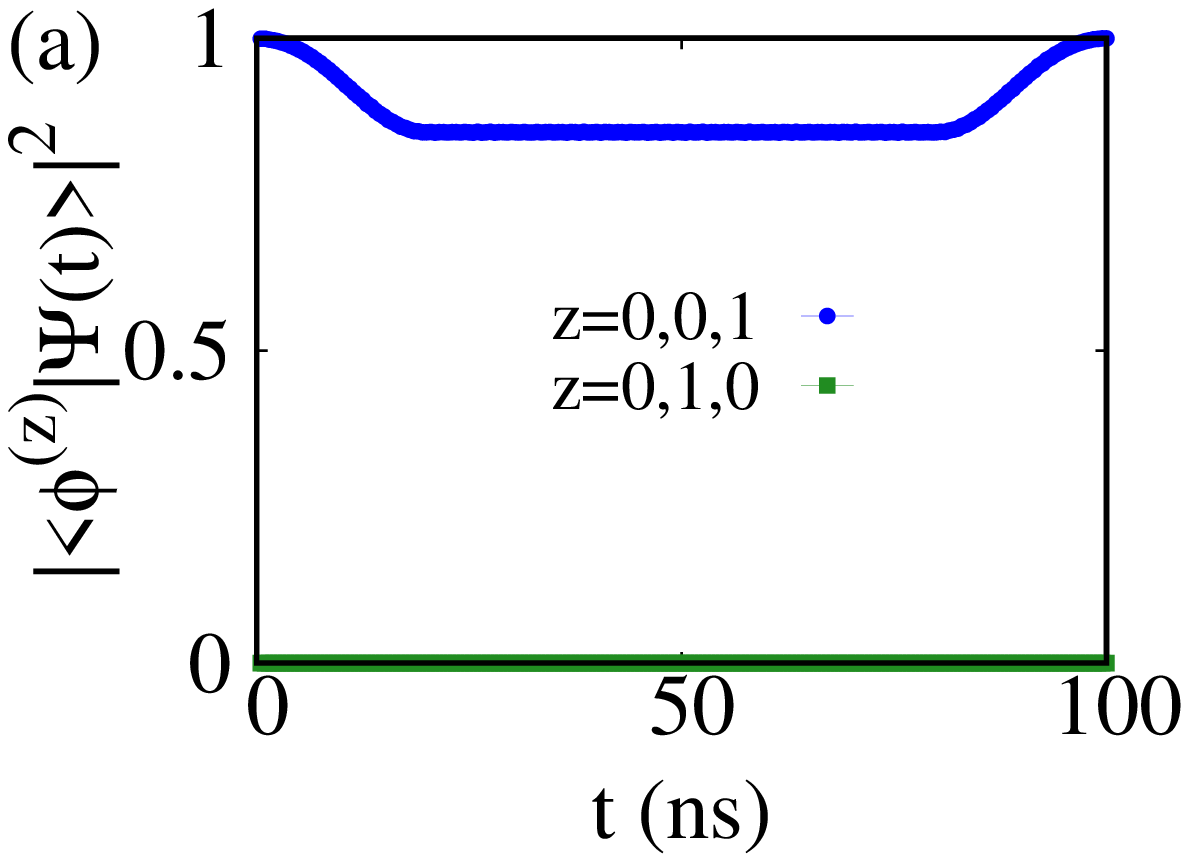} 
        \includegraphics[width=\width\textwidth]{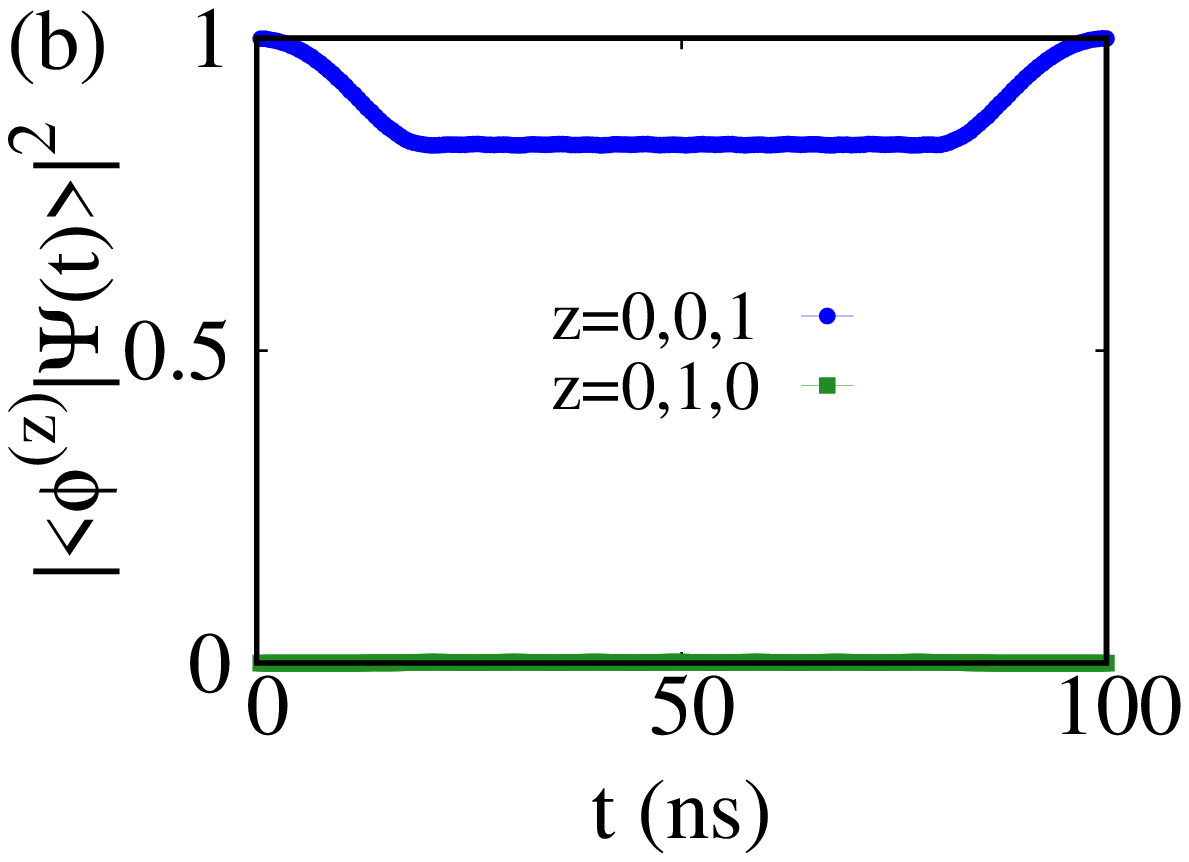} 
        \includegraphics[width=\width\textwidth]{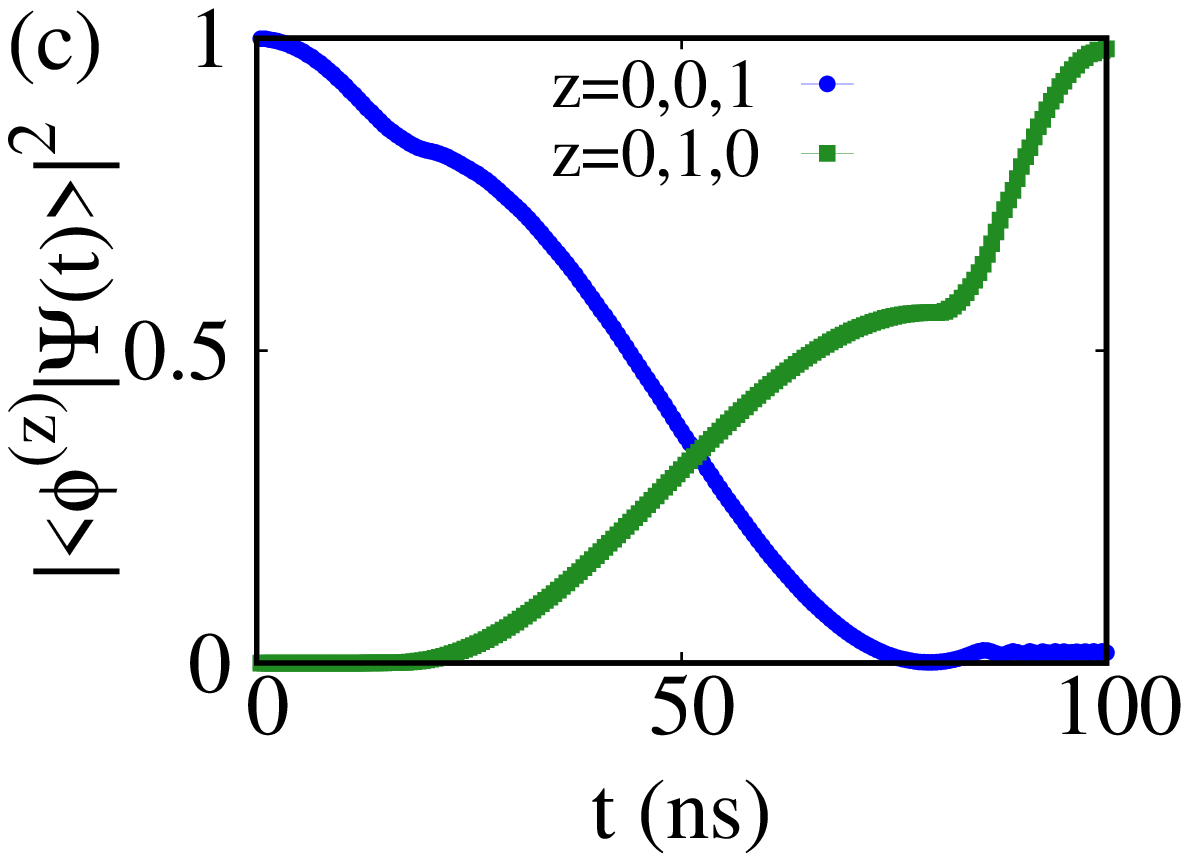} 
        \includegraphics[width=\width\textwidth]{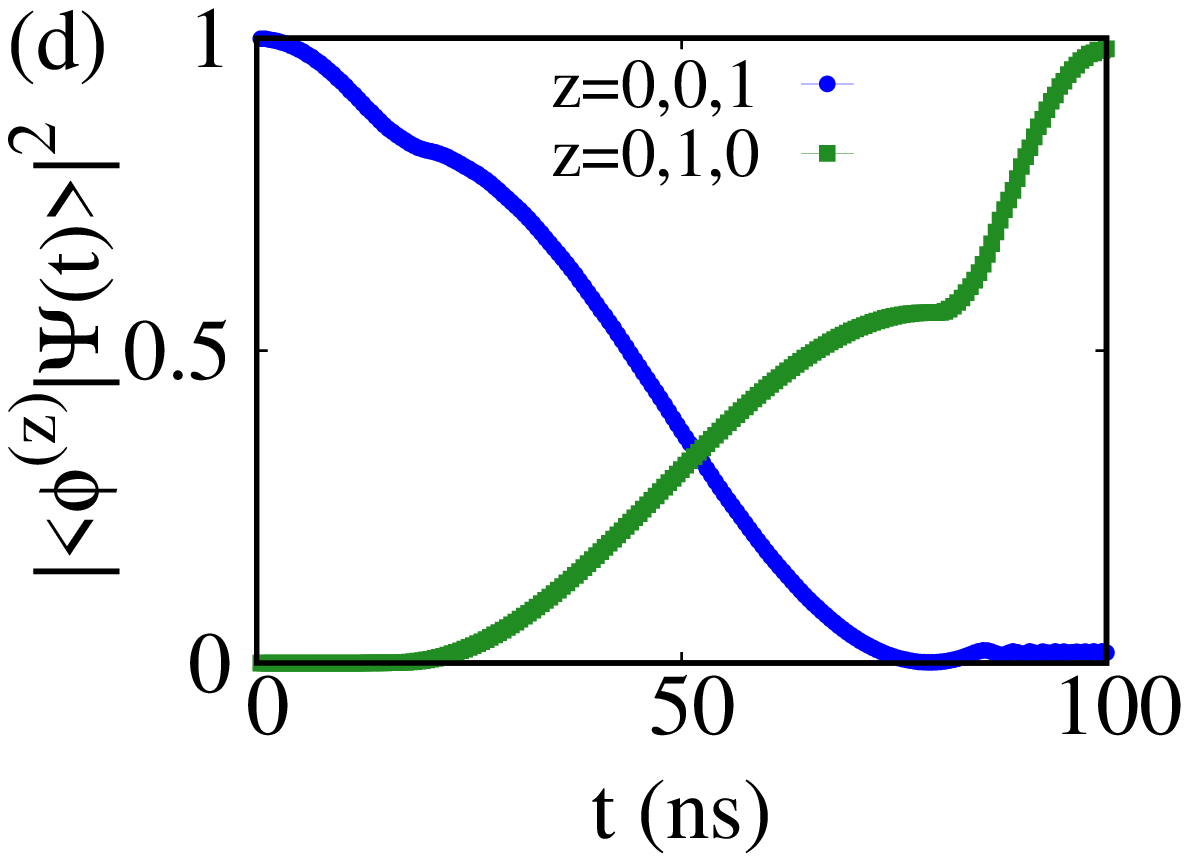} 
        \caption{(Color online) Probabilities $p^{(0,0,1)}(t)$ and $p^{(0,1,0)}(t)$ as functions of time $t$. We use $N_{m}=3$ (a), $N_{m}=4$ (b), $N_{m}=14$ (c) and $N_{m}=25$ (d) basis states to model the system. We use a control pulse of the form \equref{eq:pulse}, with the pulse parameters $\omega^{D}=0$ GHz, $T_{\mathrm{r/f}}=20$ ns and $\delta/2\pi=0.289$. The pulse duration is $T_{\mathrm{d}}=100.00$ ns. The pulse is supposed to perform an Iswap gate. The system we simulate is defined by \equref{eq:architecture II} and \tabref{tab:device_parameter_resonator_coupler_chip}. The $z=(0,0,1)\rightarrow z=(0,1,0)$ transition might be used to implement Iswap operations. Note that solutions in panel (a) and (b) do not have much in common with the reference solutions in panels (c) and (d).}
        \label{fig:iswap_res_coupler}
    \end{minipage}
\end{figure}
\renewcommand{\width}{0.45}
\begin{figure}[!tbp]
    \centering
    \begin{minipage}{0.5\textwidth}
        \centering
        \includegraphics[width=\width\textwidth]{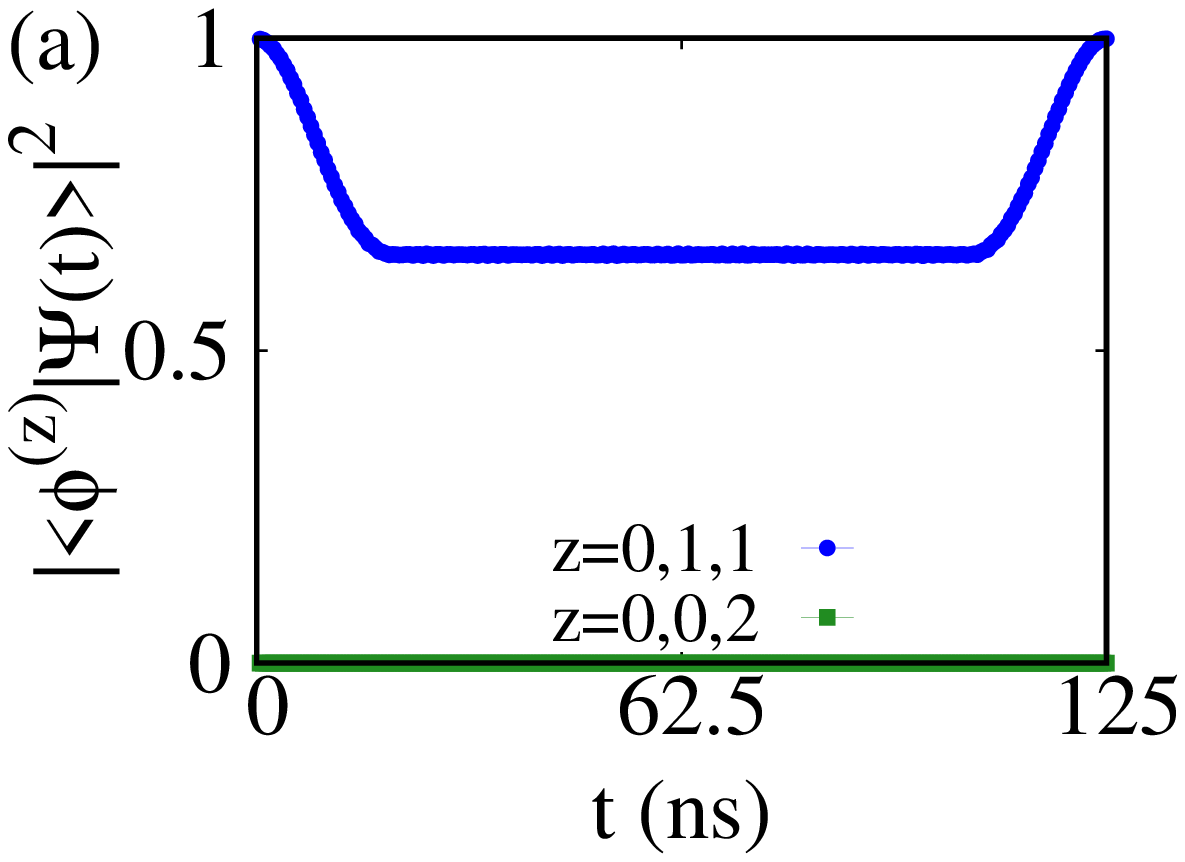}
        \includegraphics[width=\width\textwidth]{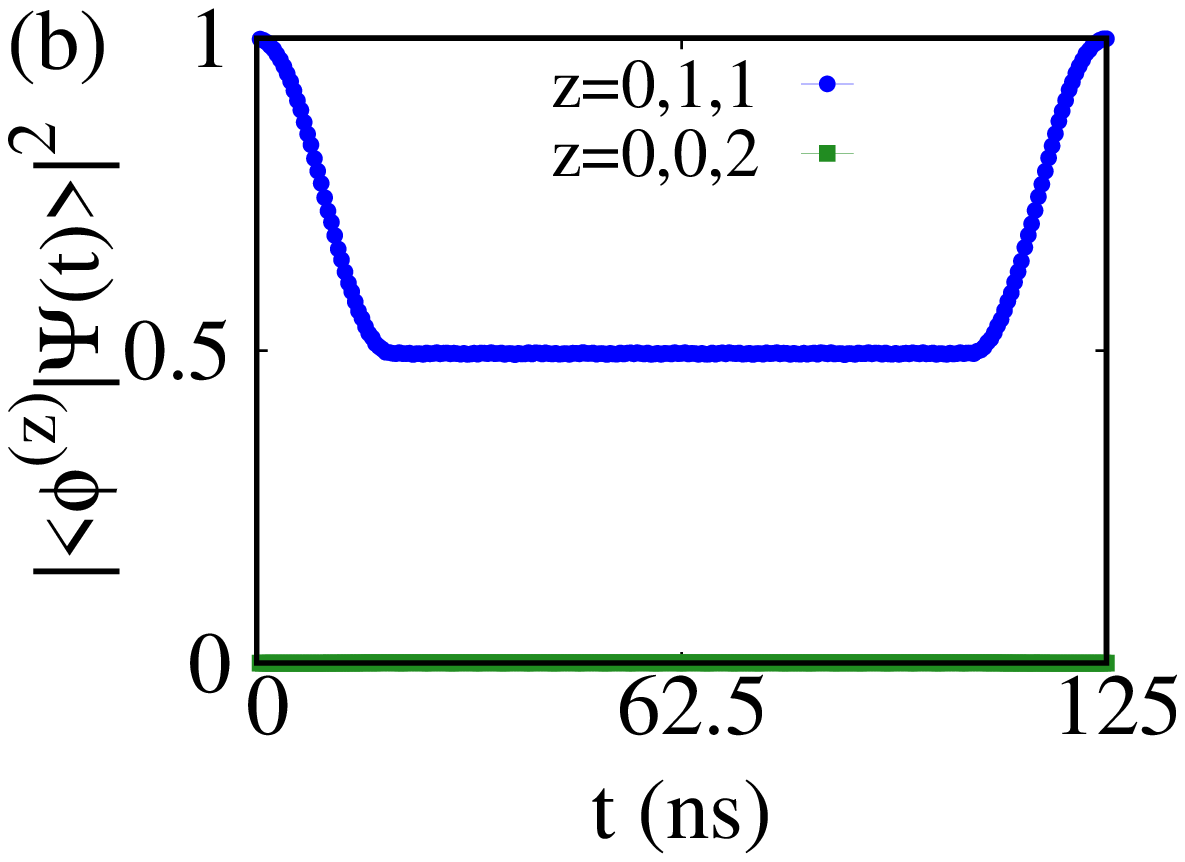}
        \includegraphics[width=\width\textwidth]{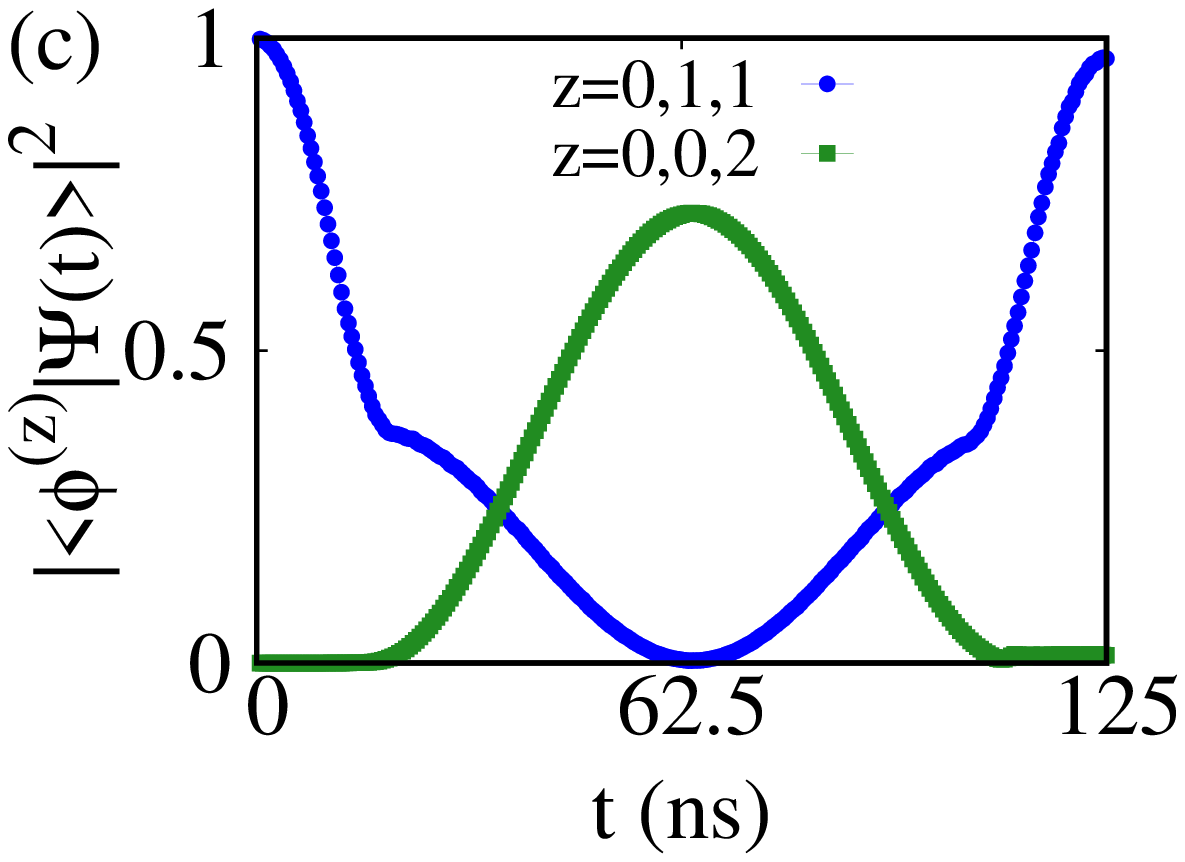}
        \includegraphics[width=\width\textwidth]{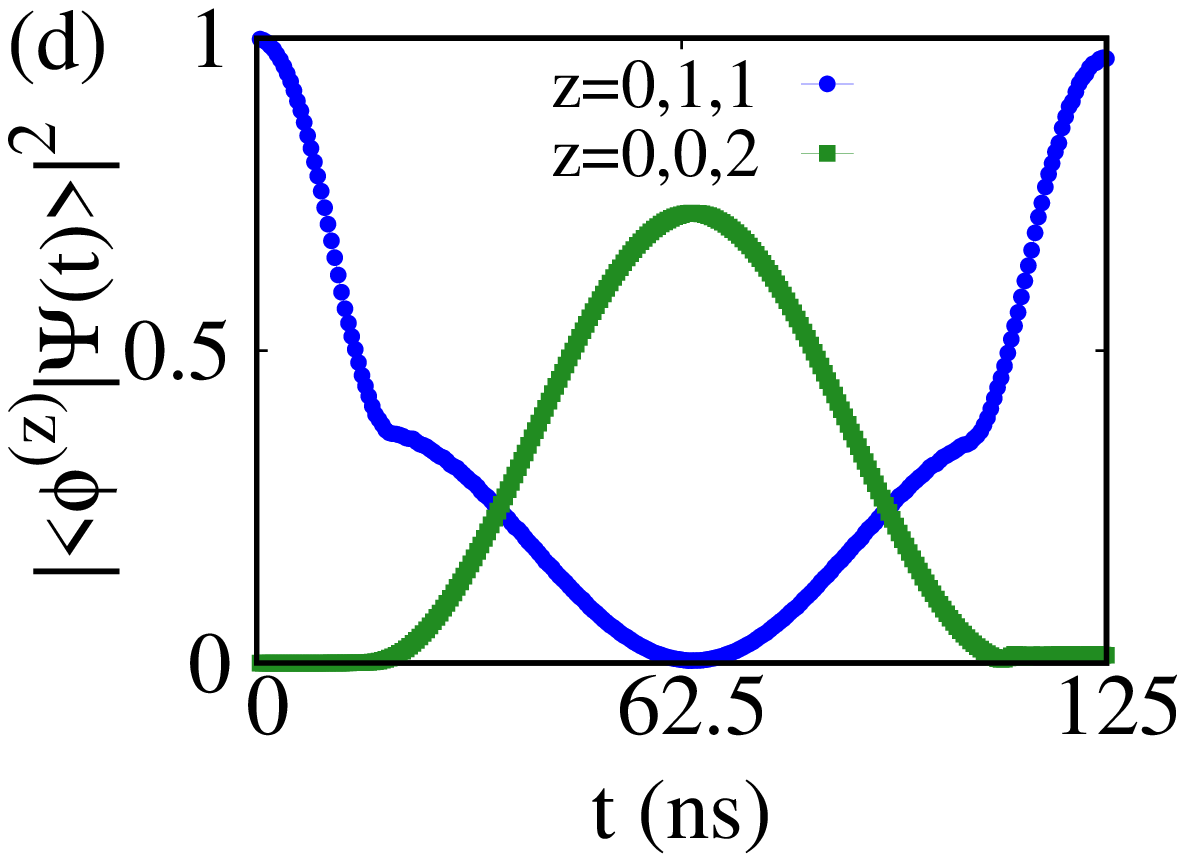}
        \caption{(Color online) Probabilities $p^{(0,1,1)}(t)$ and $p^{(0,0,2)}(t)$ as functions of time $t$. We use $N_{m}=3$ (a), $N_{m}=4$ (b), $N_{m}=16$ (c) and $N_{m}=25$ (d) basis states to model the system. We use a control pulse of the form \equref{eq:pulse}, with the pulse parameters $\omega^{\mathrm{D}}=0$ GHz, $T_{\mathrm{r/f}}=20$ ns and $\delta/2\pi=0.3335$. The pulse duration is $T_{\mathrm{d}}=125.00$ ns. The pulse is supposed to perform a Cz gate. The system we simulate is defined by \equref{eq:architecture II} and \tabref{tab:device_parameter_resonator_coupler_chip}. The $z=(0,1,1)\rightarrow z=(0,0,2)$ transition can be used to implement Cz operations, see \REF\cite{Lacroix2020}. Note that solutions in panel (a) and (b) do not have much in common with the reference solutions in panels (c) and (d).}
        \label{fig:cz_res_coupler}
    \end{minipage}
\end{figure}

Figures~\ref{fig:iswap_res_coupler}(a-d) show the time evolution of $p^{(0,0,1)}(t)$ and $p^{(0,1,0)}(t)$ as functions of time $t$, for four different numbers of basis states $N_{m}=3$ (a), $N_{m}=4$ (b), $N_{m}=14$ (c) and $N_{m}=25$ (d). We
model a transition of the Iswap type.

Obviously, \figsref{fig:iswap_res_coupler}(a,b) have not much in common with the reference solutions (c)/(d). This means that if we use three or four states to model the system, we are not able to implement an Iswap gate on this architecture. We need about fourteen states to model this operation adequately. Additionally, note that during the time evolution $ p^{(0,1,0)}(t)+p^{(0,0,1)}(t)\neq 1$ for various times $t$. The reason for this is that continuous population transfer takes place in the instantaneous basis.

The last case we study is the Cz gate, implemented on architecture II. Figures~\ref{fig:cz_res_coupler}(a-d) show the time evolution of the probabilities $p^{(0,1,1)}(t)$ and $p^{(0,0,2)}(t)$ as functions of time $t$, for $N_{m}=3$ (a), $N_{m}=4$ (b), $N_{m}=16$ (c) and $N_{m}=25$ (d).

In this case we implemented a slightly imperfect Cz operation, i.e., we implemented a pulse which ensures that $p^{(0,1,1)}(T_{\mathrm{d}})<1$. A perfect Cz gate would only change the relative phase of the state vector but not the population. Therefore, modelling the system with three basis states would yield the same result as modelling the system with 25 states (see \figref{fig:cz_res_coupler}(a)), i.e., it does not matter whether or not population exchange actually occurs. However, we want to determine the number of basis states which are needed to model the transitions $z=(0,1,1) \rightarrow (0,0,2)$ and $z=(0,0,2) \rightarrow (0,1,1)$. The easiest way to do this is to implement a slightly imperfect transition.


\bibliographystyle{apsrev4-2}
\bibliography{ms}

\end{document}